\documentclass[11pt]{article}

\usepackage{appendix}

\usepackage{epsfig}
\usepackage{graphicx}

\usepackage{caption}

\usepackage{latexsym,times}
\usepackage{amsmath}
\usepackage{amsxtra}
\usepackage{amscd}
\usepackage{amsthm}
\usepackage{amsfonts}
\usepackage{amssymb}
\usepackage{dsfont}
\usepackage{xcolor}
\usepackage{cancel}
\usepackage{framed}

\usepackage{enumitem}
\usepackage{float}

\usepackage{hyperref}
\usepackage[square,sort&compress,comma,numbers]{natbib}

\makeatletter

\@addtoreset{equation}{section} \makeatother

\newcommand\rphi{r_{\!\scriptscriptstyle{\mathbf{\phi}}}}
\newcommand\rM{r_{\!{\scriptscriptstyle{M}}}}
\newcommand\rO{r_{\!{\scriptscriptstyle{\Omega}}}}
\newcommand\sgn{\widehat{\varepsilon}_{\!\scriptscriptstyle{\mathbf{\phi\,}}}}

\newcommand\So{S_{\scriptscriptstyle{{(0)}}}}
\newcommand\To{\EM_{\scriptscriptstyle{{(0)}}}}
\newcommand\intd{{{\left({\int_{\!\vartheta}\int_{\!\varphi}\! e^{-2d\,}}\right)}}}
\newcommand\intdn{{{\dis{\int_{\!\vartheta}\int_{\!\varphi}\! e^{-2d\,}}}}}

\newcommand\rc{r_{\rm{c}}}

\newcommand\hephi{he^{-2\phio}}

\newcommand{\half}{{{\textstyle\frac{1}{2}}}}
\newcommand{\quarter}{{{\textstyle\frac{1}{4}}}}
\newcommand{\be}{\begin{equation}}
\newcommand{\ee}{\end{equation} }
\newcommand{\beqa}{\begin{eqnarray} }
\newcommand{\eeqa}{\end{eqnarray} }
\newcommand{\ba}{\begin{array}}
\newcommand{\ea}{\end{array}}
\newcommand{\bpm}{\begin{pmatrix}}
\newcommand{\epm}{\end{pmatrix}}

\newcommand{\so}{\mathbf{so}}

\newcommand{\Spin}{\mathbf{Spin}}

\newcommand{\rmc}{{\rm c}}
\newcommand{\rmd}{{\rm d}}

\newcommand{\ODD}{\mathbf{O}(D,D)}

\newcommand{\SpinD}{{\Spin(1,D{-1})}}
\newcommand{\oSpinD}{{{\Spin}(D{-1},1)}}

\newcommand{\Spint}{{\Spin(1,9)}}
\newcommand{\oSpint}{{{\Spin}(9,1)}}

\newcommand{\eleven}{{(11)}}
\newcommand{\gammae}{\gamma^{\eleven}}

\newcommand{\brcF}{\bar{\cF}}

\newcommand\EM{T}
\newcommand\EK{K}

\newcommand\diag{\mbox{diag}}

\newcommand\mPhi{m_{\scriptscriptstyle{\Phi}}}
\newcommand\mpsi{m_{\scriptscriptstyle{\psi}}}

\newcommand{\DFT}{\rm{DFT}}

\newcommand\Tr{{\rm Tr}}

\newcommand\rd{{\rm d}}
\newcommand\rD{{\rm D}}

\newcommand\bfA{{\mathbf{A}}}
\newcommand\bfF{{\mathbf{F}}}

\newcommand\cA{{\cal A}}

\newcommand\cC{{\cal C}}
\newcommand\cD{{\cal D}}

\newcommand\cF{{\cal F}}

\newcommand\cH{{\cal H}}
\newcommand\cI{{\cal I}}
\newcommand\cJ{{\cal J}}
\newcommand\cK{{\cal K}}
\newcommand\cL{{\cal L}}

\newcommand\cP{{\cal P}}
\newcommand\cQ{{\cal Q}}

\newcommand\cV{{\cal V}}
\newcommand\cW{{\cal W}}
\newcommand\cX{{\cal X}}
\newcommand\cY{{\cal Y}}
\newcommand\cZ{{\cal Z}}

\newcommand\hcW{\widehat{\cW}}

\newcommand\hcX{\widehat{\cX}}

\newcommand\brcP{{\bar{\cal{P}}}}

\newcommand\hcL{{\hat{\cal L}}}

\newcommand\fcL{{{\widetilde{\cal L}}}}

\newcommand\rhop{{\rho^{\prime}}{}}

\newcommand\psip{\psi^{\prime}}

\newcommand\varepsilonp{\varepsilon^{\prime}{}}

\newcommand\h{{h}}

\newcommand\dis{\displaystyle}

\def\tx{\tilde{x}}
\def\ty{\tilde{y}}
\def\tpartial{\tilde{\partial}}

\def\bre{\bar{e}}
\def\bri{\bar{\imath}}
\def\brj{\bar{\jmath}}

\def\breta{\bar{\eta}}
\def\bralpha{\bar{\alpha}}
\def\brbeta{\bar{\beta}}
\def\brgamma{\bar{\gamma}}
\def\brdelta{\bar{\delta}}

\def\brpsi{\bar{\psi}}

\def\brn{{\bar{n}}}
\def\brp{{\bar{p}}}
\def\brq{{\bar{q}}}
\def\brr{{\bar{r}}}
\def\brs{{\bar{s}}}

\def\bromega{{\bar{\omega}}}

\def\brPhi{{{\bar{\Phi}}}}

\def\brC{\bar{C}}

\def\brV{{\bar{V}}}

\def\brX{{\bar{X}}}
\def\brY{{\bar{Y}}}
\def\brP{{\bar{P}}}

\def\brtheta{\bar{\theta}}

\newcommand{\DO}{\mathbf{\nabla}}

\newcommand{\na}{{\nabla}}
\newcommand{\trd}{{\bigtriangledown}}

\def\phio{\phi_{\scriptscriptstyle{0}}}

\newcommand\p\partial

\begin{document}

\begin{titlepage}
\title{\bf Einstein  \,  Double\,  Field\,  Equations}
\author{\sc Stephen Angus,\,   \,Kyoungho Cho  \,\,and\,\, Jeong-Hyuck Park}

\date{}
\maketitle 
\begin{center}
Department of Physics, Sogang University, 35 Baekbeom-ro, Mapo-gu,  Seoul  04107, KOREA\\
~\\
\texttt{stephenangus4@gmail.com\,\quad khcho23@sogang.ac.kr\, \quad park@sogang.ac.kr}\\
~~~\\
\end{center}
\begin{abstract}
\noindent Upon treating   the whole  closed string massless sector   as     stringy graviton fields,  Double Field Theory may  evolve into Stringy Gravity, \textit{i.e.~}the stringy augmentation  of   General Relativity.  Equipped with an $\mathrm{O}(D,D)$ covariant  differential geometry beyond Riemann,  we spell out  the  definition of       the       Energy-Momentum tensor in Stringy Gravity  and   derive its on-shell conservation law from  doubled  general covariance.  Equating it with  the recently identified   stringy Einstein  curvature   tensor,  all the  equations of motion of the closed string massless sector are unified into a single expression, $G_{AB}=8\pi G T_{AB}$,  which we dub the \textit{Einstein Double Field Equations}.    As an example, we study    the most general ${D=4}$ static,  asymptotically flat,   spherically symmetric,   `regular' solution,   sourced by  the   stringy  Energy-Momentum tensor which is nontrivial   only  up to a finite radius from the center. Outside this radius, the solution matches  the   known  vacuum geometry which has  four constant  parameters.  We   express  these   as volume integrals of the interior stringy Energy-Momentum tensor and discuss   relevant energy conditions.

\end{abstract} 
\thispagestyle{empty}
\end{titlepage}
\newpage

\begin{flushright}
\textit{One must be prepared to follow up the consequence of theory, and feel that\\ one just has to accept the consequences no matter where they lead.}\\
Paul Dirac\\
~\\
\textit{Our mistake is not that we take our theories too seriously, but that we do not take them seriously enough.}\\
Steven Weinberg
\end{flushright}
~\\

\tableofcontents 



\newpage

\section{Introduction\label{SECintro}}
The Einstein-Hilbert action is often referred to as `pure' gravity, as  it is  formed  by the unique two-derivative scalar curvature  of   the Riemannian metric.  Minimal coupling to  matter    follows  unambiguously  through  the usual covariant derivatives,
\be
\ba{lll}
\trd_{\mu}=\partial_{\mu}+\gamma_{\mu}+\omega_{\mu}\,,~~&~~~ \gamma_{\mu\sigma}^{\rho}=
\half g^{\rho\tau}(\partial_{\mu}g_{\sigma\tau}+\partial_{\sigma}g_{\mu\tau}-\partial_{\tau}g_{\mu\sigma})\,,~~&~~~
\omega_{\mu pq}=e_{p}{}^{\nu}(\partial_{\mu}e_{\nu q}-\gamma_{\mu\nu}^{\lambda}e_{\lambda q})\,,
\ea
\label{trd}
\ee
which ensures covariance under  both diffeomorphisms and local Lorentz symmetry. In the words of Cheng-Ning Yang, \textit{symmetry dictates interaction.} The torsionless  Christoffel symbols of the connection  and the spin connection are fixed  by  the requirement of  compatibility with the metric and the vielbein.   The   existence of  Riemann normal coordinates supports  the Equivalence Principle, as  the  Christoffel symbols  vanish pointwise. Needless to say,  in General Relativity (GR), the metric is privileged  to be the only geometric and thus gravitational field,   on account of the adopted  differential geometry \textit{a la} Riemann, while all other fields are automatically categorized as additional  matter.\\

\noindent String theory may   put some twist on this Riemannian  paradigm.  First of all,  the metric is  merely   one  segment of closed string massless sector which consists of a two-form gauge potential, $B_{\mu\nu}$, and a scalar dilaton, $\phi$,  in addition to  the metric, $g_{\mu\nu}$. A genuine stringy symmetry called T-duality then converts one   to  another~\cite{Buscher:1987sk,Buscher:1987qj}. Namely, the closed string massless sector forms multiplets of $\ODD$ T-duality.  This may well hint at the existence of Stringy Gravity as an alternative to GR, which takes the entire massless sector as geometric and therefore gravitational.  In recent  years this idea has been   realized  concretely   through the developments  of so-called  Double Field Theory (DFT)~\cite{Siegel:1993xq,Siegel:1993th,Hull:2009mi,Hull:2009zb,Hohm:2010jy,Hohm:2010pp}.   The relevant covariant derivative has been identified~\cite{Jeon:2011cn,Jeon:2011vx} and reads schematically, 
\be
\cD_{A}=\partial_{A}+\Gamma_{A}+\Phi_{A}+\brPhi_{A}\,,
\ee
where $\Gamma_{A}$ is the DFT version of the Christoffel symbols for  generalized  diffeomorphisms, while $\Phi_{A}$ and $\brPhi_{A}$ are the two spin connections  for the twofold local Lorentz symmetries, $\SpinD\times\oSpinD$. They are compatible with, and  thus   formed by,   the  closed string  massless sector,  containing in particular the $H$-flux ($H=\rd B$).   The doubling of the spin group  implies the existence of   two separate locally inertial frames   for  left and  right  closed string modes, respectively~\cite{Duff:1986ne}. In a sense,  it   is  a prediction of DFT  (and also Generalized Geometry~\cite{Coimbra:2011nw}) that   there must in principle exist two distinct kinds of fermions~\cite{Choi:2015bga}.  The DFT-Christoffel symbols constitute DFT curvatures: scalar and  `Ricci'.   The scalar  curvature naturally   defines  the pure DFT Lagrangian in analogy with GR.  However,  in Stringy Gravity the Equivalence Principle is generically  broken~\cite{Choi:2015bga,Park:2017snt}:  there exist  no normal coordinates in which  the DFT-Christoffel symbols  would vanish pointwise.  This should not be a surprise since, strictly speaking,  the principle holds only  for a point particle  and does not apply to 
 an extended object like a string, which is subject to `tidal forces' via coupling to the $H$-flux.   \\

\noindent Beyond  the original  goal of reformulating supergravities in a duality-manifest framework,   DFT  turns out  to  have quite a rich spectrum.  It  describes  not only  the Riemannian supergravities but also various non-Riemannian theories in which the Riemannian metric cannot be defined~\cite{Morand:2017fnv}, such as  non-relativistic Newton--Cartan or ultra-relativistic  Carroll gravities~\cite{BergshoeffSimons}, the Gomis--Ooguri non-relativistic string~\cite{Gomis:2000bd,Ko:2015rha},  and various chiral theories including the one by Siegel~\cite{Siegel:2015axg}.  Without resorting to Riemannian variables,     supersymmetrizations have been also completed  to the full order in fermions,   both  on target spacetime~\cite{Jeon:2011sq,Jeon:2012hp} and on worldsheet~\cite{Park:2016sbw}.\\

\noindent  Combining the scalar and `Ricci' curvatures,  the DFT version of the Einstein curvature, $G_{AB}$, which is  identically conserved, $\cD_{A}G^{A}{}_{B}=0$, and generically asymmetric, $G_{AB}\neq G_{BA}$,  has been identified~\cite{Park:2015bza}.   Given this identification,  it is natural to anticipate the `Energy-Momentum' tensor in DFT, say  $\EM_{AB}$,  which should counterbalance the stringy  Einstein curvature through the  \textit{Einstein Double Field Equations,} \textit{i.e.~}the equations of motion of the entire  closed string massless sector as the stringy graviton  fields,
\be
G_{AB}=8\pi G  \, \EM_{AB}\,,
\label{EDFEo}
\ee 
where $G$ (without any subscript index) denotes Newton's constant.       For consistency, the stringy  Energy-Momentum tensor should be asymmetric, $\EM_{AB}\neq\EM_{BA}$,  and  conserved,  $\cD_{A}T^{A}{}_{B}= 0$,   especially on-shell, \textit{i.e.~}up to the equations of motion of the  additional matter fields.  \\

\noindent In order to compare the `gravitational' aspects  of  DFT  and GR,     circular geodesic motions around  the most general spherically symmetric  solution to  `${G_{AB}=0}$'  have  been studied in \cite{Ko:2016dxa} for the case of ${D=4}$.  While the solution was a re-derivation of a  previously known result in the supergravity literature~\cite{Burgess:1994kq}, the new interpretation was that it is  the `vacuum' solution to DFT,  with the  right-hand side of (\ref{EDFEo}) vanishing: it is analogous to   the Schwarzschild solution in GR.   The DFT spherical vacuum solution turns out to have  four (or three, up to a radial coordinate shift) free parameters,  in contrast to the   Schwarzschild geometry which possesses only one free parameter, \textit{i.e.~}mass.  With these  extra  free parameters,  DFT  modifies  GR at `short' scales in terms of the dimensionless parameter $R/(MG)$,  \textit{i.e.~}the radial distance  normalized by the mass times Newton's constant.  For large   $R/(MG)$,  DFT converges to GR, but   for finite $R/(MG)$ they differ generically. It is an intriguing fact that     the dark matter and dark energy problems     all arise from   astronomical observations at  smaller $R/(MG) \lesssim 10^{7}$, corresponding to  long distance divided by far heavier mass~\cite{Ko:2016dxa,Park:2017snt}.    Such a  `uroboros'  spectrum of $R/(MG)$ is listed below in natural units.
\vspace{-2pt}
\begin{table}[H]
\includegraphics[width=16.5cm]{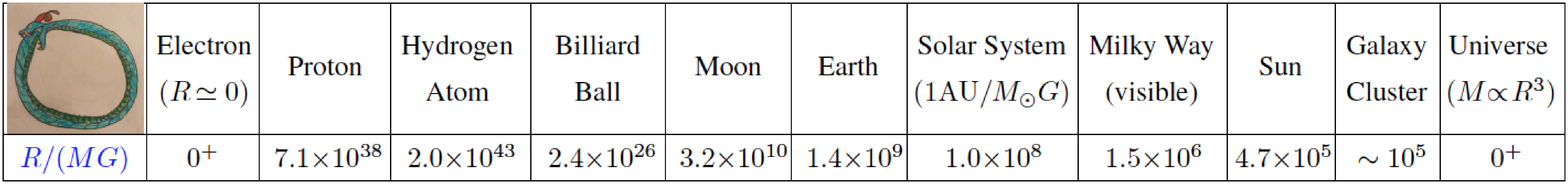}
\vspace{-2pt}
\vspace{-2pt}
\end{table}
\vspace{-20pt}


\noindent The purpose of the present paper is twofold: \textit{i)}  to propose  the definition of  the stringy  Energy-Momentum tensor which completes the Einstein Double Field Equations spelled out  in  (\ref{EDFEo}),  and  \textit{ii)} to  analyze   the  most general spherically symmetric ${D=4}$  `regular' solution  which will teach us  the physical meanings of the free parameters appearing  in the vacuum solution of \cite{Ko:2016dxa,Burgess:1994kq}.  The rest of the paper is organized as follows.
\begin{itemize}

\item[--] We start   section~\ref{SECDFTEM} by reviewing  DFT as Stringy Gravity. We  then consider  coupling  to generic matter fields,   propose  the definition of  the stringy  Energy-Momentum tensor,  and discuss its properties including the conservation law. Some examples will follow.

\item[--] In section~\ref{SECFURTHER} we devise a method to     address  isometries  in  the vielbein formulation of Stringy Gravity.   We generalize the known generalized Lie derivative one step further, to  a `further-generalized Lie derivative', which acts not only on $\ODD$ vector indices but also on all the $\SpinD\times\oSpinD$ local Lorentz indices.  

\item[--] Section~\ref{SECD4} is devoted to the study of  the most general,  asymptotically flat, spherically symmetric, static   `regular' solution to the ${D=4}$ Einstein Double Field Equations.  We postulate that  the stringy Energy-Momentum tensor is nontrivial only up to a finite  cutoff radius, $\rc$. While we recover the vacuum solution of \cite{Ko:2016dxa} for $r>\rc$, we  derive integral expressions for  its    constant parameters  in terms of  the stringy  Energy-Momentum tensor for $r<\rc$, and discuss relevant energy conditions.

\item[--] We conclude with our summary and  comments in section~\ref{SECFINAL}.

\item[--] In Appendix~\ref{SECAPPENDIX}  we collect some known features of GR, such as the general  properties of the Energy-Momentum tensor and  the most general spherically symmetric (Schwarzschild type)  regular solution to the undoubled  Einstein Field Equations, which  we  double-field-theorize  in the present paper.
\end{itemize}


\section{Einstein  Double Field Equations  \label{SECDFTEM}}

In this section we first give for completeness a self-contained review of DFT as Stringy Gravity, following which we propose the DFT, or stringy,  extensions of the Energy-Momentum tensor and the Einstein Field Equations.  \\

\subsection{Review of DFT as   Stringy Gravity}
We  review   DFT   following  the geometrically  logical---rather than historical---order: \textit{i)}  conventions,   \textit{ii)} the doubled-yet-gauged coordinate system with   associated  diffeomorphisms,  \textit{iii)}  the field content of stringy gravitons, \textit{iv)}  DFT extensions of the Christoffel symbols and spin connection, and \textit{v)}   covariant derivatives and curvatures.  For complementary  aspects,  we refer readers to   \cite{Aldazabal:2013sca,Berman:2013eva,Hohm:2013bwa} as well as \cite{Berman:2013uda,Cederwall:2014kxa}. \\

\indent  {$\bullet$ \textit{Symmetries and conventions}}\\
\noindent The built-in symmetries of  Stringy Gravity  are as follows.
\begin{itemize} 
\item[--] $\ODD$ T-duality  
\item[--] DFT diffeomorphisms
\item[--]Twofold local Lorentz symmetries,\footnote{In the most general case,  the two spin groups can  have   different dimensions~\cite{Morand:2017fnv}. }     $\SpinD\times\oSpinD$.
\end{itemize}
We shall use capital Latin letters, $A,B,\ldots, M,N,\ldots$ for the  $\ODD$ vector indices, while unbarred  small Latin letters, $p,q,\ldots$ or Greek letters, $\alpha,\beta,\ldots$ will be used for the vectorial or spinorial indices of $\SpinD$, respectively. Similarly,   barred letters denote the other $\oSpinD$ representations:  $\brp,\brq,\ldots$ (vectorial) and  $\bralpha,\brbeta,\ldots$ (spinorial).  In particular, 
each vectorial  index can be freely lowered or raised by the relevant  invariant metric,
\be
\ba{ccc}
\cJ_{AB}=\left(\ba{cc}0&1\\1&0\ea\right)\,,\quad&\quad
\eta_{pq}=\mbox{diag}(-++\cdots+)\,,\quad&\quad
\breta_{\brp\brq}=\mbox{diag}(+--\cdots-)\,.
\ea
\label{metric3}
\ee
~\\
\indent  {$\bullet$ \textit{Doubled-yet-gauged coordinates and diffeomorphisms}}\\
\noindent    By construction, functions admitted  to  Stringy Gravity are of special type.  Let  us denote the set of all the functions in Stringy Gravity by $\cF=\left\{\,\Phi_{i}\,\right\}$, which should include  not only   physical  fields   but also local symmetry parameters.   First of all,  each   function, $\Phi_{i}(x)$,  has  doubled coordinates, $x^{M}$, $M=1,2,\cdots,{D\!+\!D}$, as its  arguments. Not surprisingly,   the set   is closed under addition,  product and differentiation  such that, if $\,\Phi_{i}, \Phi_{j}\in\cF$ and  $a,b\in\mathbb{R}$,     then  
\be
\ba{lll}
 a\,\Phi_{i}+b\,\Phi_{j}\in\cF\,,\qquad&\quad
 \Phi_{i}\Phi_{j}\in\cF\,,\qquad&\quad\partial_{A}\Phi_{i}\in\cF\,,
 \ea
 \ee
and hence  $\Phi_{i}$ is $C^{\infty}$. The truly nontrivial  property of $\cF$ is that every function therein is invariant under a special class of translations: for arbitrary $\Phi_{i},\Phi_{j},\Phi_{k}\in\cF$,
\be
\ba{ll}
\Phi_{i}(x)=\Phi_{i}(x+\Delta)\,,\qquad&\qquad
\Delta^{M}=\Phi_{j}\partial^{M}\Phi_{k}\,,
\ea
\label{trinv}
\ee
where $\Delta^{M}$ is said to be {\it{derivative-index-valued}}. We emphasize that this very notion is only  possible thanks to the built-in $\ODD$  group structure,  whereby the invariant metric can raise the vector index of the partial derivative, $\partial^{M}=\cJ^{MN}\partial_{N}$. It is straightforward to show\footnote{Consider  the power series expansion of $\Phi_{i}(x+s\Delta)$  around $s=0$, where we have introduced a real parameter,  $s\in\mathbb{R}$. The linear-order term gives  $\partial_{M}\Phi_{i}\partial^{M}\Phi_{j}=0$, which in turn, after replacing  $\Phi_{i}$ and $\Phi_{j}$ by $\partial_{L}\Phi$ and $\partial_{N}\Phi$, implies that  $\partial_{L}\partial^{M}\Phi\partial_{M}\partial^{N}\Phi=0$. Consequently,  $\partial_{M}\partial^{N}\Phi$ is a nilpotent  matrix and thus must be  traceless, $\partial_{M}\partial^{M}\Phi=0$~\cite{Lee:2013hma}.} that the  above  translational invariance  is equivalent to the so-called   `section condition',   
\be
 \ba{ll}
 \partial_{M}\partial^{M}\Phi_{i}=0\,,\quad&\quad
 \partial_{M}\Phi_{i}\partial^{M}\Phi_{j}=0\,,
 \ea
 \ee
which is of practical utility.  From (\ref{trinv}),  we infer that `physics'  should be invariant under such  a shift of $\Delta^{M}=\Phi_{j}\partial^{M}\Phi_{k}$. This observation further suggests  that  the doubled coordinates may be gauged by  an equivalence relation~\cite{Park:2013mpa},
\be
\ba{ll}
x^{M}~\sim~x^{M}+\Delta^{M}\,,\qquad&\qquad\Delta^{M}\partial_{M}=0\,.
\ea
\label{xgauge}
\ee

\indent Diffeomorphisms in the doubled-yet-gauged  spacetime are then generated (actively)  by the generalized Lie derivative, $\hcL_{\xi}$, which was introduced initially  by Siegel~\cite{Siegel:1993th}, and also  later by Hull and Zwiebach~\cite{Hull:2009zb}.   Acting on an arbitrary  tensor density, $T_{M_{1}\cdots M_{n}}\in\cF$, with weight $\omega$,  it reads
\be
\hcL_{\xi}T_{M_{1}\cdots M_{n}}:=\xi^{N}\partial_{N}T_{M_{1}\cdots M_{n}}+\omega\partial_{N}\xi^{N}T_{M_{1}\cdots M_{n}}+\sum_{i=1}^{n}(\partial_{M_{i}}\xi_{N}-\partial_{N}\xi_{M_{i}})T_{M_{1}\cdots M_{i-1}}{}^{N}{}_{M_{i+1}\cdots  M_{n}}\,.
\label{hcL}
\ee
Thanks to the section condition, the generalized Lie derivative  forms a closed algebra,
\be
\left[\hcL_{\zeta},\hcL_{\xi}\right]=\hcL_{\left[\zeta,\xi\right]_{\rm{C}}}\,,
\label{closeda}
\ee
where  the so-called C-bracket is given by 
\be
\left[\zeta,\xi\right]^{M}_{\rm{C}}=\half\!\left(\hcL_{\zeta}\xi^{M}-\hcL_{\xi}\zeta^{M}\right)= \zeta^{N}\partial_{N}\xi^{M}-\xi^{N}\partial_{N}\zeta^{M}+\half \xi^{N}\partial^{M}\zeta_{N}-\half \zeta^{N}\partial^{M}\xi_{N}\,.
\label{Cbracket}
\ee
Along with this expression, it is worthwhile to note the `sum',
\be
\hcL_{\zeta}\xi^{M}+\hcL_{\xi}\zeta^{M}=\partial^{M}(\zeta_{N}\xi^{N})\,.
\label{sumexact}
\ee
Further, if the parameter of the generalized Lie derivative, $\xi^{M}$, is `derivative-index-valued', the first two terms  on the right-hand side of (\ref{hcL}) are trivial. Moreover, if this parameter is `exact' as $\xi^{M}=\partial^{M}\Phi$,   the generalized Lie derivative itself vanishes identically. Now, the closure~(\ref{closeda}) implies that the generalized Lie derivative is itself  diffeomorphism-covariant:
\be
\ba{ll}
\delta_{\xi}(\hcL_{\zeta}T_{M_{1}\cdots M_{n}})\!\!&=~
\hcL_{\zeta}(\delta_{\xi}T_{M_{1}\cdots M_{n}})+\hcL_{\delta_{\xi}\zeta}T_{M_{1}\cdots M_{n}}~=~\hcL_{\zeta}\hcL_{\xi}T_{M_{1}\cdots M_{n}}+\hcL_{\hcL_{\xi}\zeta}T_{M_{1}\cdots M_{n}}\\
{}&=~
\hcL_{\xi}\hcL_{\zeta}T_{M_{1}\cdots M_{n}}+\hcL_{[\zeta,\xi]_{{\rm{C}}}+\hcL_{\xi}\zeta}T_{M_{1}\cdots M_{n}}~=~
\hcL_{\xi}(\hcL_{\zeta}T_{M_{1}\cdots M_{n}})\,,
\ea
\label{covhcL}
\ee
where in the last step, from (\ref{Cbracket}), (\ref{sumexact}), we have used the fact that  $\,[\zeta,\xi]^{M}_{{\rm{C}}}+\hcL_{\xi}\zeta^{M}=\half\partial^{M}(\zeta_{N}\xi^{N})\,$, which is exact and hence null as a diffeomorphism parameter. However, if the tensor density carries additional $\SpinD\times\oSpinD$ indices, \textit{e.g.~}$T_{Mp\brq\alpha\bralpha}$, its generalized Lie derivative  is not local-Lorentz-covariant.  Hence the generalized Lie derivative is covariant for doubled-yet-gauged diffeomorphisms but not for local Lorentz symmetries. We shall  fix this limitation  in section~\ref{SECFURTHER} by further generalizing the generalized Lie derivative.  \\

\noindent In contrast to ordinary Riemannian geometry,  the infinitesimal one-form, $\rd x^{M}$, is not (passively)  diffeomorphism covariant in doubled-yet-gauged spacetime,
\be
\ba{ll}
\delta x^{M}=\xi^{M}\,,\qquad&\qquad \delta\rd x^{M}=\rd\xi^{M}=\rd x^{N}\partial_{N}\xi^{M}\neq
(\partial_{N}\xi^{M}-\partial^{M}\xi_{N})\rd x^{N}\,.
\ea
\ee
Furthermore, it is not invariant under the  coordinate gauge symmetry shift, $\rd x^{M}\neq\rd (x^{M}+\Delta^{M})$.  However, if we gauge $\rd x^{M}$ explicitly by introducing a derivative-index-valued gauge potential, $\cA^{M}$, 
\be
\ba{ll}
Dx^{M}:=\rd x^{M}-\cA^{M}\,,\qquad&\qquad\cA^{M}\partial_{M}=0\,,
\ea
\label{DX}
\ee
we can ensure both the  diffeomorphism covariance and the coordinate gauge symmetry invariance,
\be
\ba{llll}
\delta x^{M}=\xi^{M}\,,\quad&\quad
\delta \cA^{M}=\partial^{M}\xi_{N}(\rd x^{N}-\cA^{N})\quad&\Longrightarrow&\quad
{\delta (\rD x^{M})=(\partial_{N}\xi^{M}-\partial^{M}\xi_{N})\rD x^{N}}\,;\\
\delta x^{M}=\Delta^{M}\,,\quad&\quad
\delta\cA^{M}=\rd\Delta^{M}\quad&\Longrightarrow&\quad{\delta(\rD x^{M})=0\,}.
\ea
\ee
Utilizing the gauged infinitesimal one-form, $Dx^{M}$, it is then possible  to  define  the duality-covariant  `proper length'  in doubled-yet-gauge spacetime~\cite{Morand:2017fnv,Park:2017snt},   and  construct   associated  sigma models such as for the point particle~\cite{Ko:2016dxa,Blair:2017gwn}, bosonic string~\cite{Hull:2006va,Lee:2013hma}, Green-Schwarz superstring~\cite{Park:2016sbw} (and its coupling to the R--R sector~\cite{Sakamoto:2018krs}),  exceptional string~\cite{Arvanitakis:2017hwb,Arvanitakis:2018hfn}, \textit{etc.}\\

\noindent With the decomposition  of the doubled coordinates, $x^{M}=(\tx_{\mu},x^{\nu})$,  in accordance with the form of the $\ODD$ invariant metric, $\cJ_{MN}$~(\ref{metric3}),  the section condition reads $\tpartial^{\mu}\partial_{\mu}=0$.  Thus up to $\ODD$ rotations, the section condition is  generically solved by setting $\tpartial^{\mu}= 0$,   removing  the dependence on  $\tx_{\mu}$ coordinates.     It follows that  $\cA^{M}=A_{\lambda}\partial^{M}x^{\lambda}=(A_{\mu},0)$ and hence the $\tx_{\mu}$ coordinates are indeed gauged,  $Dx^{M}=(\rd\tx_{\mu}-A_{\mu},\rd x^{\nu})$.\\

\indent  {$\bullet$ \textit{Stringy graviton fields from the closed string massless sector}}\\
\noindent   The  $\ODD$ T-duality group is a  fundamental structure  in Stringy Gravity. All the fields therein must  assume  one  representation of it, such that  the $\ODD$ covariance is   manifest. 

The stringy graviton fields consist of the DFT dilaton, $d$, and DFT metric, $\cH_{MN}$. The former gives the integral measure in  Stringy Gravity after exponentiation, $e^{-2d}$, which is a scalar density of unit weight. The latter is then, by definition,  a  symmetric $\ODD$ element:
\be
\ba{ll}
\cH_{MN}=\cH_{NM}\,,\qquad&\qquad\cH_{K}{}^{L}\cH_{M}{}^{N}\cJ_{LN}=\cJ_{KM}\,.
\ea
\label{cHdef}
\ee
Combining $\cJ_{MN}$ and $\cH_{MN}$, we acquire a pair of symmetric projection matrices,
\be
\ba{ll}
P_{MN}=P_{NM}=\half(\cJ_{MN}+\cH_{MN})\,,\quad&\quad P_{L}{}^{M}P_{M}{}^{N}=P_{L}{}^{N}\,,\\
\brP_{MN}=\brP_{NM}=\half(\cJ_{MN}-\cH_{MN})\,,\quad&\quad \brP_{L}{}^{M}\brP_{M}{}^{N}=\brP_{L}{}^{N}\,,
\ea
\label{P1}
\ee
which are orthogonal and complete,
\be
\ba{ll}
 P_{L}{}^{M}\brP_{M}{}^{N}=0\,,\qquad&\qquad
P_{M}{}^{N}+\brP_{M}{}^{N}=\delta_{M}{}^{N}\,.
\ea
\label{P2}
\ee
It follows that  the  infinitesimal variations   of the projection matrices  satisfy 
\be
\ba{ll}
\delta P_{M}{}^{N}=-
\delta\brP_{M}{}^{N}=(P\delta P\brP)_{M}{}^{N}+(\brP\delta PP)_{M}{}^{N}\,,\quad&\quad
P_{L}{}^{M}\delta P_{M}{}^{N}=\delta P_{L}{}^{M}\brP_{M}{}^{N}\,.
\ea
\label{deltaPbrP}
\ee
Further, taking the ``square roots" of the  projectors,  
\be
\ba{ll}
P_{MN}=V_{M}{}^{p}V_{N}{}^{q}\eta_{pq}\,,\quad&\quad
\brP_{MN}=\brV_{M}{}^{\brp}\brV_{N}{}^{\brq}\breta_{\brp\brq}\,,
\ea
\ee
we acquire  a pair of DFT vielbeins, which satisfy   four defining properties:
\be
\ba{llll}
V_{Mp}V^{M}{}_{q}=\eta_{pq}\,,\quad&\quad
\brV_{M\brp}\brV^{M}{}_{\brq}=\breta_{\brp\brq}\,,\quad&\quad
V_{Mp}\brV^{M}{}_{\brq}=0\,,\quad&\quad
V_{M}{}^{p}V_{Np}+\brV_{M}{}^{\brp}\brV_{N\brp}=\cJ_{MN}\,,
\ea
\label{VVdef}
\ee
such that  (\ref{P1}) and (\ref{P2}) hold. Essentially, $(V_{M}{}^{p},\brV_{M}{}^{\brp})$, when viewed as a $({D+D})\times({D+D})$ matrix, diagonalizes $\cJ_{MN}$ and $\cH_{MN}$ simultaneously into  `$\diag(\eta,+\breta)$' and `$\diag(\eta,-\breta)$', respectively.    The presence of  twofold  vielbeins as well as spin groups are a truly   stringy feature, as it indicates    two distinct locally inertial frames existing separately for the left-moving and right-moving closed string  sectors~\cite{Duff:1986ne}, and may be a  testable   prediction of Stringy Gravity in itself~\cite{Choi:2015bga}.\\

\noindent It is absolutely crucial to note that DFT~\cite{Siegel:1993xq,Siegel:1993th,Hohm:2010pp} and its supersymmetric extensions~\cite{Jeon:2011sq,Jeon:2012hp,Park:2016sbw} are   formulatable  in terms of nothing but the very fields satisfying precisely the defining relations~(\ref{cHdef}), (\ref{VVdef}).  The most general   solutions to the defining  equations turn out to be  classified  by  two non-negative integers, $(n,\brn)$. With  $1\leq i,j\leq n$ and $1\leq \bri,\brj\leq\brn$,   the DFT metric is of  the  most general form~\cite{Morand:2017fnv},
\be
\cH_{MN}=\left(\ba{cc}\cH^{\mu\nu}&
-\cH^{\mu\sigma}B_{\sigma\lambda}+Y_{i}^{\mu}X^{i}_{\lambda}-
\brY_{\bri}^{\mu}\brX^{\bri}_{\lambda}\\
B_{\kappa\rho}\cH^{\rho\nu}+X^{i}_{\kappa}Y_{i}^{\nu}
-\brX^{\bri}_{\kappa}\brY_{\bri}^{\nu}\quad&~~
~~\cK_{\kappa\lambda}-B_{\kappa\rho}\cH^{\rho\sigma}B_{\sigma\lambda}
+2X^{i}_{(\kappa}B_{\lambda)\rho}Y_{i}^{\rho}
-2\brX^{\bri}_{(\kappa}B_{\lambda)\rho}\brY_{\bri}^{\rho}
\ea\right)
\label{cHFINAL}
\ee
where\, \textit{i)}  $\cH$ and $\cK$ are symmetric,  but  $B$ is skew-symmetric,  \textit{i.e.}~$\cH^{\mu\nu}=\cH^{\nu\mu}$, $\cK_{\mu\nu}=\cK_{\nu\mu}$, $B_{\mu\nu}=-B_{\nu\mu}$\,;\\
\indent\indent\!\!\!\!\!\, \textit{ii)} $\cH$ and $\cK$ admit kernels,  
$~\cH^{\mu\nu}X^{i}_{\nu}=\cH^{\mu\nu}\brX^{\bri}_{\nu}=0$, 
$~\cK_{\mu\nu}Y_{j}^{\nu}=\cK_{\mu\nu}\brY_{\brj}^{\nu}=0$\,;\\
\indent\indent\!\!\!\!\! \textit{iii)} a completeness relation must be met, $~\cH^{\mu\rho}\cK_{\rho\nu}
+Y_{i}^{\mu}X^{i}_{\nu}+\brY_{\bri}^{\mu}\brX^{\bri}_{\nu}
=\delta^{\mu}{}_{\nu}$\,.\\
\noindent  It follows from the linear independence of the kernel eigenvectors   that  
\[
\ba{lllll}
Y^{\mu}_{i}X_{\mu}^{j}=\delta_{i}{}^{j}\,,&\,\,
\brY^{\mu}_{\bri}\brX_{\mu}^{\brj}=\delta_{\bri}{}^{\brj}\,,&\,\,
Y^{\mu}_{i}\brX_{\mu}^{\brj}=
\brY^{\mu}_{\bri}X_{\mu}^{j}=0\,,&\,\,
\cH^{\rho\mu}\cK_{\mu\nu}\cH^{\nu\sigma}=\cH^{\rho\sigma}\,,&\,\,
\cK_{\rho\mu}\cH^{\mu\nu}\cK_{\nu\sigma}=\cK_{\rho\sigma}\,.
\ea
\]
With the  section choice ${\tpartial^{\mu}= 0}$ and the parameter decomposition $\xi^{A}=(\tilde{\xi}_{\mu},\xi^{\nu})$,  the generalized Lie derivative, $\hcL_{\xi}\cH_{MN}$,   reduces to the ordinary (\textit{i.e.~}undoubled)  Lie derivative, $\cL_{\xi}$,  plus $B$-field gauge symmetry, 
\be
\ba{llll}
\quad\delta X^{i}_{\mu}=\cL_{\xi}X^{i}_{\mu}\,,\qquad&~~\quad
\delta \brX^{\bri}_{\mu}=\cL_{\xi}\brX^{\bri}_{\mu}\,,\qquad&~~\quad
\delta Y_{j}^{\nu}=\cL_{\xi}Y_{j}^{\nu}\,,\qquad&~~\quad
\delta \brY_{\brj}^{\nu}=\cL_{\xi}\brY_{\brj}^{\nu}\,,\\
\multicolumn{4}{c}{\delta \cH^{\mu\nu}=\cL_{\xi}\cH^{\mu\nu}\,,\qquad\quad
\delta \cK_{\mu\nu}=\cL_{\xi}\cK_{\mu\nu}\,,\qquad\quad
\delta B_{\mu\nu}=\cL_{\xi}B_{\mu\nu}+\partial_{\mu}\tilde{\xi}_{\nu}-\partial_{\nu}\tilde{\xi}_{\mu}\,.}
\ea
\label{deltaXYHK}
\ee
Only in the case of $(n,\brn)=(0,0)$ can   $\cK_{\mu\nu}$ and  $\cH^{\mu\nu}$ be identified with the (invertible) Riemannian metric and its inverse. The $(0,0)$  Riemannian     DFT metric then  takes the rather well-known form,
\be
\cH_{MN}=\left(\ba{cc}
\quad g^{\mu\nu}\quad&\quad -g^{\mu\lambda}B_{\lambda\tau}\quad\\
\quad B_{\sigma\kappa}g^{\kappa\nu}\quad&\quad
g_{\sigma\tau}-B_{\sigma\kappa}g^{\kappa\lambda}B_{\lambda\tau}\quad
\ea\right)\,,
\label{00H}
\ee
and   the corresponding DFT vielbeins  read 
\be
\ba{ll}
V_{Mp}=\frac{1}{\sqrt{2}}\left(\ba{c}e_{p}{}^{\mu}\\
e_{\nu}{}^{q}\eta_{qp}+B_{\nu\sigma}e_{p}{}^{\sigma}\ea\right)\,,\quad&\quad
\brV_{M\brp}=\frac{1}{\sqrt{2}}\left(\ba{c}\bre_{\brp}{}^{\mu}\\
\bre_{\nu}{}^{\brq}\breta_{\brq\brp}+B_{\nu\sigma}\bre_{\brp}{}^{\sigma}\ea\right)\,,
\ea
\label{00V}
\ee
where $e_{\mu}{}^{p}$ and $\bre_{\mu}{}^{\brp}$ are a pair of Riemannian vielbeins for the common Riemannian metric, 
\be
e_{\mu}{}^{p}e_{\nu p}=-\bre_{\mu}{}^{\brp}\bre_{\nu\brp}=g_{\mu\nu}\,.
\label{eeg}
\ee
With the non-vanishing determinant, $g=\det g_{\mu\nu}\neq0$, the DFT dilaton can be further parametrized    by
\be
e^{-2d}=\sqrt{-g}\,e^{-2\phi}\,.
\label{00d}
\ee
In this way, the stringy gravitons may represent   the  conventional  closed string massless sector, $\{g_{\mu\nu},B_{\mu\nu},\phi\}$.

Other cases of $(n,\brn)\neq(0,0)$ are then generically non-Riemannian, as the Riemannian metric cannot be defined. They include   $(1,0)$ or $({D-1},0)$ for  non-relativistic Newton--Cartan or ultra-relativistic  Carroll gravities~\cite{BergshoeffSimons},  $(1,1)$ for the Gomis--Ooguri non-relativistic string~\cite{Gomis:2000bd,Ko:2015rha},  and various chiral theories, \textit{e.g.~}\cite{Siegel:2015axg}.\\

For later use, it is worth noting that the two-indexed projectors  generate in turn a pair of multi-indexed  projectors,
\be
\ba{l}
\cP_{ABC}{}^{DEF}:=P_{A}{}^{D}P_{[B}{}^{[E}P_{C]}{}^{F]}+\textstyle{\frac{2}{P_{M}{}^{M}-1}}P_{A[B}P_{C]}{}^{[E}P^{F]D}\,,\\
\bar{\cP}_{ABC}{}^{DEF}:=\brP_{A}{}^{D}\brP_{[B}{}^{[E}\brP_{C]}{}^{F]}+\textstyle{\frac{2}{\brP_{M}{}^{M}-1}}\brP_{A[B}\brP_{C]}{}^{[E}\brP^{F]D}\,,
\ea
\label{P6}
\ee
satisfying
\be
\ba{ll}
\cP_{ABC}{}^{DEF}\cP_{DEF}{}^{GHI}=\cP_{ABC}{}^{GHI}\,,\qquad&\qquad
\brcP_{ABC}{}^{DEF}\brcP_{DEF}{}^{GHI}=\brcP_{ABC}{}^{GHI}\,.
\ea
\ee
They are symmetric and traceless in the following sense:
\be
\ba{lll}
\cP_{ABCDEF}=\cP_{DEFABC}\,,\quad&\quad\cP_{ABCDEF}=\cP_{A[BC]D[EF]}\,,
\quad&\quad P^{AB}\cP_{ABCDEF}=0\,,\\
\brcP_{ABCDEF}=\brcP_{DEFABC}\,,\quad&\quad\brcP_{ABCDEF}=\brcP_{A[BC]D[EF]}\,,
\quad&\quad \brP^{AB}\brcP_{ABCDEF}=0\,.
\ea
\label{symP6}
\ee
~\\

\indent  {$\bullet$ \textit{Covariant derivatives with stringy Christoffel  symbols and spin connections}}\\
The `master' covariant derivative in Stringy Gravity,
\be
\cD_{A}=\partial_{A}+\Gamma_{A}+\Phi_{A}+\brPhi_{A}\,,
\label{MastercD}
\ee
is equipped with  the  stringy Christoffel symbols of the diffeomorphism connection~\cite{Jeon:2011cn},
\be
\ba{ll}
\Gamma_{CAB}=&2\left(P\partial_{C}P\brP\right)_{[AB]}
+2\left({{\brP}_{[A}{}^{D}{\brP}_{B]}{}^{E}}-{P_{[A}{}^{D}P_{B]}{}^{E}}\right)\partial_{D}P_{EC}\\
{}&-4\left(\textstyle{\frac{1}{P_{M}{}^{M}-1}}P_{C[A}P_{B]}{}^{D}+\textstyle{\frac{1}{\brP_{M}{}^{M}-1}}\brP_{C[A}\brP_{B]}{}^{D}\right)\!\left(\partial_{D}d+(P\partial^{E}P\brP)_{[ED]}\right)\,,
\ea
\label{Gammao}
\ee
and the spin connections for the twofold local Lorentz symmetries~\cite{Jeon:2011vx},
\be
\ba{ll}
\Phi_{Apq}=\Phi_{A[pq]}=V^{B}{}_{p}\na_{A}V_{Bq}\,,\quad&\quad
\brPhi_{A\brp\brq}=\brPhi_{A[\brp\brq]}=\brV^{B}{}_{\brp}\na_{A}\brV_{B\brq}\,.
\ea
\label{PhibrPhi}
\ee
In the above, we  set
\be
\na_{A}:=\partial_{A}+\Gamma_{A}\,,
\ee
which, ignoring any local Lorentz  indices, acts explicitly  on a tensor density with weight $\omega$  as
\be
\na_{C}T_{A_{1}A_{2}\cdots A_{n}}
:=\partial_{C}T_{A_{1}A_{2}\cdots A_{n}}-\omega_{{\scriptscriptstyle{T\,}}}\Gamma^{B}{}_{BC}T_{A_{1}A_{2}\cdots A_{n}}+
\sum_{i=1}^{n}\,\Gamma_{CA_{i}}{}^{B}T_{A_{1}\cdots A_{i-1}BA_{i+1}\cdots A_{n}}\,.
\label{asemicov}
\ee
The stringy Christoffel symbols~(\ref{Gammao}) can be uniquely determined by requiring three properties:
\begin{itemize}
\item[\textit{i)}] full  compatibility with all the stringy graviton fields, 
\be
\ba{ll}
\cD_{A}P_{BC}=\na_{A}P_{BC}=0\,,\qquad&\qquad \cD_{A}\brP_{BC}=\na_{A}\brP_{BC}=0\,,\\
\multicolumn{2}{c}{
\cD_{A}d=\na_{A}d=-\half e^{2d}\na_{A}(e^{-2d})=\partial_{A}d+\half\Gamma^{B}{}_{BA}=0\,,}
\ea
\label{Gcomp}
\ee
which implies, in particular,
\be
\ba{ll}
\cD_{A}\cJ_{BC}=\na_{A}\cJ_{BC}=0\,,\qquad&\qquad
\Gamma_{ABC}=-\Gamma_{ACB}\,;
\ea
\ee
\item[\textit{ii)}] a cyclic property (traceless condition), 
\be
\Gamma_{ABC}+\Gamma_{BCA}+\Gamma_{CAB}=0\,,
\label{GABC}
\ee
which makes  $\na_{A}$    compatible  with  the generalized Lie derivative~(\ref{hcL}) as well as  the C-bracket~(\ref{Cbracket}),  such that we may freely replace the ordinary derivatives therein by $\na_{A}$, 
\be
\ba{ll}
\hcL_{\xi}(\partial)=\hcL_{\xi}(\na)\,,\qquad&\qquad
[\zeta,\xi]_{\rm{C}}(\partial)=[\zeta,\xi]_{\rm{C}}(\na)\,;
\ea
\label{untwtorsionless}
\ee   
\item[\textit{iii)}] projection constraints,
\be
\ba{ll}
\cP_{ABC}{}^{DEF}\Gamma_{DEF}=0\,,\quad&\quad\bar{\cP}_{ABC}{}^{DEF}\Gamma_{DEF}=0\,,
\label{kernel}
\ea
\ee
which  ensure the uniqueness.  
\end{itemize}
Unlike the  Christoffel symbols in GR, there exist  no normal coordinates where the stringy Christoffel symbols would vanish pointwise. The Equivalence Principle holds   for the point particle but not for the string~\cite{Choi:2015bga,Park:2017snt}.

Once the stringy Christoffel symbols are fixed, the spin connections~(\ref{PhibrPhi}) follow  immediately  from the compatibility with the DFT vielbeins,
\be
\ba{l}
\cD_{A}V_{Bp}=\na_{A}V_{Bp}+\Phi_{Ap}{}^{q}V_{Bq}=\partial_{A}V_{Bp}+\Gamma_{AB}{}^{C}V_{Cp}+\Phi_{Ap}{}^{q}V_{Bq}=0\,,\\
\cD_{A}\brV_{B\brp}=\na_{A}\brV_{B\brp}+\brPhi_{A\brp}{}^{\brq}\brV_{B\brq}=\partial_{A}\brV_{B\brp}+\Gamma_{AB}{}^{C}\brV_{C\brp}+\brPhi_{A\brp}{}^{\brq}\brV_{B\brq}=0\,.
\ea
\label{cDVV}
\ee
The master derivative is also compatible with the two sets of   local Lorentz metrics and gamma matrices, 
\be
\ba{llll}
\cD_{A}\eta_{pq}=0\,,\quad&\quad
\cD_{A}\breta_{\brp\brq}=0\,,\quad&\quad\cD_{A}(\gamma^{p})^{\alpha}{}_{\beta}=0\,,\quad&\quad
\cD_{A}(\brgamma^{\brp})^{\bralpha}{}_{\brbeta}=0\,,
\ea
\ee
such that, as in GR,
\be
\ba{llll}
\Phi_{Apq}=-\Phi_{Aqp}\,,\quad&\quad
\brPhi_{A\brp\brq}=-\brPhi_{A\brq\brp}\,,\quad&\quad
\Phi_{A}{}^{\alpha}{}_{\beta}=\quarter\Phi_{Apq}(\gamma^{pq})^{\alpha}{}_{\beta}\,,\quad&\quad 
\brPhi_{A}{}^{\bralpha}{}_{\brbeta}=\quarter\brPhi_{A\brp\brq}(\brgamma^{\brp\brq})^{\bralpha}{}_{\brbeta}\,.
\ea
\label{Phisv}
\ee
The master  derivative~(\ref{MastercD}) acts explicitly as  
\be
\ba{lll}
\cD_{N}T_{Mp}{}^{\alpha}{}_{\brp}{}^{\bralpha}&=&
\na_{N}T_{Mp}{}^{\alpha}{}_{\brp}{}^{\bralpha}+
\Phi_{Np}{}^{q}
T_{Mq}{}^{\alpha}{}_{\brp}{}^{\bralpha}+\Phi_{N}{}^{\alpha}{}_{\beta}T_{Mp}{}^{\beta}{}_{\brp}{}^{\bralpha}+\brPhi_{N\brp}{}^{\brq}
T_{Mp}{}^{\alpha}{}_{\brq}{}^{\bralpha}+\brPhi_{N}{}^{\bralpha}{}_{\brbeta}T_{Mp}{}^{\alpha}{}_{\brp}{}^{\brbeta}\\
{}&=&\partial_{N}T_{Mp}{}^{\alpha}{}_{\brp}{}^{\bralpha}-\omega
\Gamma^{L}{}_{LN}T_{Mp}{}^{\alpha}{}_{\brp}{}^{\bralpha}
+\Gamma_{NM}{}^{L}T_{Lp}{}^{\alpha}{}_{\brp}{}^{\bralpha}
+\Phi_{Np}{}^{q}
T_{Mq}{}^{\alpha}{}_{\brp}{}^{\bralpha}+\Phi_{N}{}^{\alpha}{}_{\beta}T_{Mp}{}^{\beta}{}_{\brp}{}^{\bralpha}\\
{}&{}&+\brPhi_{N\brp}{}^{\brq}
T_{Mp}{}^{\alpha}{}_{\brq}{}^{\bralpha}+\brPhi_{N}{}^{\bralpha}{}_{\brbeta}T_{Mp}{}^{\alpha}{}_{\brp}{}^{\brbeta}\,.
\ea
\label{MastercDexplicit}
\ee
Unsurprisingly the master derivative is   completely covariant for the twofold   local Lorentz symmetries.  The characteristic  of the master derivative, $\cD_{A}$, as well as  $\na_{A}$  is that  they are actually `semi-covariant'  under doubled-yet-gauged  diffeomorphisms:  the stringy Christoffel symbols   transform as
\be 
\ba{ll}
\multicolumn{2}{c}{
\delta_{\xi}\Gamma_{CAB}=\hcL_{\xi}\Gamma_{CAB}+2\Big[(\cP+\brcP)_{CAB}{}^{FDE}-\delta_{C}^{~F}\delta_{A}^{~D}
\delta_{B}^{~E}\Big]\partial_{F}\partial_{[D}\xi_{E]}\,,}\\
\delta_{\xi}\Phi_{Apq}=\hcL_{\xi}\Phi_{Apq}+2\cP_{Apq}{}^{DEF}\partial_{D}\partial_{[E}\xi_{F]}\,,\quad&\quad
\delta_{\xi}\brPhi_{A\brp\brq}=\hcL_{\xi}\brPhi_{A\brp\brq}+2\brcP_{A\brp\brq}{}^{DEF}\partial_{D}\partial_{[E}\xi_{F]}\,,
\ea
\label{GPbrP}
\ee
such that $\cD_{A}$ and $\na_{A}$ are not automatically diffeomorphism-covariant, \textit{e.g.}
\be
\delta_{\xi}\big(\na_{C}T_{A_{1}\cdots A_{n}}\big)=\hcL_{\xi}\big(\na_{C}T_{A_{1}\cdots A_{n}}\big)+
\dis{\sum_{i=1}^{n}2(\cP{+\brcP})_{CA_{i}}{}^{BDEF}
\partial_{D}\partial_{E}\xi_{F}\,T_{A_{1}\cdots A_{i-1} BA_{i+1}\cdots A_{n}}\,.}
\label{diffeoanormalous}
\ee
Nevertheless, the potentially anomalous terms are uniquely given, or controlled, by the multi-indexed projectors, as seen in (\ref{GPbrP}) and (\ref{diffeoanormalous}), such that they can be easily projected out. Completely covariantized  derivatives include~\cite{Jeon:2011cn}
\be
\ba{ll}
P_{C}{}^{D}{\brP}_{A_{1}}{}^{B_{1}}\cdots{\brP}_{A_{n}}{}^{B_{n}}
\DO_{D}T_{B_{1}\cdots B_{n}}\,,\quad&~
{\brP}_{C}{}^{D}P_{A_{1}}{}^{B_{1}}\cdots P_{A_{n}}{}^{B_{n}}
\DO_{D}T_{B_{1}\cdots B_{n}}\,,\\
P^{AB}{\brP}_{C_{1}}{}^{D_{1}}\cdots{\brP}_{C_{n}}{}^{D_{n}}\DO_{A}T_{BD_{1}\cdots D_{n}}\,,\quad&~
\brP^{AB}{P}_{C_{1}}{}^{D_{1}}\cdots{P}_{C_{n}}{}^{D_{n}}\DO_{A}T_{BD_{1}\cdots D_{n}}\quad\quad\!(\mbox{divergences})\,,\\
P^{AB}{\brP}_{C_{1}}{}^{D_{1}}\cdots{\brP}_{C_{n}}{}^{D_{n}}
\DO_{A}\DO_{B}T_{D_{1}\cdots D_{n}}\,,\quad&~
{\brP}^{AB}P_{C_{1}}{}^{D_{1}}\cdots P_{C_{n}}{}^{D_{n}}
\DO_{A}\DO_{B}T_{D_{1}\cdots D_{n}}\quad(\mbox{Laplacians})\,,
\ea
\label{coco1}
\ee
which  can be freely pulled back by the DFT vielbeins, with  $\cD_{p}=V^{A}{}_{p}\cD_{A}$ and  $\cD_{\brp}=\brV^{A}{}_{\brp}\cD_{A}$,  to
\be
\ba{llllll}
\cD_{p}T_{\brq_{1}\cdots \brq_{n}}\,,\quad&
\cD_{\brp}T_{q_{1}\cdots q_{n}}\,,\quad&
\cD_{p}T^{p}{}_{\brq_{1}\cdots\brq_{n}}\,,\quad&
\cD_{\brp}T^{\brp}{}_{q_{1}\cdots q_{n}}\,,\quad&
\cD_{p}\cD^{p}T_{\brq_{1}\cdots \brq_{n}}\,,\quad&
\cD_{\brp}\cD^{\brp}T_{q_{1}\cdots q_{n}}\,.
\ea
\label{coco2}
\ee
In particular, for a weightless vector, $J^{A}$, it is useful to note
\be
\partial_{A}\!\left(e^{-2d}J^{A}\right)=\na_{A}\!\left(e^{-2d}J^{A}\right)=e^{-2d}\na_{A}J^{A}\,.
\label{divJ}
\ee
Furthermore, from  (\ref{GPbrP}),  the following    modules of the spin connections are completely   covariant under diffeomorphisms:
\be
\ba{llllll}
\brP_{A}{}^{B}\Phi_{Bpq}\,,~&~P_{A}{}^{B}\brPhi_{B\brp\brq}\,,~&~\Phi_{A[pq}V^{A}{}_{r]}\,,~&~
\brPhi_{A[\brp\brq}\brV^{A}{}_{\brr]}\,,~&~\Phi_{Apq}V^{Ap}\,,~&~
\brPhi_{A\brp\brq}\brV^{A\brp}\,.
\ea
\label{covPhi}
\ee 
Consequently,  acting on $\SpinD$ spinors, $\rho^{\alpha}$, $\psi_{\brp}^{\alpha}$, or $\oSpinD$ spinors, $\rho^{\prime\bralpha}$, $\psi^{\prime\bralpha}_{p}$, 
the completely covariant Dirac operators are, with respect to both diffeomorphisms and local Lorentz symmetries~\cite{Jeon:2011vx,Jeon:2011sq},
\be
\ba{llllllll}
\gamma^{p}\cD_{p}\rho\,,\quad&~
\gamma^{p}\cD_{p}\psi_{\brp}\,,\quad&~
\cD_{\brp}\rho\,,\quad&~
\cD_{\brp}\psi^{\brp}\,,\quad&~
\brgamma^{\brp}\cD_{\brp}\rhop\,,\quad&~
\brgamma^{\brp}\cD_{\brp}\psi^{\prime}_{p}\,,\quad&~
\cD_{p}\rhop\,,\quad&\quad
\cD_{p}\psip{}^{p}\,.
\ea
\label{covDirac}
\ee
For  a $\SpinD\times\oSpinD$ bi-fundamental spinorial field or the Ramond--Ramond potential, $\cC^{\alpha}{}_{\bralpha}$, a pair of completely covariant  nilpotent derivatives, $\cD_{+}$ and $\cD_{-}$, can be defined~\cite{Jeon:2012kd} (\textit{c.f.~}\cite{Rocen:2010bk}),
\be
\ba{ll}
\cD_{\pm}\cC:=\gamma^{p}\cD_{p}\cC\pm\gamma^{(D+1)}\cD_{\brp}\cC\brgamma^{\brp}\,,\quad&\quad \cD_{\pm}^{2}\cC=0\,,
\ea
\label{Dpm}
\ee
where, with (\ref{Phisv}),  $\cD_{A}\cC=\partial_{A}\cC+\Phi_{A}\cC-\cC\brPhi_{A}$. Specifically, the R--R field strength is given by  $\cF=\cD_{+}\cC$.\\

Finally, for a Yang--Mills potential, $\bfA_{M}$, the completely covariant field strength reads~\cite{Jeon:2011kp}
\be
\bfF_{p\brq}:=V^{M}{}_{p}\brV^{N}{}_{\brq}\big(\na_{M}\bfA_{N}-\na_{N}\bfA_{M}-i\left[\bfA_{M},\bfA_{N}\right]\big)\,.
\label{YMF}
\ee
In order to recover the standard  (undoubled) physical degrees of freedom,  one should impose    additional   ``section conditions" on   the doubled  Yang--Mills gauge potential~\cite{Choi:2015bga},
\be
\ba{ll}
\bfA^{M}\partial_{M}=0\,,\quad&\quad
\bfA^{M}\bfA_{M}=0\,.
\ea
\label{AAsection}
\ee
It turns out that the standard field strength, 
\be
F_{MN}:=\partial_{M}\bfA_{N}-\partial_{N}\bfA_{M}-i[\bfA_{M},\bfA_{N}]\,,
\label{standardF}
\ee
then becomes  completely covariant, and (\ref{YMF}) reduces to 
\be
\bfF_{p\brq}=V^{M}{}_{p}\brV^{N}{}_{\brq}F_{MN}\,.
\ee
~\\

Upon Riemannian backgrounds~(\ref{00V}), (\ref{00d}),    the  spin connections~(\ref{covPhi}) reduce explicitly to
\be
\ba{ll}
\brV^{A}{}_{\brp}\Phi_{Apq}=\frac{1}{\sqrt{2}}\bre_{\brp}{}^{\mu}\left(\omega_{\mu pq}+\half H_{\mu pq}\right)\,,\quad&\quad
V^{A}{}_{p}\brPhi_{A\brp\brq}=\frac{1}{\sqrt{2}}e_{p}{}^{\mu}\left(\bromega_{\mu\brp\brq}+\half H_{\mu \brp\brq}\right)\,,\\

\Phi_{A[pq}V^{A}{}_{r]}=\frac{1}{\sqrt{2}}\left(\omega_{[pqr]}+\textstyle{\frac{1}{6}} H_{pqr}\right)\,,\quad&\quad
\brPhi_{A[\brp\brq}\brV^{A}{}_{\brr]}=\frac{1}{\sqrt{2}}\left(\bromega_{[\brp\brq\brr]}+\textstyle{\frac{1}{6}} H_{\brp\brq\brr}\right)\,,\\
\Phi_{Apq}V^{Ap}=\frac{1}{\sqrt{2}}\left(e^{p\mu}\omega_{\mu pq}-2e_{q}{}^{\nu}\partial_{\nu}\phi\right)\,,\quad&\quad
\brPhi_{A\brp\brq}\brV^{A\brp}=\frac{1}{\sqrt{2}}\left(\bre^{\brp\mu}\bromega_{\mu\brp\brq}-2\bre_{\brq}{}^{\nu}\partial_{\nu}\phi\right)\,,
\ea
\label{covPhi2}
\ee 
where,  generalizing  (\ref{trd}),  we have  $\omega_{\mu pq}=e_{p}{}^{\nu}(\partial_{\mu}e_{\nu q}-\gamma_{\mu}^{\lambda}{}_{\nu}e_{\lambda q})$, $~\bromega_{\mu \brp\brq}=\bre_{\brp}{}^{\nu}(\partial_{\mu}\bre_{\nu\brq}-\gamma_{\mu}^{\lambda}{}_{\nu}\bre_{\lambda\brq})$, and 
\be
\ba{llllll}
\trd_{\mu}:=\partial_{\mu}+\gamma_{\mu}+\omega_{\mu}+\bromega_{\mu}\,,&~
\trd_{\mu} e_{\nu}{}^{p}=0\,,&~
\trd_{\mu}\eta_{pq}=0\,,&~
\trd_{\mu}\bre_{\nu}{}^{\brq}=0,&~
\trd_{\mu}\breta_{\brp\brq}=0\,,&~
\trd_{\lambda} g_{\mu\nu}=0\,.
\ea
\label{trdmaster}
\ee

\indent  {$\bullet$ \textit{Curvatures: stringy Einstein tensor}}\\
\noindent The semi-covariant Riemann curvature in Stringy Gravity  is defined by~\cite{Jeon:2011cn}
\be
S_{ABCD}:=\half\left(R_{ABCD}+R_{CDAB}-\Gamma^{E}{}_{AB}\Gamma_{ECD}\right)\,,
\label{RiemannS}
\ee
where $\Gamma_{ABC}$ are the stringy Christoffel symbols~(\ref{Gammao}) and  $R_{ABCD}$ denotes their ``field strength",
\be
R_{CDAB}=\partial_{A}\Gamma_{BCD}-\partial_{B}\Gamma_{ACD}+\Gamma_{AC}{}^{E}\Gamma_{BED}-\Gamma_{BC}{}^{E}\Gamma_{AED}\,.
\ee
Crucially, by construction, it satisfies symmetric properties and an algebraic ``Bianchi" identity,\footnote{As an alternative to direct verification, the Bianchi identity can also be shown   using    (\ref{closeda}),  (\ref{untwtorsionless}) and the relation~\cite{Jeon:2010rw}
\[
0=\left.\left(\left[\hcL_{\zeta},\hcL_{\xi}\right]-\hcL_{\left[\zeta,\xi\right]_{\rm{C}}}\right)\right|_{\partial\rightarrow\na}T_{A_{1}A_{2}\cdots A_{n}}=\sum_{i=1}^{n}\,6S_{A_{i}[BCD]}\zeta^{B}\xi^{C}T_{A_{1}\cdots A_{i-1}}{}^{D}{}_{A_{i+1}\cdots A_{n}}\,.
\]
}
\be
\ba{ll}
S_{ABCD}=S_{CDAB}=S_{[AB][CD]}\,,\qquad&\quad
S_{A[BCD]}=0\,.
\ea
\label{aB}
\ee
Furthermore,  just like the Riemann curvature in GR~(\ref{deltaRiemann}), it   transforms as 	`total' derivatives under  the arbitrary variation of the   stringy Christoffel symbols,\footnote{Eq.(\ref{deltaS})  can  be generalized to include torsion, such that     the `1.5' formalism works  in the full-order supersymmetric extensions of DFT~\cite{Jeon:2011sq,Jeon:2012hp}, where the connection becomes  torsionful, $\Gamma_{[ABC]}\neq0$.}
\be
\delta S_{ABCD}=\na_{[A}\delta\Gamma_{B]CD}+\na_{[C}\delta\Gamma_{D]AB}\,.
\label{deltaS}
\ee
In particular, it is `semi-covariant' under doubled-yet-gauged diffeomorphisms,
\be
\delta_{\xi}S_{ABCD}=\hcL_{\xi}S_{ABCD}+
2\na_{[A}\Big(\!(\cP{+\brcP})_{B][CD]}{}^{EFG}\partial_{E}\partial_{F}\xi_{G}\Big)
+2\na_{[C}\Big(\!(\cP{+\brcP})_{D][AB]}{}^{EFG}\partial_{E}\partial_{F}\xi_{G}\Big)\,.
\ee
In DFT there is no completely covariant four-indexed `Riemann' curvature~\cite{Jeon:2011cn,Hohm:2011si}, which is in a sense consistent with the absence of `normal' coordinates for  strings. 
The completely covariant `Ricci' and scalar curvatures are then, with $S_{AB}=S_{BA}=S^{C}{}_{ACB}$,
\be
\ba{ll}
S_{p\brq}:=V^{A}{}_{p}\brV^{B}{}_{\brq}S_{AB}\,,\quad&\quad
\So:=\left(P^{AC}P^{BD}-\brP^{AC}\brP^{CD}\right)S_{ABCD}=S_{pq}{}^{pq}-S_{\brp\brq}{}^{\brp\brq}\,.
\ea
\label{RicciSc}
\ee
These completely covariant curvatures contain  both $\cH_{MN}$ and $d$, as is the case for the connection, $\Gamma_{LMN}$~(\ref{Gammao}).  The DFT metric alone cannot generate any  covariant curvature~\textit{c.f.~}\cite{Jeon:2010rw}.  

It is worth noting the identities
\be
\ba{llll}
S_{pr\brq}{}^{r}=S_{p\brr\brq}{}^{\brr}=\half S_{p\brq}\,,\quad&\quad S_{pq}{}^{pq}+S_{\brp\brq}{}^{\brp\brq}=0\,,\quad&\quad S_{pq\brp\brq}=0\,,\quad&\quad S_{p\brp q\brq}=0\,,
\ea
\ee
\be
\ba{ll}
(\gamma^{p}\cD_{p})^{2}\varepsilon
+\cD_{\brp}\cD^{\brp}\varepsilon=-\quarter S_{pq}{}^{pq}\varepsilon=-\textstyle{\frac{1}{8}}\So\varepsilon\,,\quad&\quad
(\brgamma^{\brp}\cD_{\brp})^{2}\varepsilonp
+\cD_{p}\cD^{p}\varepsilonp=-\quarter S_{\brp\brq}{}^{\brp\brq}\varepsilonp=\textstyle{\frac{1}{8}}\So\varepsilonp\,,
\ea
\ee
and  the commutation relations~\cite{Coimbra:2011nw,Cho:2015lha}
\be
\ba{ll}
[\cD_{p},\cD_{\brq}]T^{p}=S_{p\brq}T^{p}\,,\quad&\qquad
[\cD_{\brq},\cD_{p}]T^{\brq}=S_{p\brq}T^{\brq}\,,\\
{}[\gamma^{p}\cD_{p},\cD_{\brq}]\varepsilon=\half S_{p\brq}\gamma^{p}\varepsilon\,,\quad&\qquad
{}[\brgamma^{\brq}\cD_{\brq},\cD_{p}]\varepsilon^{\prime}=\half S_{p\brq}\gamma^{\brq}\varepsilonp\,.
\ea
\ee

Combining the `Ricci' and the scalar curvatures, it is possible to construct the stringy `Einstein' tensor which is covariantly conserved~\cite{Park:2015bza}, 
\be
\ba{ll}
G_{AB}:=4V_{[A}{}^{p}\brV_{B]}{}^{\brq}S_{p\brq}-\half\cJ_{AB}\So\,,\qquad&\quad
\na_{A}G^{AB}=0\,.
\ea
\label{sE}
\ee
From (\ref{coco1}), this conservation law is completely covariant. Note also that in general,   $G_{AB}\neq G_{BA}\,$ and $\,\na_{B}G^{AB}\neq 0\,$. However, we may symmetrize the stringy  Einstein tensor,   still preserving the conservation law,  by multiplying the DFT metric from the right,   
\be
\ba{ll}
(G\cH)_{AB}=(G\cH)_{BA}:=G_{AC}\cH^{C}{}_{B}=-4 V_{(A}{}^{p}\brV_{B)}{}^{\brq}S_{p\brq}-\half\cH_{AB}\So\,,
\quad&\quad
\na_{A}(G\cH)^{AB}=0\,.
\ea
\label{GH}
\ee
Since $G_{A}{}^{A}=-D\So$, the vanishing of the stringy Einstein tensor, ${G_{AB}\equiv 0}$, is equivalent to  the separate vanishing of the `Ricci'  and the scalar curvatures, ${S_{p\brq}\equiv 0}$ and  ${\So\equiv 0}$, respectively, which correspond to the original DFT equations of motion~\cite{Siegel:1993xq,Hohm:2010pp}.

Restricting to  Riemannian backgrounds~(\ref{00V}), (\ref{00d}),  we have explicitly, 
\be
\ba{l}
S_{p\brq}=\half e_{p}{}^{\mu}\bre_{\brq}{}^{\nu}\Big[
R_{\mu\nu}+2\trd_{\mu}(\partial_{\nu}\phi)-\quarter H_{\mu\rho\sigma}H_{\nu}{}^{\rho\sigma}
+\half\trd^{\rho}H_{\rho\mu\nu}-(\partial^{\rho}\phi)H_{\rho\mu\nu}
\Big]\,,\\
\So=R+4\Box\phi-4\partial_{\mu}\phi\partial^{\mu}\phi-\textstyle{\frac{1}{12}}H_{\lambda\mu\nu}H^{\lambda\mu\nu}\,.
\ea
\label{RiemannS}
\ee
In particular,  the upper left $D{\times D}$  diagonal block of $(G\cH)_{AB}$  contains the  undoubled Einstein tensor in GR,
\be
(G\cH)^{\mu\nu}=R^{\mu\nu}-\half g^{\mu\nu}R+2\trd^{\mu}(\partial^{\nu}\phi)
-2g^{\mu\nu}(\Box\phi-\partial_{\sigma}\phi\partial^{\sigma}\phi)-\quarter H^{\mu\rho\sigma}H^{\nu}{}_{\rho\sigma}
+\textstyle{\frac{1}{24}} g^{\mu\nu}H_{\rho\sigma\tau}H^{\rho\sigma\tau}\,.
\label{GcHuu}
\ee
~\\

\subsection{Stringy Energy-Momentum tensor \& Einstein Double Field Equations}
We now consider Stringy Gravity coupled to generic matter fields, in analogy to GR~(\ref{GRaction}),
\be
\dis{\int_{\Sigma}e^{-2d}\Big[\,\textstyle{\frac{1}{16\pi G}}\So+L_{\rm{matter}}\,\Big]\,,}
\label{SGaction}
\ee
where $L_{\rm{matter}}$ is the $\ODD$ symmetric Lagrangian of the matter fields, $\Upsilon_{a}$, equipped with  the completely covariantized  derivatives, $\cD_{M}$. Some examples will follow below in  subsection~\ref{SECexample},   including  cases~(\ref{chichi}), (\ref{particleL}), (\ref{stringL})  where  the Lagrangian density, $\cL_{\rm{matter}}\equiv e^{-2d}L_{\rm{matter}}$,  does not  contain, and hence decouples from,  the DFT dilaton, $d$.   The integral is taken over a $D$-dimensional section, $\Sigma$, corresponding to a `gauge slice',  \textit{c.f.~}(\ref{xgauge}). We seek the  variation of the above  action which is induced by  the arbitrary  transformations  of all the fields, $\delta d$, $\delta P_{AB}$, $\delta\brP_{AB}$, $\delta V_{Ap}$, $\delta\brV_{A\brp}$, and   $\delta\Upsilon_{a}$. They are subject to   the following algebraic relations,  originating  from the defining properties the stringy graviton fields, (\ref{cHdef}), (\ref{P1}), (\ref{VVdef}),
\be
\ba{c}
\delta P_{AB}=-\delta\brP_{AB}=\half\delta\cH_{AB}=2P_{(A}{}^{C}\brP_{B)}{}^{D}\delta P_{CD}
=2\brV_{(A}{}^{\brp}V_{B)}{}^{q}\,\brV^{C}{}_{\brp}\delta V_{Cq}\,,\quad\,~~\,~
\brV^{C}{}_{\brq}\delta V_{C p}=-V^{C}{}_{p}\delta\brV_{C\brq}\,,\\
\delta V_{Ap}=\brV_{A}{}^{\brq}\brV^{C}{}_{\brq}\delta V_{C p}+(\delta V_{C[p}V^{C}{}_{q]})V_{A}{}^{q}\,,\quad\qquad
\delta\brV_{A\brp}=V_{A}{}^{q}V^{C}{}_{q}\delta \brV_{C\brp}+(\delta \brV_{C[\brp}\brV^{C}{}_{\brq]})\brV_{A}{}^{\brq}\,.
\label{deltaP}
\ea
\ee

Firstly, as is known \cite{Jeon:2011cn}, the pure Stringy Gravity term transforms, from (\ref{Gcomp}), (\ref{aB}), (\ref{deltaS}), (\ref{RicciSc}), as
\be
\ba{lll}
\delta\left(e^{-2d}\So\right)&=&4e^{-2d}\left(\delta P^{AB}V_{A}{}^{p}\brV_{B}{}^{\brq}S_{p\brq}-\half\delta d\So\right)+\partial_{A}\left[2e^{-2d}\left(P^{AC}P^{BD}-\brP^{AC}\brP^{BD}\right)
\delta\Gamma_{BCD}\right]\\
{}&=&4e^{-2d}\left( \brV^{A\brq}\delta V_{A}{}^{p}S_{p\brq}-\half\delta d\,\So\right)~+~\mbox{total~derivative}\,,
\ea
\label{deltaSo}
\ee 
of which the total derivative can be  ignored in the variation of the action. 

Secondly, in a similar  fashion to  (\ref{deltamatter}),   the  (local-Lorentz-symmetric)  matter Lagrangian transforms, up to  total derivatives ($\simeq$), as
\be
\ba{lll}
\delta L_{\rm{matter}}&\simeq&\dis{
\delta V_{A}{}^{p}\frac{\delta L_{\rm{matter}}}{\delta V_{A}{}^{p}}+
\delta \brV_{A}{}^{\brp}\frac{\delta L_{\rm{matter}}}{\delta \brV_{A}{}^{\brp}}+
\delta d \frac{\delta L_{\rm{matter}}}{\delta d}+
\delta\Upsilon_{a}\frac{\delta L_{\rm{matter}}}{\delta \Upsilon_{a}}}\\
&\simeq&\dis{\brV^{B\brq}\delta V_{B}{}^{p}\left(\brV_{A\brq}
\frac{\delta L_{\rm{matter}}}{\delta V_{A}{}^{p}}-
V_{Ap}\frac{\delta L_{\rm{matter}}}{\delta \brV_{A}{}^{\brq}}\right)+
\delta d \frac{\delta L_{\rm{matter}}}{\delta d}+
\delta^{\prime}\Upsilon_{a}\frac{\delta L_{\rm{matter}}}{\delta \Upsilon_{a}}}\,,
\ea
\label{deltaLmatter}
\ee
where  $\frac{\delta L_{\rm{matter}}}{\delta \Upsilon_{a}}$ corresponds to the Euler--Lagrange equation  for each  matter field, $\Upsilon_{a}$, and   $\delta^{\prime}\Upsilon_{a}$ is the  arbitrary variation  of the matter field, \textit{i.e.~}$\delta\Upsilon_{a}$,  supplemented by  the  infinitesimal local Lorentz rotations set by the parameters  $\delta V_{C[p}V^{C}{}_{q]}$ and $\delta \brV_{C[\brp}\brV^{C}{}_{\brq]}$.  Eq.(\ref{deltaLmatter})  holds since   
  $L_{\rm{matter}}$ is supposed to be    $\SpinD\times\oSpinD$  local Lorentz symmetric and therefore 
    the second terms in the variations of the DFT vielbeins in (\ref{deltaP})  can be  inversely traded with the  $\SpinD\times\oSpinD$ local Lorentz transformations of the matter fields, which justifies to the change $\delta\Upsilon_{a}\rightarrow\delta^{\prime}\Upsilon_{a}$.

The variation~(\ref{deltaLmatter}) suggests the following two definitions, 
\be
\ba{ll}
\EK_{p\brq}:=\dis{\frac{1}{2} \left(V_{Ap}\frac{\delta L_{\rm{matter}}}{\delta \brV_{A}{}^{\brq}}-\brV_{A\brq}\frac{\delta L_{\rm{matter}}}{\delta V_{A}{}^{p}}\right)\,,}\qquad&\qquad
\To:= e^{2d}\times\dis{\frac{\delta\left(e^{-2d} L_{\rm{matter}}\right)}{\delta d}\,,}
\ea
\label{cK}
\ee
both of which will constitute the conserved  Energy-Momentum tensor in Stringy Gravity, see (\ref{EMSG}).  We stress that to avoid any ambiguity, the functional derivatives are best computed from the infinitesimal variation of the Lagrangian.

Eq.(\ref{deltaLmatter}) then reads
\be
\delta\!\left(e^{-2d} L_{\rm{matter}}\right)\simeq e^{-2d}\left(-2\brV^{A\brq}\delta V_{A}{}^{p}\EK_{p\brq}+\delta d\To+\delta^{\prime}\Upsilon_{a}\frac{\delta L_{\rm{matter}}}{\delta \Upsilon_{a}}\right)\,.
\label{deltaLmatter2}
\ee
It is worthwhile to note that,  for the restricted cases   of  $L_{\rm{matter}}$ in which  the DFT vielbeins are absent and only the projectors are present,   we have
\be
\delta L_{\rm{matter}}=\dis{
\delta P_{AB}\frac{\delta L_{\rm{matter}}}{\delta P_{AB}}+
\delta \brP_{AB}\frac{\delta L_{\rm{matter}}}{\delta \brP_{AB}}+
\delta d \frac{\delta L_{\rm{matter}}}{\delta d}+
\delta\Upsilon_{a}\frac{\delta L_{\rm{matter}}}{\delta \Upsilon_{a}}}\,,
\ee
and, from (\ref{deltaP}),   the above definition of $\EK_{p\brq}$ reduces to
\be
\EK_{p\brq}=V_{Ap}\brV_{B\brq}\left(\frac{\delta L_{\rm{matter}}}{\delta \brP_{AB}}-\frac{\delta L_{\rm{matter}}}{\delta P_{AB}}\right)\,.
\ee
Now, collecting  all the results of (\ref{deltaP}), (\ref{deltaSo}) and (\ref{cK}), the variation of the action~(\ref{SGaction})  reads, disregarding any surface integrals, 
\be
\ba{l}
\,~~\delta\dis{\int_{\Sigma}}e^{-2d}\Big[\,\textstyle{\frac{1}{16\pi G}}\So+L_{\rm{matter}}\,\Big]\\
=\dis{\int_{\Sigma}}e^{-2d}\left[\textstyle{\frac{1}{4\pi G}}
\brV^{A\brq}\delta V_{A}{}^{p}(S_{p\brq}-8\pi G\EK_{p\brq})-\textstyle{\frac{1}{8\pi G}}\delta d(\So-8\pi G\To)+
\delta^{\prime}\Upsilon_{a}\dis{\frac{\delta L_{\rm{matter}}}{\delta \Upsilon_{a}}}\right]\,.
\ea
\label{deltaSG}
\ee
All the equations of motion are then given by
\be
\ba{lll}
S_{p\brq}=8\pi G\EK_{p\brq}\,,\qquad&\quad \So=8\pi G\To\,,\qquad&\quad
\dis{\frac{\delta L_{\rm{matter}}}{\delta \Upsilon_{a}}}\equiv 0\,,
\ea
\label{EOMSG}
\ee
where `$\equiv$' is used to denote the on-shell equations for the matter fields.
Specifically, when the variation is generated by doubled-yet-gauged diffeomorphisms, we have
\be
\ba{ll}
\delta_{\xi} d=-\half e^{2d}\hcL_{\xi}\left(e^{-2d}\right)=-\half\cD_{A}\xi^{A}\,,\quad&\quad
\delta_{\xi}\Upsilon_{a}=\hcL_{\xi}\Upsilon_{a}\,,
\ea
\label{deltaxid}
\ee
and, from $\cD_{B}V_{Ap}=0$~(\ref{cDVV}),
\be
\delta_{\xi} V_{Ap}=\hcL_{\xi}V_{Ap}=\xi^{B}\na_{B}V_{Ap}+2\na_{[A}\xi_{B]}V^{B}{}_{p}=
-\xi^{B}\Phi_{Bp}{}^{q}V_{Aq}+2\na_{[A}\xi_{B]}V^{B}{}_{p}\,,
\ee
which implies
\be
\ba{ll}
\brV^{A\brq}\delta_{\xi} V_{A}{}^{p}=2\cD_{[A}\xi_{B]}\brV^{A\brq}V^{Bp}\,,\qquad&\quad
\delta_{\xi}P_{AB}=\hcL_{\xi}P_{AB}=4\brP_{(A}{}^{C}P_{B)}{}^{D}\cD_{[C}\xi_{D]}\,.
\ea
\label{deltaxiP}
\ee
Substituting these results into (\ref{deltaSG}), utilizing the invariance of the action under doubled-yet-gauged diffeomorphisms while neglecting  surface terms,  we achieve a crucial result,
\be
0=\dis{\int_{\Sigma}}e^{-2d}\left[
\textstyle{\frac{1}{8\pi G}}\xi^{B}\cD^{A}\left\{4V_{[A}{}^{p}\brV_{B]}{}^{\brq}
(S_{p\brq}-8\pi G\EK_{p\brq})-\half\cJ_{AB}(\So-8\pi G\To)\right\}+
\delta^{\prime}\Upsilon_{a}\dis{\frac{\delta L_{\rm{matter}}}{\delta \Upsilon_{a}}}\right]\,.
\ee
This   leads to   the  definitions of the off-shell conserved stringy Einstein curvature  tensor~(\ref{sE}) from \cite{Park:2015bza},
\be
\ba{ll}
G_{AB}=4V_{[A}{}^{p}\brV_{B]}{}^{\brq}S_{p\brq}-\half\cJ_{AB}\So\,,\qquad&\quad
\cD_{A}G^{AB}=0\qquad(\mbox{off-shell})\,,
\ea
\label{sE2}
\ee
and  separately  the on-shell conserved  \textit{Energy-Momentum tensor in Stringy Gravity},
\be
\ba{ll}
\EM_{AB}:=4V_{[A}{}^{p}\brV_{B]}{}^{\brq}\EK_{p\brq}-\half\cJ_{AB}\To\,,\qquad&\quad
\cD_{A}\EM^{AB}\equiv 0\qquad(\mbox{on-shell})\,.
\ea
\label{EMSG}
\ee
Note\footnote{Although we use the same conventional letter symbols, no component of  $G_{AB}$ or $\EM_{AB}$  coincides precisely with that of the  undoubled  Einstein and  Energy-Momentum tensors  in GR,  \textit{c.f.}~(\ref{TGRdef}),   (\ref{GcHuu}).}   $\EM_{AB}\neq \EM_{BA}\,$ and $\,\cD_{B}\EM^{AB}\neq 0\,$. However, like $(G\cH)_{AB}=(G\cH)_{BA}$~(\ref{GH}),  we may symmetrize the stringy Energy-Momentum tensor,  
\be
\ba{ll}
(\EM\cH)_{AB}=(\EM\cH)_{BA}:=\EM_{AC}\cH^{C}{}_{B}=-4 V_{(A}{}^{p}\brV_{B)}{}^{\brq}\EK_{p\brq}-\half\cH_{AB}\To\,,
\quad&\quad
\cD_{A}(\EM\cH)^{AB}\equiv 0\,.
\ea
\label{TH}
\ee
$G_{AB}$ and  $\EM_{AB}$ each have ${D^{2}+1}$ components, given by
\be
\ba{llll}
V^{A}{}_{p}\brV^{B}{}_{\brq}G_{AB}=2S_{p\brq}\,,
\quad&\quad
G^{A}{}_{A}=-D\So\,,\quad&\quad
V^{A}{}_{p}\brV^{B}{}_{\brq}\EM_{AB}=2\EK_{p\brq}\,,\quad&\quad
\EM^{A}{}_{A}=-D\To\,,
\ea
\ee
respectively. The equations of motion of the DFT vielbeins and the DFT dilaton are unified into a single expression, the \textit{Einstein Double Field Equations},
\be
G_{AB}=8\pi G\EM_{AB}\,,
\label{EDFE}
\ee
which is naturally  consistent with the central idea that Stringy Gravity treats the entire closed string massless sector as geometrical stringy graviton fields.\\

From (\ref{deltaxid}), (\ref{deltaxiP}) and (\ref{EMSG}),  if we contract the stringy Energy-Momentum tensor with an $\ODD$ vector,   its divergence  reads
\be
\cD_{A}\!\left(\EM^{A}{}_{B}\xi^{B}\right)\equiv
\EM_{AB}\cD^{A}\xi^{B}=
-2V^{Ap}\brV^{B\brq}\EK_{p\brq}\left(\hcL_{\xi}P_{AB}\right)+\To\left(\hcL_{\xi}d\right)\,.
\ee
Therefore, if  $\xi^{A}$ is a DFT-Killing vector satisfying the DFT-Killing equations~\cite{Park:2015bza},
\be
\ba{ll}
\hcL_{\xi}P_{AB}=4\brP_{(A}{}^{C}P_{B)}{}^{D}\na_{[C}\xi_{D]}=0\,,\quad&\qquad
\hcL_{\xi}d=-\half\na_{A}\xi^{A}=0\,,
\ea
\label{KillingPd}
\ee
the contraction $\EM^{A}{}_{B}\xi^{B}$ gives an on-shell conserved Noether current (from  (\ref{divJ})), 
\be
\partial_{A}\left(e^{-2d}\EM^{A}{}_{B}\xi^{B}\right)=e^{-2d}\cD_{A}\left(\EM^{A}{}_{B}\xi^{B}\right)\equiv0\,,
\ee
and  the corresponding  Noether charge,
\be
\cQ[\xi]=\dis{\int_{\Sigma}}~e^{-2d}\EM^{t}{}_{A}\xi^{A}\,,
\label{cQxi}
\ee
where the superscript index, $t$, denotes the time component  for  a   chosen section. It is worthwhile to note that the alternative contraction with the symmetrized Energy-Momentum  tensor,  $(\EM\cH)^{A}{}_{B}\xi^{B}$,  is not conserved even if $\xi^{A}$ is a Killing vector.

Through contraction with the DFT vielbeins, the conservation law of the Energy-Momentum tensor decomposes into two separate formulae,
\be
\ba{ll}
\cD_{A}\EM^{AB}V_{Bp}=-2\cD_{\brq}\EK_{p}{}^{\brq}-\half\cD_{p}\To\equiv0\,,\quad&\qquad
\cD_{A}\EM^{AB}\brV_{B\brq}=2\cD_{p}\EK^{p}{}_{\brq}-\half\cD_{\brq}\To\equiv0\,.
\ea
\ee
Restricting to Riemannian  backgrounds~(\ref{00V}), (\ref{00d}),   with $\trd_{\mu}=\partial_{\mu}+\gamma_{\mu}+\omega_{\mu}+\bromega_{\mu}$~(\ref{trdmaster}), we have
\be
\ba{lll}
0~\equiv~\cD_{A}\EM^{A}{}_{p}&=&-\sqrt{2}\bre^{\brq\mu}\!\left(\trd_{\mu}\EK_{p\brq}-2\partial_{\mu}\phi\,\EK_{p\brq}+\half H_{\mu p}{}^{q}\EK_{q\brq}\right)-\textstyle{\frac{1}{2\sqrt{2}}}e_{p}{}^{\mu}\partial_{\mu}\To\\
{}&=&\textstyle{\frac{1}{\sqrt{2}}}e_{p}{}^{\nu}\!\left(\trd_{\mu}\EK_{\nu}{}^{\mu}-2\partial_{\mu}\phi\,\EK_{\nu}{}^{\mu}+\half H_{\nu\lambda\mu}\EK^{\lambda\mu}-\half\partial_{\nu}\To\right)\,,
\ea
\ee
and
\be
\ba{lll}
0~\equiv~\cD_{A}\EM^{A}{}_{\brq}&=&\sqrt{2}e^{p\mu}\!\left(\trd_{\mu}\EK_{p\brq}-2\partial_{\mu}\phi\,\EK_{p\brq}+\half H_{\mu\brq}{}^{\brr}\EK_{p\brr}\right)-\textstyle{\frac{1}{2\sqrt{2}}}\bre_{\brq}{}^{\mu}\partial_{\mu}\To~~~\\
{}&=&\textstyle{\frac{1}{\sqrt{2}}}\bre_{\brq}{}^{\nu}\!\left(\trd_{\mu}\EK^{\mu}{}_{\nu}-2\partial_{\mu}\phi\,\EK^{\mu}{}_{\nu}+\half H_{\nu\lambda\mu}\EK^{\lambda\mu}-\half\partial_{\nu}\To\right)\,.
\ea
\ee
Thus, the conservation law reduces to the following two sets of equations, 
\begin{eqnarray}
\na^{\mu}\EK_{(\mu\nu)}-2\partial^{\mu}\phi\,\EK_{(\mu\nu)}+\half H_{\nu}{}^{\lambda\mu}\EK_{[\lambda\mu]}-\half\partial_{\nu}\To\equiv0\,,&&\label{conKT1}\\
\na^{\mu}\!\left(e^{-2\phi}\EK_{[\mu\nu]}\right)\equiv0\,.&&\label{conKT2}
\end{eqnarray}
In fact, for the above computations,  we first  put, \textit{c.f.~}(\ref{RiemannS}), 
\be
\EK_{p\brq}\equiv \half e_{p}{}^{\mu}\bre_{\brq}{}^{\nu}\EK_{\mu\nu}\qquad\Longleftrightarrow\qquad\EK_{\mu\nu}\equiv 2e_{\mu}{}^{p}\bre_{\nu}{}^{\brq}\EK_{p\brq}\,,
\label{cKpmu}
\ee
and then  let the  Greek indices of $\EK_{\mu\nu}$ be raised  by  the Riemannian metric~(\ref{eeg}),  $g^{\mu\nu}=e_{p}{}^{\mu}e^{p\nu}=-\bre_{\brp}{}^{\mu}\bre^{\brp\nu}$:
\be
\ba{ll}
\EK^{\mu}{}_{\nu}= g^{\mu\rho}\EK_{\rho\nu}=2e^{p\mu}\bre_{\nu}{}^{\brq}\EK_{p\brq}\,;\qquad&\qquad
\EK_{\mu}{}^{\nu}= g^{\nu\rho}\EK_{\mu\rho}=-2e_{\mu}{}^{p}\bre^{\nu\brq}\EK_{p\brq}\,.
\ea
\label{EKEKEK}
\ee
\noindent It follows that
\be
V_{A}{}^{p}\brV_{B}{}^{\brq}\EK_{p\brq}
=\quarter\left(\ba{cc}
-\EK^{\mu\nu}\quad&\quad \EK^{\mu}{}_{\sigma}+\EK^{\mu\lambda}B_{\lambda\sigma}\\
-\EK_{\rho}{}^{\nu}-B_{\rho\kappa}\EK^{\kappa\nu}\quad&\quad
\EK_{\rho\sigma}+B_{\rho\kappa}\EK^{\kappa\lambda}B_{\lambda\sigma}+B_{\rho}{}^{\kappa}\EK_{\kappa\sigma}+\EK_{\rho\lambda}B^{\lambda}{}_{\sigma}
\ea\right)\,,
\ee
and 
\be
\ba{lll}
\!\EM_{AB}\!\!&=&4V_{[A}{}^{p}\brV_{B]}{}^{\brq}\EK_{p\brq}-\half\cJ_{AB}\To\\
{}&=&\!\!\!\left(\ba{cc}
-\EK^{[\mu\nu]}& \EK^{(\mu\lambda)}g_{\lambda\sigma}
+\EK^{[\mu\lambda]}B_{\lambda\sigma}-\half\delta^{\mu}{}_{\sigma}\To\\
\!\!-g_{\rho\kappa}\EK^{(\kappa\nu)}-B_{\rho\kappa}\EK^{[\kappa\nu]}-\half\delta_{\rho}{}^{\nu}\To&\quad
\EK_{[\rho\sigma]}+B_{\rho\kappa}\EK^{[\kappa\lambda]}B_{\lambda\sigma}+B_{\rho}{}^{\kappa}\EK_{(\kappa\sigma)}+\EK_{(\rho\lambda)}B^{\lambda}{}_{\sigma}
\ea\right).
\ea
\label{TcK}
\ee

\noindent The Einstein Double Field Equations~(\ref{EDFE}) reduce, upon Riemannian backgrounds~(\ref{00V}),   (\ref{RiemannS}), to
\begin{eqnarray}
R_{\mu\nu}+2\trd_{\mu}(\partial_{\nu}\phi)-\quarter H_{\mu\rho\sigma}H_{\nu}{}^{\rho\sigma}
&=&8\pi G\EK_{(\mu\nu)}\,,\label{EDFER1}\\
\trd^{\rho}\!\left(e^{-2\phi}H_{\rho\mu\nu}\right)&=&16\pi Ge^{-2\phi}\EK_{[\mu\nu]}\,,\label{EDFER2}\\
R+4\Box\phi-4\partial_{\mu}\phi\partial^{\mu}\phi-\textstyle{\frac{1}{12}}H_{\lambda\mu\nu}H^{\lambda\mu\nu}&=&8\pi G\To\,.\label{EDFER3}
\end{eqnarray}
These also imply the two reduced conservation laws, (\ref{conKT1}) and (\ref{conKT2}), as ${\cD_{A}G^{AB}}=0$ is an off-shell  identity.  Explicitly, we have
\be
\trd_{\mu}\trd_{\nu}\!\left(e^{-2\phi}H^{\lambda\mu\nu}\right)=
\half\left[\trd_{\mu},\trd_{\nu}\right]\!\left(e^{-2\phi}H^{\lambda\mu\nu}\right)=
\half R^{\lambda}{}_{[\rho\mu\nu]}e^{-2\phi}H^{\rho\mu\nu}+R_{[\mu\nu]}e^{-2\phi}H^{\lambda\mu\nu}=0\,,
\ee
which implies the second conservation law~(\ref{conKT2}). On the other hand,   solving $R_{\mu\nu}$ and $R\,$ from (\ref{EDFER1}) and (\ref{EDFER3}) respectively,  we get
\be
\ba{lll}
0\,\,=\,\,\trd^{\mu}(R_{\mu\nu}-\half g_{\mu\nu}R)&=&8\pi G\left(\trd^{\mu}\EK_{(\mu\nu)}-2\partial^{\mu}\phi\EK_{(\mu\nu)}+\half H_{\nu}{}^{\rho\sigma}\EK_{[\rho\sigma]}
-\half\partial_{\nu}\To\right)\\
{}&{}&-2\left(R_{\mu\nu}+2\trd_{\mu}\partial_{\nu}\phi-\quarter H_{\mu\rho\sigma}H_{\nu}{}^{\rho\sigma}-8\pi G\EK_{(\mu\nu)}\right)\partial^{\mu}\phi\\
{}&{}&+\quarter e^{2\phi}\left[
\trd^{\mu}\left(e^{-2\phi}H_{\mu\rho\sigma}\right)-16\pi Ge^{-2\phi}\EK_{[\rho\sigma]}\right]\\
{}&{}&+\textstyle{\frac{1}{3}}H^{\mu\rho\sigma}\partial_{[\mu}H_{\rho\sigma\nu]}\,,
\ea
\label{conconserv}
\ee
where we have used the identity  $\Box\partial_{\nu}\phi-\trd_{\nu}\Box\phi=R_{\nu\rho}\partial^{\rho}\phi\,$.  The last three lines in (\ref{conconserv}) vanish separately up to (\ref{EDFER1}), (\ref{EDFER2}), and the closedness of the $H$-flux. Therefore, we recover the first conservation law~(\ref{conKT1}) as the vanishing of the first line on the right-hand side of the equality above.

\subsubsection{Examples\label{SECexample}}
Here we list various matter fields coupled to Stringy Gravity and  write down their contributions to  the stringy  Energy-Momentum tensor~(\ref{EMSG}),  (\ref{TcK}).

\begin{itemize}
\item \textbf{Cosmological constant}\\
In Stringy Gravity, the cosmological constant term is given by a constant,  $\Lambda_{\DFT}$,  times the integral measure, $e^{-2d}$~\cite{Jeon:2011cn},  
\be
\textstyle{\frac{1}{16\pi G}}e^{-2d}\left(\So-2\Lambda_{\DFT}\right).
\ee
The corresponding Energy-Momentum tensor is 
\be
\EM_{AB}=-\textstyle{\frac{1}{8\pi G}}\cJ_{AB}\Lambda_{\DFT}\,,
\ee
such that $\EK_{p\brq}=0$ and $\To=\frac{1}{4\pi G}\Lambda_{\DFT}$.
~\\
\item\textbf{Scalar field}\\
A free scalar field  in Stringy Gravity  is described by, \textit{e.g.~}\cite{Choi:2015bga}, 
\be
L_{\Phi}=-\half\cH^{MN}\partial_{M}\Phi\partial_{N}\Phi-\half\mPhi^{2}\Phi^{2}
=\half(\brP^{MN}-P^{MN})\partial_{M}\Phi\partial_{N}\Phi-\half\mPhi^{2}\Phi^{2}\,.
\ee
It is straightforward to see, with  $\partial_{p}\equiv V^{A}{}_{p}\partial_{A}$,  $\partial_{\brq}\equiv\brV^{A}{}_{\brq}\partial_{A}$, 
\be
\ba{ll}
\EK_{p\brq}=\partial_{p}\Phi\partial_{\brq}\Phi\,,\qquad&\qquad\To=\cH^{MN}\partial_{M}\Phi\partial_{N}\Phi+\mPhi^{2}\Phi^{2}
=-2L_{\Phi}\,.
\ea
\ee
For Riemannian backgrounds~(\ref{00V}), (\ref{00d}),  and the section choice ${\tpartial^{\mu}=0}$, we have $\partial_{p}=\frac{1}{\sqrt{2}}e_{p}{}^{\mu}\partial_{\mu}$, $\partial_{\brq}=\frac{1}{\sqrt{2}}\bre_{\brq}{}^{\mu}\partial_{\mu}$, and 
\be
\ba{ll}
\EK_{\mu\nu}=\EK_{(\mu\nu)}=\partial_{\mu}\Phi\partial_{\nu}\Phi\,,\quad&\quad
\EK_{[\mu\nu]}=0\,.
\ea
\ee
In particular, each diagonal component of $\EK_{\mu\nu}$ is non-negative, as  $\EK_{\mu\mu}=(\partial_{\mu}\Phi)^{2}$.\\

\item \textbf{Spinor field}\\
Fermionic spinor fields are described by~\cite{Jeon:2011sq}
\be
L_{\psi}=\brpsi\gamma^{p}\cD_{p}\psi+\mpsi\brpsi\psi\,,
\label{fermionL}
\ee
where\footnote{
In the specific case of  four-dimensional spacetime  with Minkowskian signature $[-+++]$,  we may set
\[
\ba{llll}
\left(\gamma^{p}\right)^{\dagger}=-A\gamma^{p}A^{-1}\,,&\quad A^{\dagger}=A&\quad\Longrightarrow\quad&\left(A\gamma^{p_{1}p_{2}\cdots p_{n}}\right)^{\dagger}=(-1)^{\frac{1}{2}n(n+1)}A\gamma^{p_{1}p_{2}\cdots p_{n}}\,,\\
\left(\gamma^{p}\right)^{T}=-C\gamma^{p}C^{-1}\,,&\quad C^{T}=-C&\quad\Longrightarrow\quad&\left(C\gamma^{p_{1}p_{2}\cdots p_{n}}\right)^{T}=-(-1)^{\frac{1}{2}n(n+1)}C\gamma^{p_{1}p_{2}\cdots p_{n}}\,.
\ea
\]
We may  also,  if desired,  identify $A$ with $C$ and use the Majorana, \textit{i.e.~}real representation of the gamma matrices.   Our  analysis  also  holds for   Majorana spinors satisfying   $\psi^{\dagger}A=\psi^{T}C$\,.} $\brpsi=\psi^{\dagger}A$, $\,A=A^{\dagger}$, and $(\gamma^{p})^{\dagger}=-A\gamma^{p}A^{-1}$.\\

Under arbitrary variations of the stringy graviton fields  and the spinor,   the fermionic kinetic term transforms, up to total derivatives (`$\simeq$'), as (\textit{c.f.~}\cite{Jeon:2011sq,Jeon:2012hp})
\be
\ba{lll}
\delta\left(e^{-2d}\brpsi\gamma^{A}\cD_{A}\psi\right)&\simeq&
e^{-2d}\delta V_{Ap}\brV^{A\brq}\,\brpsi\gamma^{p}\cD_{\brq}\psi
+\quarter e^{-2d}\delta\Gamma_{ABC}\,\brpsi\gamma^{A}\gamma^{BC}\psi
\\
{}&{}&
+e^{-2d}\left(\delta\brpsi-\quarter V_{Ap}\delta V^{A}{}_{q}\brpsi\gamma^{pq}-2\delta d\,\brpsi\right)\gamma^{B}\cD_{B}\psi\\
{}&{}&
-e^{-2d}\cD_{B}\brpsi\gamma^{B}\left(\delta\psi+\quarter
V_{Ap}\delta V^{A}{}_{q}\gamma^{pq}\psi\right)\,.
\ea
\label{vF1}
\ee
In the full-order supersymmetric extensions of DFT~\cite{Jeon:2011sq,Jeon:2012hp}, the variation of the stringy Christoffel symbols, $\delta\Gamma_{ABC}$, vanishes automatically, which realizes the `1.5 formalism'. However, in the present example, we do not consider  any supersymmetry nor quartic fermionic terms. Instead, we  proceed, with $\delta\Gamma_{[ABC]}=0$, to obtain
\be
\ba{ll}
\!\!\quarter e^{-2d}\delta\Gamma_{ABC}\,\brpsi\gamma^{A}\gamma^{BC}\psi
\!\!&=\half e^{-2d} P^{AB}\delta\Gamma_{ABC}\,\brpsi\gamma^{C}\psi\\
{}\!\!&=
\half e^{-2d}\brpsi\gamma^{A}\psi\left(\cD_{p}\delta V_{A}{}^{p}-2\partial_{A}\delta d\right)+\half e^{-2d}\brpsi\gamma^{p}\psi\,\cD_{A}\delta V^{A}{}_{p}\\
{}&\!\!\!\!\!\!\!\!\simeq e^{-2d}\delta d\left(\cD_{A}\brpsi\gamma^{A}\psi
+\brpsi\gamma^{A}\cD_{A}\psi\right)-e^{-2d}\delta V^{Ap}\left(\cD_{(A}\brpsi\gamma_{p)}\psi+
\brpsi\gamma_{(A}\cD_{p)}\psi\right),
\ea
\ee
and  derive the final form of the variation of the fermionic part of the Lagrangian, \textit{c.f.~}an analogous expression   in GR~(\ref{GRspinor}),
\be
\ba{lll}
\delta\left[e^{-2d}\left(\brpsi\gamma^{A}\cD_{A}\psi+\mpsi\brpsi\psi\right)\right]&\simeq&
\half e^{-2d}\delta V_{A}{}^{p}\brV^{A\brq}\left(\brpsi\gamma_{p}\cD_{\brq}\psi
-\cD_{\brq}\brpsi\gamma_{p}\psi\right)
\\
{}&{}&
+e^{-2d}\left(\delta\brpsi-\delta d\,\brpsi-\quarter V_{A[p}\delta V^{A}{}_{q]}\brpsi\gamma^{pq}\right)\left(\gamma^{B}\cD_{B}\psi+\mpsi\psi\right)\\
{}&{}&
-e^{-2d}\left(\cD_{B}\brpsi\gamma^{B}-\mpsi\brpsi\right)\left(\delta\psi-\delta d\,\psi+\quarter
V_{A[p}\delta V^{A}{}_{q]}\gamma^{pq}\psi\right)\,.
\ea
\label{vF2}
\ee
This result is quite satisfactory: unlike (\ref{vF1}), the hermiticity is now manifest, as the first line on the right-hand side is by itself  real while the second and the third are hermitian conjugate to each other. Further, as discussed in the general setup~(\ref{deltaLmatter}), the infinitesimal  local Lorentz rotation of the spinor field by  $V_{A[p}\delta V^{A}{}_{q]}$ can be absorbed into the equation of motion for the matter  field through $\delta\psi^{\prime}=\delta\psi+\quarter V_{A[p}\delta V^{A}{}_{q]}\gamma^{pq}\psi$. The variation of the DFT dilaton can be also absorbed in the same manner.  Comparing  (\ref{deltaLmatter2}) and (\ref{vF2}), we obtain 
\be
\ba{ll}
\EK_{p\brq}=-\quarter(\brpsi\gamma_{p}\cD_{\brq}\psi
-\cD_{\brq}\brpsi\gamma_{p}\psi)\,,\qquad&\qquad
\To\equiv 0\,.
\ea
\ee
Upon Riemannian backgrounds~(\ref{00V}), fermions thus provide a nontrivial example of asymmetric $\EK_{\mu\nu}$,
\be
\EK_{\mu\nu}:=2e_{\mu}{}^{p}\bre_{\nu}{}^{\brq}\EK_{p\brq}=
-\textstyle{\frac{1}{2\sqrt{2}}}(\brpsi\gamma_{\mu}\trd_{\nu}\psi
-\trd_{\nu}\brpsi\gamma_{\mu}\psi)\neq\EK_{\nu\mu}\,.
\label{spinorEK}
\ee
Finally, if we redefine the field in terms of a spinor  density, $\chi:=e^{-d}\psi$, with weight $\omega=\half$, the DFT dilaton decouples from the Lagrangian completely,
\be
e^{-2d}L_{\psi}=e^{-2d}\left(\brpsi\gamma^{p}\cD_{p}\psi+\mpsi\brpsi\psi\right)=\bar{\chi}\gamma^{p}\cD_{p}\chi+\mpsi\bar{\chi}\chi\,.
\label{chichi}
\ee
Like fundamental strings, \textit{c.f.~}(\ref{stringL}), the weightful  spinor field $\chi$  couples only to  the DFT vielbeins (or  $g_{\mu\nu}$ and $B_{\mu\nu}$  for Riemannian backgrounds)~\cite{Choi:2015bga}. In this case, $\To=0$ holds off-shell.\\

\item \textbf{Yang--Mills}\\
With the field strength~(\ref{YMF}),  the Yang--Mills theory is coupled to Stringy Gravity   by~\cite{Choi:2015bga} (\textit{c.f.~}\cite{Jeon:2011kp})
\be
L_{\rm{YM}}=\Tr\left[\bfF_{p\brq}\bfF^{p\brq}\right]\,.
\ee
The corresponding stringy  Energy-Momentum tensor is given, from \cite{Park:2015bza}, by
\be
\ba{ll}
\EK_{p\brq}=-\Tr\left[\bfF_{pr}\bfF^{r}{}_{\brq}-\bfF_{p\brr}\bfF^{\brr}{}_{\brq}+\cD^{M}\left({{\bfF_{p\brq}\bfA_{M}}}\right)\right]\,,\quad&\quad
\To=-2\Tr\left[\bfF_{p\brq}\bfF^{p\brq}\right]\,.
\ea
\ee
When the doubled  Yang--Mills gauge potential   satisfies   the extra condition, $\bfA^{M}\partial_{M}=0$~(\ref{AAsection}),   the above expression  reduces to, with (\ref{standardF}), 
\be
\EK_{p\brq}=\Tr\left[\bfF_{p}{}^{\brr}\left(\bfF_{\brr\brq}+\bfA^{M}\brPhi_{M\brr\brq}\right)
-\left(\bfF_{pr}+\bfA^{M}\Phi_{Mpr}\right)\bfF^{r}{}_{\brq}\right]
=-\Tr\left[F_{KL}\cH^{LM}F_{MN}V^{K}{}_{p}\brV^{N}{}_{\brq}\right]\,.
\ee

\item \textbf{Ramond--Ramond sector}\\
The R--R sector of the critical superstring has the kinetic term~\cite{Jeon:2012kd,Jeon:2012hp} (\textit{c.f.~}\cite{Rocen:2010bk,Hohm:2011zr,Hohm:2011dv})
\be
L_{\rm{RR}}=\half\Tr(\cF\brcF)\,,
\label{RRL}
\ee
where $\cF=\cD_{+}\cC$ is the R--R field strength given by  the nilpotent differential operator $\cD_{+}$~(\ref{Dpm}) acting on the  $\Spint\times\oSpint$ bi-spinorial R--R potential $\cC^{\alpha}{}_{\bralpha}$;   $\brcF=\brC^{-1}\cF^{T}C$ is the  charge conjugation of $\cF$; and the trace is taken over the $\Spint$ spinorial indices.  This formalism  is `democratic',  as in \cite{Bergshoeff:2001pv}, and needs to be supplemented by a self-duality relation,
\be
\gammae\cF\equiv\cF\,.
\ee
The R--R sector contributes to the stringy  Energy-Momentum tensor, from (25) of \cite{Jeon:2012hp} as well as  (3.3) of \cite{Jeon:2012kd},  by
\be
\ba{ll}
\EK_{p\brq}=-\quarter\Tr(\gamma_{p}\cF\brgamma_{\brq}\brcF)\,,\qquad&\qquad\To\equiv0\,.
\ea
\ee
Upon Riemannian reduction, $\EK_{\mu\nu}=2e_{\mu}{}^{p}\bre_{\nu}{}^{\brq}\EK_{p\brq}$ is generically asymmetric, $\EK_{\mu\nu}\neq\EK_{\nu\mu}$, which can be verified explicitly   after taking   the diagonal gauge  of the twofold local Lorentz symmetries and then  expanding the R--R potential $\cC^{\alpha}{}_{\brbeta}$ in terms of conventional $p$-form fields~\cite{Jeon:2012kd}.  The asymmetry is also consistent with the observation that the  $\ODD$-covariant nilpotent differential operator $\cD_{+}$ (\ref{Dpm})  reduces to the $H$-twisted exterior derivative, $\rd_{H}=\rd+ H_{\scriptscriptstyle{(3)}}\wedge~$, such that the $B$-field contributes  to  the R--R kinetic term~(\ref{RRL}) in a nontrivial manner.\\


\item\textbf{Point particle}\\
We consider the doubled-yet-gauged particle action from \cite{Ko:2016dxa} and  write   the corresponding Lagrangian density using a Dirac delta function,
\be
e^{-2d}L_{\rm{particle}}=
\int\rd\tau~\big[\,e^{-1\,}\rD_{\tau}y^{A}\rD_{\tau}y^{B}\cH_{AB}(x)-\quarter m^{2}e\,\big]\delta^{D\!}\big(x-y(\tau)\big)\,,
\label{particleL}
\ee
where  $D_{\tau}y^{M}=\frac{\rd~}{\rd\tau}{y}^{M}(\tau)-\cA^{M}$ is the gauged infinitesimal one-form~(\ref{DX}). Integrating the above  over a section, $\int_{\Sigma}e^{-2d}L_{\rm{particle}}$, one can  recover  precisely the action in  \cite{Ko:2016dxa}.   Note that in Stringy Gravity, the Dirac delta function itself should satisfy the section condition and meet the  defining property
\be
\dis{\int_{\Sigma}\Phi(x)\delta^{D\!}(x-y)=\Phi(y)\,.}
\ee
It follows  straightforwardly that
\be
\ba{ll}
\EK_{p\brq}=
-\dis{\int\rd\tau}~2e^{-1\,}(\rD_{\tau}y)_{p}(\rD_{\tau}y)_{\brq}\,e^{2d(x)}\delta^{D\!}\big(x-y(\tau)\big)\,,
\quad&\quad\To=0\,,
\ea
\ee
where, naturally,  $(\rD_{\tau}y)_{p}=\rD_{\tau}y^{A}V_{Ap}$ and $\,(\rD_{\tau}y)_{\brq}=\rD_{\tau}y^{A}\brV_{A\brq}$.\\

Upon Riemannian backgrounds~(\ref{00V}), (\ref{00d}),  with  ${\tpartial^{\mu}\equiv0}$ and the on-shell value of the gauge connection, $\cA^{M}=(A_{\mu},0)\equiv(\frac{\rd~}{\rd\tau}\ty_{\mu}-B_{\mu\nu}\frac{\rd~}{\rd\tau} y^{\nu},0)$~\cite{Ko:2016dxa},  we have
\be
\EK_{\mu\nu}=2e_{\mu}{}^{p}\bre_{\nu}{}^{\brq}\EK_{p\brq}=
{\dis{\int\rd\tau}}~2e^{-1\,}g_{\mu\rho} g_{\nu\sigma}\frac{\rd y^{\rho}}{\rd\tau}\frac{\rd y^{\sigma}}{\rd\tau}\,\frac{\delta^{D\!}\big(x-y(\tau)\big)e^{2\phi}}{\sqrt{-g}}\,,
\ee
which is symmetric, $\EK_{\mu\nu}=\EK_{\nu\mu}$,  as one may well expect for the point particle. Each diagonal component is non-negative, $\EK_{\mu\mu}\geq0$,  as
$\big(g_{\mu\rho}\frac{\rd y^{\rho}}{\rd\tau}\big)^{2}\geq 0$.\\

\item\textbf{String}\\
In a similar fashion,  the doubled-yet-gauged  bosonic string action~\cite{Hull:2006va}\cite{Lee:2013hma} gives
\be
e^{-2d}L_{\rm{string}}={\textstyle{\frac{1}{4\pi\alpha^{\prime}}}}{\dis{\int}}\rd^{2}\sigma\left[-\half\sqrt{-h}h^{\alpha\beta}\rD_{\alpha}y^{A}\rD_{\beta}y^{B}\cH_{AB}(x)-\epsilon^{\alpha\beta}\rD_{\alpha}x^{A}\cA_{\beta A}\right]\delta^{D\!}\big(x-y(\sigma)\big)\,,
\label{stringL}
\ee
and hence
\be
\ba{ll}
\EK_{p\brq}={\textstyle{\frac{1}{4\pi\alpha^{\prime}}}}{\dis{\int}}\rd^{2}\sigma\sqrt{-h}h^{\alpha\beta}(\rD_{\alpha}y)_{p}(\rD_{\beta}y)_{\brq}\,e^{2d(x)}\delta^{D\!}\big(x-y(\sigma)\big)\,,
\quad&\quad\To=0\,.
\ea
\ee
Upon reduction to Riemannian backgrounds~(\ref{00V}), (\ref{00d}), the on-shell value of the gauge connection is $\cA_{\alpha}^{M}=(A_{\alpha\mu},0)\equiv(\partial_{\alpha}\ty_{\mu}-B_{\mu\nu}\partial_{\alpha}y^{\nu}+\frac{1}{\sqrt{-h}}\epsilon_{\alpha}{}^{\beta}g_{\mu\nu}\partial_{\beta}y^{\nu},0)$~\cite{Lee:2013hma},  and we have
\be
\EK_{\mu\nu}=2e_{\mu}{}^{p}\bre_{\nu}{}^{\brq}\EK_{p\brq}=\,-\,{\textstyle{\frac{1}{2\pi\alpha^{\prime}}}}{\dis{\int}}\rd^{2}\sigma~g_{\mu\rho}g_{\nu\sigma}
\left(\sqrt{-h}h^{\alpha\beta}
+\epsilon^{\alpha\beta}\right)\partial_{\alpha}y^{\rho}\partial_{\beta}y^{\sigma}
\,\frac{\delta^{D\!}\big(x-y(\tau)\big)e^{2\phi}}{\sqrt{-g}}\,,
\label{stringEK}
\ee
which is generically asymmetric, $\EK_{\mu\nu}\neq\EK_{\nu\mu}$, due to the $B$-field.  It is easy to see in  lightcone gauge  that the  diagonal  components of   $\EK_{\mu\nu}$, or $g_{\mu\rho}\partial_{+}y^{\rho}g_{\mu\sigma}\partial_{-}y^{\sigma}$,  are not necessarily positive.\\

\noindent The above analysis  further generalizes to the doubled-yet-gauged  Green-Schwarz superstring~\cite{Park:2016sbw},
\be
\ba{ll}
\EK_{p\brq}={\textstyle{\frac{1}{4\pi\alpha^{\prime}}}}{\dis{\int}}\rd^{2}\sigma\sqrt{-h}h^{\alpha\beta}\Pi_{\alpha p}\Pi_{\beta\brq}\,e^{2d(x)}\delta^{D\!}\big(x-y(\sigma)\big)\,,
\quad&\quad\To=0\,,
\ea
\ee
where $\Pi_{\alpha}^{M}=\partial_{\alpha}y^{M}-\cA_{\alpha}^{M}-i\brtheta\gamma^{M}\partial_{\alpha}\theta-i\brtheta^{\prime}\brgamma^{M}\partial_{\alpha}\theta^{\prime}$ is the supersymmetric extension of $D_{\alpha}y^{M}$. 

Upon Riemannian reduction with  $\Pi_{\alpha}^{M}=(\widetilde{\Pi}_{\alpha\mu},\Pi_{\alpha}^{\nu})$, the on-shell value of the gauge connection $\cA_{\alpha}^{M}$ sets $\widetilde{\Pi}_{\alpha\mu}-B_{\mu\nu}\Pi_{\alpha}^{\nu}+\frac{1}{\sqrt{-h}}\epsilon_{\alpha}{}^{\beta}g_{\mu\nu}\Pi_{\beta}^{\nu}\equiv0$~\cite{Park:2016sbw}, and  the corresponding $\EK_{\mu\nu}$ is similarly asymmetric,  
\be
\EK_{\mu\nu}=2e_{\mu}{}^{p}\bre_{\nu}{}^{\brq}\EK_{p\brq}=\,-\,{\textstyle{\frac{1}{2\pi\alpha^{\prime}}}}{\dis{\int}}\rd^{2}\sigma~g_{\mu\rho}g_{\nu\sigma}
\left(\sqrt{-h}h^{\alpha\beta}
+\epsilon^{\alpha\beta}\right)\Pi_{\alpha}^{\rho}\Pi_{\beta}^{\sigma}
\,\frac{\delta^{D\!}\big(x-y(\tau)\big)e^{2\phi}}{\sqrt{-g}}\,.
\ee
The asymmetry, $\EK_{\mu\nu}\neq\EK_{\nu\mu}$, is a genuine stringy property.
\end{itemize}


\section{Further-generalized Lie derivative, $\fcL_{\xi}$ \label{SECFURTHER}}
In analogy with GR, the notion of isometries in Stringy Gravity  can be naturally addressed through the generalized Lie derivative~(\ref{hcL}), which can also, from  (\ref{untwtorsionless}),   be expressed using DFT-covariant derivatives~(\ref{asemicov}),  leading to the  DFT-Killing equations~(\ref{KillingPd})~\cite{Park:2015bza},
\be
\ba{ll}
\hcL^{\partial}_{\xi}P_{AB}=\hcL^{\na}_{\xi}P_{AB}=4\brP_{(A}{}^{C}P_{B)}{}^{D}\na_{[C}\xi_{D]}\stackrel{!}{=} 0\,,\quad&\qquad
\hcL^{\partial}_{\xi}d=\hcL^{\na}_{\xi}d=-\half\na_{A}\xi^{A}\stackrel{!}{=} 0\,,
\ea
\label{KillingPd2}
\ee
where the final equalities with `$\,!\,$' hold only in the case of an isometry.  However, in the vielbein formulation of Stringy Gravity this result needs to be further generalized, as one should  be able to  construct Killing equations for the DFT vielbeins. 

\noindent  Using   $V^{A}{}_{p}\brV^{B}{}_{\brq\,}\hcL_{\xi}\brP_{AB}=-V^{A}{}_{p}\brV^{B}{}_{\brq\,}\hcL_{\xi}P_{AB}=2\cD_{[p}\xi_{\brq]}$ and the fact that the DFT vielbeins are covariantly constant,  $\cD_{A}V_{Bp}=\cD_{A}\brV_{B\brp}=0$ (\ref{cDVV}),  their generalized Lie derivatives can be   related to the generalized Lie derivatives of the projectors,  
\be
\ba{l}
\hcL_{\xi}V_{Ap}=\brP_{A}{}^{B}V^{C}{}_{p}\left(\hcL_{\xi}P_{BC}\right)-\left(\xi^{B}\Phi_{Bpq}+2\cD_{[p}\xi_{q]}\right)V_{A}{}^{q}\,,\\
\hcL_{\xi}\brV_{A\brp}=P_{A}{}^{B}\brV^{C}{}_{\brp}\left(\hcL_{\xi}\brP_{BC}\right)-\left(\xi^{B}\brPhi_{B\brp\brq}
+2\cD_{[\brp}\xi_{\brq]}\right)\brV_{A}{}^{\brq}\,.
\ea
\ee
These expressions are quite instructive, as we can arrange them as
\be
\ba{l}
\xi^{B}\cD_{B}V_{Ap}+2\cD_{[A}\xi_{B]}V^{B}{}_{p}+2
\cD_{[p}\xi_{q]}V_{A}{}^{q}=\brP_{A}{}^{B}\!\left(\hcL_{\xi}P_{BC}\right)\!V^{C}{}_{p}\,,\\
\xi^{B}\cD_{B}\brV_{A\brp}+2\cD_{[A}\xi_{B]}\brV^{B}{}_{\brp}+2\cD_{[\brp}\xi_{\brq]}\brV_{A}{}^{\brq}=P_{A}{}^{B}\!\left(\hcL_{\xi}\brP_{BC}\right)\!\brV^{C}{}_{\brp}\,.
\ea
\label{design}
\ee
Motivated by the expressions of the left-hand sides above, we propose to generalize the generalized Lie derivative one step further by constructing a   \textit{further-generalized Lie derivative,} $\fcL_{\xi}$, which can  act on an arbitrary tensor density  carrying  $\ODD$ and $\SpinD\times\oSpinD$  indices\footnote{\textit{.c.f.~}\cite{Coimbra:2011nw,Hohm:2011zr,Hohm:2011dv}  where the generalized Lie derivative was extended to act not on local Lorentz  but on  $\ODD$ spinors.   Our further-generalized Lie derivative acts on both $\ODD$ and  $\SpinD\times\oSpinD$ indices.  } as 
\be
\ba{lll}
\fcL_{\xi}T_{Mp\brp}{}^{\alpha}{}_{\beta}{}^{\bralpha}{}_{\brbeta}&:=&\xi^{N}\cD_{N}T_{Mp\brp}{}^{\alpha}{}_{\beta}{}^{\bralpha}{}_{\brbeta}+\omega\cD_{N}\xi^{N}T_{Mp\brp}{}^{\alpha}{}_{\beta}{}^{\bralpha}{}_{\brbeta}+2\cD_{[M}\xi_{N]}T^{N}{}_{p\brp}{}^{\alpha}{}_{\beta}{}^{\bralpha}{}_{\brbeta}\\
{}&{}&+2\cD_{[p}\xi_{q]}T_{M}{}^{q}{}_{\brp}{}^{\alpha}{}_{\beta}{}^{\bralpha}{}_{\brbeta}+
\half\cD_{[r}\xi_{s]}(\gamma^{rs})^{\alpha}{}_{\delta}T_{Mp\brp}{}^{\delta}{}_{\beta}{}^{\bralpha}{}_{\brbeta}- \half\cD_{[r}\xi_{s]}(\gamma^{rs})^{\delta}{}_{\beta}T_{Mp\brp}{}^{\alpha}{}_{\delta}{}^{\bralpha}{}_{\brbeta}\\
{}&{}&+2\cD_{[\brp}\xi_{\brq]}T_{Mp}{}^{\brq\alpha}{}_{\beta}{}^{\bralpha}{}_{\brbeta}+
\half\cD_{[\brr}\xi_{\brs]}(\brgamma^{\brr\brs})^{\bralpha}{}_{\brdelta}T_{Mp\brp}{}^{\alpha}{}_{\beta}{}^{\brdelta}{}_{\brbeta}-
\half\cD_{[\brr}\xi_{\brs]}(\brgamma^{\brr\brs})^{\brdelta}{}_{\brbeta}T_{Mp\brp}{}^{\alpha}{}_{\beta}{}^{\bralpha}{}_{\brdelta}\,.
\ea
\label{fgL}
\ee
In short,  the further-generalized Lie derivative comprises    the original generalized Lie derivative and  additional    infinitesimal  local Lorentz  rotations  given by the  terms 
\be
\ba{l}
\xi^{A}\Phi_{Apq}+2\cD_{[p}\xi_{q]}=2\partial_{[p}\xi_{q]}+\Phi_{\brr pq}\xi^{\brr}+3\Phi_{[pqr]}\xi^{r}\,,\\
\xi^{A}\brPhi_{A\brp\brq}+2\cD_{[\brp}\xi_{\brq]}=2\partial_{[\brp}\xi_{\brq]}+\brPhi_{r\brp\brq}\xi^{r}
+3\brPhi_{[\brp\brq\brr]}\xi^{\brr}\,,
\ea
\label{additional}
\ee
such that, for example,
\be
\ba{lll}
\fcL_{\xi}T_{Mp\brp}{}^{\alpha\bralpha}&\!\!=\!\!&\hcL_{\xi}T_{Mp\brp}{}^{\alpha\bralpha}
+\left(\xi^{N}\Phi_{Npq}+2\cD_{[p}\xi_{q]}\right)T_{M}{}^{q}{}_{\brp}{}^{\alpha\bralpha}+
\quarter\left(\xi^{N}\Phi_{Nrs}+2\cD_{[r}\xi_{s]}\right)(\gamma^{rs})^{\alpha}{}_{\beta}T_{Mp\brp}{}^{\beta\bralpha}\\
{}&{}&\quad+\left(\xi^{N}\brPhi_{N\brp\brq}+2\cD_{[\brp}\xi_{\brq]}\right)T_{Mp}{}^{\brq\alpha\bralpha}+
\quarter\left(\xi^{N}\brPhi_{N\brr\brs}+2\cD_{[\brr}\xi_{\brs]}\right)
(\brgamma^{\brr\brs})^{\bralpha}{}_{\brbeta}T_{Mp\brp}{}^{\alpha\brbeta}\,.
\ea
\label{fgL2}
\ee
The further-generalization is also equivalent to replacing the ordinary  derivatives in the original generalized Lie derivative by  master derivatives, $\hcL_{\xi}^{\partial}\rightarrow\hcL_{\xi}^{\cD}$, and adding the local Lorentz rotations, $2\cD_{[p}\xi_{q]}$ and $2\cD_{[\brp}\xi_{\brq]}$.

It is then crucial to note that the further-generalized Lie derivative is completely covariant  for  both  the  twofold  local Lorentz symmetries and the doubled-yet-gauged  diffeomorphisms, as follows. The local Lorentz covariance is  guaranteed  by the use of the master derivatives everywhere in the  definition~(\ref{fgL}).  The diffeomorphism covariance also holds,  since the original generalized Lie derivative and the additional local Lorentz rotations~(\ref{additional}) are separately diffeomorphism covariant,   from (\ref{covhcL})  and (\ref{covPhi}).   For example, the further-generalized Lie derivative acting on a $\SpinD$ spinor field, $\psi^{\alpha}$,  reads
\be
\fcL_{\xi}\psi=\xi^{M}\cD_{M}\psi+\half\cD_{[p}\xi_{q]}\gamma^{pq}\psi
= \xi^{\brp}\cD_{\brp}\psi+\half\gamma^{p}\cD_{p}\!\left(\gamma^{q}\xi_{q}\psi\right)+
\half\gamma^{q}\xi_{q}\!\left(\gamma^{p}\cD_{p}\psi\right)-\half\left(\cD_{p}\xi^{p}\right)\!\psi\,.
\ee
As expected, or directly seen from (\ref{covDirac}), this expression is completely covariant under both the  local Lorentz  and the doubled-yet-gauged diffeomorphism transformations. It is advantageous to use $\fcL_{\xi}$ because of its local Lorentz covariance; whereas, in contrast, $\hcL_{\xi}$ is not locally Lorentz covariant.

The isometry of the DFT vielbeins is then characterized by the vanishing of  their  further-generalized Lie derivatives, which is then,   with  (\ref{deltaPbrP}), (\ref{KillingPd2}), (\ref{design}),  equivalent to nothing but the isometry of the projectors, as
\be
\ba{ll}
\fcL_{\xi}V_{Ap}=\brP_{A}{}^{B}\!\left(\hcL_{\xi}P_{BC}\right)\!V^{C}{}_{p}
=\left(\hcL_{\xi}P_{AC}\right)\!V^{C}{}_{p}\,,\quad&\quad\!
\fcL_{\xi}\brV_{A\brp}=P_{A}{}^{B}\!\left(\hcL_{\xi}\brP_{BC}\right)\!\brV^{C}{}_{\brp}=\left(\hcL_{\xi}\brP_{AC}\right)\!\brV^{C}{}_{\brp}\,.
\ea
\label{fcLVV}
\ee
Moreover,   gamma and charge conjugation matrices, $(\gamma^{p})^{\alpha}{}_{\beta}$, $(\brgamma^{\brp})^{\bralpha}{}_{\brbeta}$,  $C_{\alpha\beta}$, $\brC_{\bralpha\brbeta}$, are all compatible  with  $\fcL_{\xi}$,
\be
\ba{rll}
\fcL_{\xi}\gamma^{p}&\!=\!&
\xi^{N}\cD_{N}\gamma^{p}+2\cD^{[p}\xi^{q]}\gamma_{q}
+\half\cD_{[r}\xi_{s]}\left[\gamma^{rs},\gamma^{p}\right]\,=\,0\,,\\
\fcL_{\xi}\brgamma^{\brp}&\!=\!&
\xi^{N}\cD_{N}\brgamma^{\brp}+2\cD^{[\brp}\xi^{\brq]}\brgamma_{\brq}
+\half\cD_{[\brr}\xi_{\brs]}\left[\brgamma^{\brr\brs},\brgamma^{\brp}\right]\,=\,0\,,
\ea
\ee
\be
\ba{rll}
\fcL_{\xi}C_{\alpha\beta}&\!=\!&- \half\cD_{[p}\xi_{q]}\left(C\gamma^{pq}+(\gamma^{pq})^{T}C\right)_{\alpha\beta}\,=\,0\,,\\
\fcL_{\xi}\brC_{\bralpha\brbeta}&\!=\!&- \half\cD_{[\brp}\xi_{\brq]}\left(\brC\brgamma^{\brp\brq}+
(\brgamma^{\brp\brq})^{T}\brC\right)_{\bralpha\brbeta}\,=\,0\,.
\ea
\ee
~\\
The further-generalized Lie derivative  is closed by the C-bracket~(\ref{Cbracket}) and  twofold  local Lorentz rotations,
\be
\left[\fcL_{\zeta},\fcL_{\xi}\right]=\fcL_{\left[\zeta,\xi\right]_{\rm{C}}}+\omega_{pq}(\zeta,\xi)+\bromega_{\brp\brq}(\zeta,\xi)\,,
\label{fclosed}
\ee
of which the two infinitesimal local Lorentz rotation parameters are specified  by $\zeta^{M}$ and $\,\xi^{M}$ as
\be
\ba{l}
\omega_{pq}(\zeta,\xi)=-\omega_{qp}(\zeta,\xi)=
(\cD_{\brp}\zeta_{p}-\cD_{p}\zeta_{\brp})(\cD^{\brp}\xi_{q}-\cD_{q}\xi^{\brp})-
(\cD_{\brp}\zeta_{q}-\cD_{q}\zeta_{\brp})(\cD^{\brp}\xi_{p}-\cD_{p}\xi^{\brp})\,,\\
\bromega_{\brp\brq}(\zeta,\xi)=-\bromega_{\brq\brp}(\zeta,\xi)=
(\cD_{p}\zeta_{\brp}-\cD_{\brp}\zeta_{p})(\cD^{p}\xi_{\brq}-\cD_{\brq}\xi^{p})-
(\cD_{p}\zeta_{\brq}-\cD_{\brq}\zeta_{p})(\cD^{p}\xi_{\brp}-\cD_{\brp}\xi^{p})\,.
\ea
\ee
The closure essentially boils down to\footnote{Note, \textit{e.g.} with $T_{p}=T^{A}V_{Ap}$, 
\[
\ba{lll}
\left[\fcL_{\zeta},\fcL_{\xi}\right]T_{p}&\!=\!&\left[\fcL_{\zeta},\fcL_{\xi}\right]\left(T^{A}V_{Ap}\right)\,=\,\left(\left[\fcL_{\zeta},\fcL_{\xi}\right]T^{A}\right)V_{Ap}+
T^{A}\left[\fcL_{\zeta},\fcL_{\xi}\right]V_{Ap}\\
{}&\!=\!&\left(\fcL_{\left[\zeta,\xi\right]_{\rm{C}}}T^{A}\right)V_{Ap}+
T^{A}\left(\fcL_{\left[\zeta,\xi\right]_{\rm{C}}}V_{Ap}+\omega_{pq}(\zeta,\xi)V_{A}{}^{q}\right)\,=\,\fcL_{\left[\zeta,\xi\right]_{\rm{C}}}\left(T^{A}V_{Ap}\right)+\omega_{pq}(\zeta,\xi)T^{q}\\
{}&\!=\!&\fcL_{\left[\zeta,\xi\right]_{\rm{C}}}T_{p}+\omega_{pq}(\zeta,\xi)T^{q}\,.
\ea
\]} 
\be
\ba{ll}
\left[\fcL_{\zeta},\fcL_{\xi}\right]V_{Mp}=\fcL_{\left[\zeta,\xi\right]_{\rm{C}}}V_{Mp}+\omega_{pq}(\zeta,\xi)V_{M}{}^{q}\,,\quad&\quad
\left[\fcL_{\zeta},\fcL_{\xi}\right]\brV_{M\brp}=\fcL_{\left[\zeta,\xi\right]_{\rm{C}}}\brV_{M\brp}+\bromega_{\brp\brq}(\zeta,\xi)\brV_{M}{}^{\brq}\,,
\ea
\ee
which can easily be verified using   (\ref{closeda}) and  (\ref{fcLVV}).\\
~\\
~\\
~\\


\section{Regular spherical  solution to  Einstein Double Field Equations  \label{SECD4}}
In this section we derive the most general, spherically symmetric,  asymptotically flat,   static, Riemannian, regular solution to the ${D=4}$ Einstein Double Field Equations. Hereafter, we fix the section as ${\tpartial^{\mu}\equiv0}$,  adopt spherical coordinates,   $\{t,r,\vartheta,\varphi\}$, and focus on the Riemannian parametrizations~(\ref{00H})--(\ref{00d}).  We shall encounter  various parameters and variables, as listed in Table~\ref{Tablepv}.  

\begin{table}[H]
\centering
\begin{tabular}{rcl}
\hline
$A(r),B(r),C(r)$&:& Stringy graviton field components 
(\ref{ansatzgB}),  (\ref{SOL2})\\
$D(r),E(r),F(r)$&:& Stringy Energy-Momentum tensor components   (\ref{ansatzcK})\\
$\cV(r), \cW(r), \cX(r), \cY(r), \cZ(r)$&:& Integrals of the stringy E-M tensor  (\ref{cVdef}), (\ref{WXY}), (\ref{Zdef})\\
$\alpha,\beta,a,b,h$ &:&  Spherical vacuum parameters~\cite{Ko:2016dxa}   (\ref{ansatzgB}),  (\ref{alphabeta}), (\ref{fixa}), (\ref{ababh})\\
\hline
\end{tabular}
\caption{Variables and parameters for the spherical solution to the ${D=4}$ Einstein Double Field Equations.  The final form of the solution is summarized in (\ref{SOL2}), (\ref{ababh}), and (\ref{Zdef}),  where the  constant parameters of the  spherical vacuum geometry, $\{\alpha,\beta,a,b,h\}$, are all identified as  integrals of the stringy Energy-Momentum tensor, $T_{AB}$, localized at the center.   }
\label{Tablepv}
\end{table}

\subsection{Most general ${D=4}$ spherical  ansatz}
The spherical symmetry in ${D=4}$ Stringy Gravity is characterized by  three Killing vectors,  $\xi^{N}_{a}$, $a=1,2,3$, which form an  $\so(3)$ algebra through the $\mathbf{C}$-bracket,
\be
\left[\xi_{a},\xi_{b}\right]_{\mathbf{C}}=\sum_{c}\epsilon_{abc}\xi_{c}\,.
\label{so3Hd}
\ee
With  ${\tpartial^{\mu}\equiv0}$, the $\so(3)$ Killing vectors, 
  $\xi^{N}_{a}=(\tilde{\xi}_{a\mu},\xi_{a}^{\nu})$,   decompose explicitly  into   $\tilde{\xi}_{a}=\tilde{\xi}_{a\mu}\rd x^{\mu}$ and   $\xi_{a}=\xi_{a}^{\nu}\partial_{\nu}$~\cite{Ko:2016dxa},
\be
\ba{ll}
\tilde{\xi}_{1}=\frac{\cos\varphi}{\sin\vartheta}\big[h\rd t+B(r)\rd r\big]\,,\quad&\qquad\xi_{1}=\sin\varphi\partial_{\vartheta}
+\cot\vartheta\cos\varphi\partial_{\varphi}\,,\\
\tilde{\xi}_{2}=\frac{\sin\varphi}{\sin\vartheta}\big[h\rd t+B(r)\rd r\big]\,,\quad&\qquad\xi_{2}=-\cos\varphi\partial_{\vartheta}+
\cot\vartheta\sin\varphi\partial_{\varphi}\,,\\
\tilde{\xi}_{3}=0\,,\quad&\qquad\xi_{3}=-\partial_{\varphi}\,,
\ea
\label{Killing}
\ee
where $h$ is constant and $B(r)$ is an arbitrary function of the radius.  The three of  $\xi^{\mu}_{a}\partial_{\mu}$'s are the standard (undoubled) $\so(3)$ angular momentum operators.  In terms of the further-generalized Lie derivative, the spherical symmetry of the  stringy graviton fields  implies
\be
\ba{lllll} 
\fcL_{\xi_{a}}V_{Ap}=0\,,\quad&\quad
\fcL_{\xi_{a}}\brV_{A\brp}=0\,,\quad&\quad
\fcL_{\xi_{a}}P_{AB}=0\,,\quad&\quad
\fcL_{\xi_{a}}\brP_{AB}=0\,,\quad&\quad
\fcL_{\xi_{a}}d=0\,,
\ea
\label{so3VV}
\ee
such that the DFT-Killing equations~(\ref{KillingPd2}) are satisfied,
\be
\ba{ll}
P_{A}{}^{C}\brP_{B}{}^{D}(\na_{C}\xi_{aD}-\na_{D}\xi_{aC})= 0\quad\Longleftrightarrow\quad
(P\na)_{A}(\brP\xi_{a})_{B}=(\brP\na)_{B}(P\xi_{a})_{A}\,,\quad&\quad
\na_{A}\xi_{a}^{A}= 0\,.
\ea
\ee
On Riemannian backgrounds~(\ref{00H}), (\ref{00V}), (\ref{00d}), the above DFT-Killing equations   reduce to
\be
\ba{lll}
\cL_{\xi_{a}}g_{\mu\nu}=0\,,\quad&\qquad
\cL_{\xi_{a}}B_{\mu\nu}
+\partial_{\mu}\tilde{\xi}_{a\nu}-\partial_{\nu}\tilde{\xi}_{a\mu}=0\,,\quad&\qquad\cL_{\xi_{a}}\phi=0\,.
\ea
\label{00KillingPd}
\ee
In addition to the  spherical symmetry we  also require  the static condition, such that all the fields are time-independent, $\partial_{t}\equiv0$.  By utilizing  ordinary diffeomorphisms  we can set  $g_{tr}\equiv0$~\cite{Weinberg:1972kfs},   and hence without loss of generality we can  put the Riemannian  metric into the diagonal form~\cite{Ko:2016dxa}
\be
\ba{l}
\rd s^{2}=e^{2\phi(r)}\left[-A(r)\rd t^{2} +A^{-1}(r)\rd r^{2}+A^{-1}(r)C(r)\,\rd\Omega^{2}\right],\\
B_{\scriptscriptstyle{(2)}}=
B(r)\cos\vartheta\,\rd r\wedge\rd\varphi+ h\cos\vartheta\,\rd t\wedge\rd\varphi\,,
\ea
\label{ansatzgB}
\ee
where  $\rd\Omega^{2}=\rd\vartheta^{2}+\sin^{2}\vartheta\rd\varphi^{2}$ and $B_{\scriptscriptstyle{(2)}}=\half B_{\mu\nu}\rd x^{\mu}\wedge\rd x^{\nu}$.    This ansatz solves the spherical DFT-Killing equations, or (\ref{00KillingPd}), with    four   unknown radial  functions, $A(r)$, $B(r)$, $C(r)$, $\phi(r)$, and one free constant, $h$.  It differs slightly  from the rather well known ansatz in GR~(\ref{metric}), but  accords with the analytic solution in  \cite{Ko:2016dxa}.  The   $H$-flux then  corresponds to the most general spherically symmetric three-form,\footnote{In terms of  Cartesian coordinates, $x^{1}=r\sin\vartheta\cos\varphi, \,x^{2}=r\sin\vartheta\sin\varphi, \,x^{3}=r\cos\vartheta$, we have
\[
\sin\vartheta\,\rd\vartheta\wedge\rd\varphi=\half\epsilon_{ijk}(x^{i}/r^{3})\,\rd x^{j}\wedge\rd x^{k}\,.
\]}
\be
\ba{ll}
H_{\scriptscriptstyle{(3)}}=\rd B_{\scriptscriptstyle{(2)}}=
B(r)\sin\vartheta\,\rd r\wedge\rd\vartheta\wedge\rd\varphi+ h\sin\vartheta\,\rd t\wedge\rd\vartheta\wedge\rd\varphi\,,\quad&\quad\cL_{\xi_{a}}H_{\scriptscriptstyle{(3)}}=0\,.
\ea
\label{Hflux}
\ee
The above ansatz  reduces to the flat Minkowskian spacetime  if and only if $A=1$,  $C=r^{2}$, $B=\phi=h=0$. \\
It is worth expanding the DFT  integral measure,  and its integration over $0\leq\vartheta<\pi$ and $0\leq\varphi<2\pi$,  on the Riemannian background,
\be
\ba{c}
e^{-2d}=e^{-2\phi}\sqrt{-g}=e^{2\phi}A^{-1}C\sin\vartheta=R^{2}\sin\vartheta\,,\\
\intdn\equiv\dis{\int_{0}^{\pi}\rd\vartheta\int_{0}^{2\pi}\rd\varphi}~e^{-2d}=
4\pi e^{2\phi}A^{-1}C=4\pi R^{2}\,,
\ea
\label{emd}
\ee
where  $R$ denotes the so-called  `areal radius',
\be
R:=e^{\phi}\sqrt{C/A}\,.
\label{arealR}
\ee

We  further let the Energy-Momentum tensor, \textit{i.e.~}matter, be spherically symmetric,
\be
\fcL_{\xi_{a}}\EM_{AB}=0\,,
\ee
which, with (\ref{cK}) and (\ref{so3VV}),   decomposes into
\be
\ba{ll}
\fcL_{\xi_{a}}\EK_{p\brq}=0\,,\qquad&\qquad
\fcL_{\xi_{a}}\To=0\,.
\ea
\ee
The latter implies  that $\To(r)$ is another radial function, while the former gives, 
 using the convention ${\EK_{p\brq}=\half e_{p}{}^{\mu}\bre_{\brq}{}^{\nu}\EK_{\mu\nu}}$~(\ref{cKpmu}), 
\be
\cL_{\xi_{a}}\EK_{\mu\nu}=0\,,
\ee
which follows from the generic expression of the  further-generalized Lie derivative acting on $\EK_{p\brq}$,
\be
\fcL_{\xi}\EK_{p\brq}=\quarter e_{p}{}^{\mu}\bre_{\brq}{}^{\nu}\!\left[2\cL_{\xi}\EK_{\mu\nu}
+\left\{2\partial_{[\mu}\tilde{\xi}_{\rho]}+\cL_{\xi}({B-g})_{\mu\rho}\right\}\!g^{\rho\sigma}\EK_{\sigma\nu}
-\left\{2\partial_{[\nu}\tilde{\xi}_{\rho]}+\cL_{\xi}({B+g})_{\nu\rho}\right\}\!
g^{\rho\sigma}\EK_{\mu\sigma}\right],
\ee
together with the isometry condition~(\ref{00KillingPd}).

Combining these results, we arrive at the final form of $\EK_{\mu\nu}$,
\be
\EK_{\mu\nu}=\left(\ba{cccc}
~\EK_{tt}(r)~&~D(r)+E(r)~&~0~&~0~\\
~D(r)-E(r)~&~\EK_{rr}(r)~&~0~&~0~\\
~0~&~0~&~\EK_{\vartheta\vartheta}(r)~&~F(r)\sin\vartheta~\\
~0~&~0~&~-F(r)\sin\vartheta~&~\EK_{\vartheta\vartheta}(r)\sin^{2}\vartheta
\ea\right)\,,
\label{ansatzcK}
\ee
such that there are  six  radial functions, $\EK_{tt}(r)$, $\EK_{rr}(r)$, $\EK_{\vartheta\vartheta}(r)$, $D(r)$, $E(r)$ and $F(r)$, with 
\be
\ba{lll}
\EK_{(tr)}=D(r)\,,\qquad&\qquad
\EK_{[tr]}=E(r)\,,\qquad&\qquad
\EK_{\vartheta\varphi}=-\EK_{\varphi\vartheta}=F(r)\sin\vartheta\,.
\ea
\ee
In particular,  it includes anti-symmetric components which induce a two-form,
\be
\EK_{\scriptscriptstyle{(2)}}:=\half\EK_{[\mu\nu]}\rd x^{\mu}\wedge\rd x^{\nu}=
E(r)\rd t\wedge\rd r+F(r)\sin\vartheta\rd\vartheta\wedge\rd\varphi\,.
\label{cKform}
\ee
This is a novel feature of Stringy Gravity which is not present in GR.
For later use, it is worthwhile to note, from (\ref{cQxi}), (\ref{TcK}), that
\be
\ba{lll}
\EM^{t}{}_{A}\xi^{A}
&=&(\EK_{t}{}^{t}-\half\To)\xi^{t}+g^{tt}\EK_{(tr)}\xi^{r}+\EK^{[tr]}B_{r\varphi}\xi^{\varphi}-\EK^{[tr]}\tilde{\xi}_{r}\\
{}&=&\xi^{t}(\EK_{t}{}^{t}-\half\To)-\xi^{r}e^{-2\phi}A^{-1}D-e^{-4\phi}E(\xi^{\varphi}B\cos\vartheta-\tilde{\xi}_{r})\,,
\ea
\label{EMxi}
\ee
and 
\be
\ba{ll}
\EK_{\vartheta}{}^{\vartheta}=
e^{-2\phi}AC^{-1}\EK_{\vartheta\vartheta}=\EK_{\varphi}{}^{\varphi}\,,\qquad&\qquad
e^{-2d}\EK^{\vartheta\varphi}=e^{-2\phi}AC^{-1}F\,,
\ea
\ee
both of which depend on the radius, $r$, only. \\


\subsection{Solving the Einstein Double Field Equations\label{SECsolve}}
Having prepared the most general spherically symmetric static ansatz,  (\ref{ansatzgB}), (\ref{ansatzcK}),  we now proceed to solve the Einstein Double Field Equations~(\ref{EDFER1}), (\ref{EDFER2}), (\ref{EDFER3}).  Subtracting the `trace' of (\ref{EDFER1}) from (\ref{EDFER3}) and employing the differential form notation of (\ref{Hflux}) and (\ref{cKform}), we focus on the three equivalent equations
\begin{eqnarray}
\Box\phi-2\partial_{\mu}\phi\partial^{\mu}\phi+
\textstyle{\frac{1}{12}}H_{\mu\nu\rho}H^{\mu\nu\rho}=
4\pi G(\To-\EK_{\mu}{}^{\mu})\,,&&\label{EDFER3r}\\
R_{\mu\nu}+2\trd_{\mu}(\partial_{\nu}\phi)-\quarter H_{\mu\rho\sigma}H_{\nu}{}^{\rho\sigma}
=8\pi G\EK_{(\mu\nu)}\,,\label{EDFER1r}&&\\
-\star\,\rd\star\left(e^{-2\phi}H_{\scriptscriptstyle{(3)}}\right)=16\pi Ge^{-2\phi}\EK_{\scriptscriptstyle{(2)}}\,.&&\label{EDFER2r}
\end{eqnarray}
We will assume that the stringy Energy-Momentum tensor is nontrivial  only up to  a finite radius, $\rc$, and thus vanishes outside this radius:
\be
\EM_{AB}=0\qquad\mbox{\bf{if}}\qquad r\geq\rc\,.
\label{BCT}
\ee
That is to say,  matter is localized only up to the finite `cutoff' radius, $\rc$, in a spherically symmetric manner.  We emphasize  that  we  never force the $H$-flux nor the gradient of  the string dilaton to be trivial outside a finite radius: this would have been the case if we had viewed  them as extra matter, but in the current framework of  Stringy Gravity, they are part of the stringy graviton fields, on the same footing as the Riemannian metric, $g_{\mu\nu}$. Their profiles are dictated by    the Einstein Double Field Equations only.   

The strict  localization of the matter~(\ref{BCT}) motivates us to restrict spacetime to be  asymptotically `flat' (Minkowskian)  at infinity,  by imposing the following boundary conditions~\cite{Ko:2016dxa},  
\be
\ba{lll}
\dis{\lim_{r\rightarrow\infty}A=1\,,}\quad&\quad 
\dis{\lim_{r\rightarrow\infty}{A^{\prime}}=0\,,}\quad&\quad
\dis{\lim_{r\rightarrow\infty}{A^{\prime\prime}}=0\,,}\\
\dis{\lim_{r\rightarrow\infty}r^{-2}C=1\,,}\quad&\quad
\dis{\lim_{r\rightarrow\infty}{C^{\prime}C^{-1/2}}=2\,,}\quad&\quad\dis{\lim_{r\rightarrow\infty}{C^{\prime\prime}}=2\,,}\\
\dis{\lim_{r\rightarrow\infty}\phi=0\,,}\quad&\quad
\dis{\lim_{r\rightarrow\infty}{\phi^{\prime}}=0\,,}\quad&\quad
\dis{\lim_{r\rightarrow\infty}{\phi^{\prime\prime}}=0\,.}
\ea
\label{BCinfty}
\ee
The vacuum expectation  value of the string dilaton at infinity,  or $\dis{\lim_{r\rightarrow\infty}{e^{-2\phi} =1}}$,    is our conventional    normalization, as we have  the Newton constant, $G$, at our disposal as a separate  free parameter in  the   master action of Stringy Gravity   coupled to matter~(\ref{SGaction}).  The  conditions of  (\ref{BCT}) and (\ref{BCinfty})   should enable us to  recover  the previously acquired,  most general, spherically symmetric, asymptotically flat, static vacuum solution  to ${D=4}$ Stringy Gravity~\cite{Ko:2016dxa}   (\textit{c.f.~}\cite{Burgess:1994kq})   outside the cutoff radius, $r\geq\rc$. \\

\noindent  In addition,   we   postulate that  matter and hence the spacetime geometry  are `regular'  and `non-singular' at the origin, ${r=0}$. We require  
\be
\ba{lll}
\dis{\lim_{r\rightarrow 0}C=0\,,}\qquad&\qquad
\dis{\lim_{r\rightarrow 0}{A^{\prime}}C=0\,,}\qquad&\qquad
\dis{\lim_{r\rightarrow 0}\phi^{\prime}C=0\,,}
\ea
\label{BCo}
\ee
of which the first is a natural condition  for  the consistency of the spherical coordinate system at the origin~(\ref{ansatzgB}). The second and third can then be satisfied easily as long as  $A^{\prime}$ and $\phi^{\prime}$ are finite at ${r=0}$.  Note also that the areal radius, $R=e^{\phi}\sqrt{C/A}$~(\ref{arealR}),  vanishes at the origin. \\

\noindent All the nontrivial (Riemannian) Christoffel symbols of  the  metric ansatz~(\ref{ansatzgB}) are, exhaustively~\cite{Ko:2016dxa}, 
\be
\ba{lll}
\gamma^{t}_{tr}=\gamma^{t}_{rt}
=\half{A^{\prime}}A^{-1}+{\phi^{\prime}}\,,\quad&\quad\!
\gamma^{r}_{tt}=\half{A^{\prime}}A+{\phi^{\prime}}A^{2}\,,\quad&\quad\!
\gamma^{r}_{\vartheta\vartheta}=\half{A^{\prime}}A^{-1}C-\half{C^{\prime}}-C{\phi^{\prime}}\,,\\
\gamma^{r}_{rr}=-\half{A^{\prime}}A^{-1}+{\phi^{\prime}}\,,\quad&\quad\!
\gamma^{r}_{\varphi\varphi}
=\sin^{2}\vartheta\gamma^{r}_{\vartheta\vartheta}\,,\quad&\quad\!
\gamma^{\vartheta}_{r\vartheta}=\gamma^{\vartheta}_{\vartheta r}=-\half{A^{\prime}}A^{-1}
+\half{C^{\prime}}C^{-1}+{\phi^{\prime}}\,,\\
\gamma^{\vartheta}_{\varphi\varphi}=-\sin\vartheta\cos\vartheta\,,\quad&\quad\!
\gamma^{\varphi}_{\vartheta\varphi}=\gamma^{\varphi}_{\varphi\vartheta}
=\cot\vartheta\,,\quad&\quad\!
\gamma^{\varphi}_{r\varphi}=\gamma^{\varphi}_{\varphi r}=
-\half{A^{\prime}}A^{-1}+\half{C^{\prime}}C^{-1}+{\phi^{\prime}}\,.
\ea
\label{RCS}
\ee
From the off-shell conservation of the stringy Einstein tensor, the three equations~(\ref{EDFER3r}), (\ref{EDFER1r}), (\ref{EDFER2r})  must imply the on-shell conservation of the stringy Energy-Momentum tensor as in (\ref{conKT1}) and (\ref{conKT2}).   For the present spherical and static ansatz, the nontrivial components of (\ref{conKT1})  come from `${\nu=t}$' and `${\nu=r}$' only. They are, respectively,  
\be
he^{-2\phi}AC^{-1}F=-\frac{\rd~}{\rd r}(CD)\,,\label{conKT11}
\ee
and
\be
\ba{lll}
\frac{\rd~}{\rd r}(\EK_{r}{}^{r}-\half\To)&=&\half A^{\prime}A^{-1}(\EK_{t}{}^{t}+\EK_{r}{}^{r}-2\EK_{\vartheta}{}^{\vartheta})+\phi^{\prime}(\EK_{t}{}^{t}-\EK_{r}{}^{r}+2\EK_{\vartheta}{}^{\vartheta})\\
{}&{}&\,-C^{\prime}C^{-1}(\EK_{r}{}^{r}-\EK_{\vartheta}{}^{\vartheta})-e^{-4\phi}A^{2}BC^{-2}F\,.
\ea
\label{conKT12}
\ee
On the other hand, for  (\ref{conKT2}),   there appears  only one nontrivial relation  from the choice of  `${\nu=t}$', 
\be
\ba{l}
\frac{\rd~}{\rd r}\!\left(e^{-2\phi}A^{-1}CE\right)=0\,.
\ea
\label{conKT21}
\ee
We will confirm that these relations are  indeed satisfied automatically by the three equations~(\ref{EDFER3r}), (\ref{EDFER1r}), (\ref{EDFER2r}) which reduce, with the Christoffel symbols~(\ref{RCS}),    as follows. Firstly,  the scalar equation~(\ref{EDFER3r}) becomes 
\be
4\pi G(\To-\EK_{\mu}{}^{\mu})=
e^{-2\phi}{\phi^{\prime}}A\frac{\rd~}{\rd r}\ln({\phi^{\prime}}C)+\half e^{-6\phi}(A^{3}B^{2}C^{-2}-h^{2}AC^{-2})\,.
\label{scalarphipp}
\ee
The Ricci curvature, $R_{\mu\nu}$, and  the two derivatives of the string dilaton, $\trd_{\mu}\partial_{\nu}\phi$, are automatically diagonal, such that the tensorial equation~(\ref{EDFER1r})   is  almost diagonal,
\be
\ba{lll}
\!\!8\pi G\EK_{tt}=R_{tt}+2\trd_{t}\partial_{t}\phi-\quarter H_{t\rho\sigma}H_{t}{}^{\rho\sigma}&\!\!\!\!\!\!\!=\!\!\!\!&\!\!\!\!\!\half {A^{\prime}}A\frac{\rd~}{\rd r}\ln({A^{\prime}}A^{-1}C)+{\phi^{\prime}}A^{2}\frac{\rd~}{\rd r}\ln({\phi^{\prime}}C)-\half h^{2}A^{2}C^{-2}e^{-4\phi}\,,\\
\!\!8\pi G\EK_{rr}=R_{rr}+2\trd_{r}\partial_{r}\phi-\quarter H_{r\rho\sigma}H_{r}{}^{\rho\sigma}&\!\!\!\!\!=\!\!\!\!\!&\!\!\!\half {A^{\prime}}A^{-1}\frac{\rd~}{\rd r}\ln({A^{\prime}}A^{-1}C)-\half{A^{\prime}}^{2}A^{-2}
-C^{-1/2}\frac{\rd~}{\rd r}\left({C^{\prime}}C^{-1/2}\right)\\
{}&{}&-2{\phi^{\prime}}^{2}-{\phi^{\prime}}\frac{\rd~}{\rd r}\ln({\phi^{\prime}}C)-\half A^{2}B^{2}C^{-2}e^{-4\phi}\,,\\
\!\!8\pi G\EK_{\vartheta\vartheta}\!=\!R_{\vartheta\vartheta\!}+2\trd_{\vartheta}\partial_{\vartheta}\phi-\quarter H_{\vartheta\rho\sigma}H_{\vartheta}{}^{\rho\sigma}\!\!&\!\!=\!\!\!&\!
1-\half C^{\prime\prime\!}+\frac{\rd~}{\rd r}\!\left(\half A^{\prime}A^{-1}C
-{\phi^{\prime}}C\right)\!-\half(A^{2}B^{2}-h^{2})C^{-1}e^{-4\phi}\,,\\
\multicolumn{3}{c}{\!\!\!
8\pi G\EK_{\varphi\varphi}=8\pi G\sin^{2\!}\vartheta\EK_{\vartheta\vartheta}=R_{\varphi\varphi}+2\trd_{\varphi}\partial_{\varphi}\phi
-\quarter H_{\varphi\rho\sigma}H_{\varphi}{}^{\rho\sigma}=\sin^{2\!}\vartheta\!\left(R_{\vartheta\vartheta}+2\trd_{\vartheta}\partial_{\vartheta}\phi-\quarter H_{\vartheta\rho\sigma}H_{\vartheta}{}^{\rho\sigma}\right),}
\ea
\label{dgn}
\ee
with  one  exception, an  off-diagonal component,
\be
8\pi G\EK_{(tr)}=8\pi G D(r)=
R_{tr}+2\trd_{t}\partial_{r}\phi-\quarter H_{t\rho\sigma}H_{r}{}^{\rho\sigma}=-\quarter H_{t\rho\sigma}H_{r}{}^{\rho\sigma}=-\half\, hBe^{-4\phi}A^{2}C^{-2}\,.
\label{eorm}
\ee
The last equation for the $H$-flux~(\ref{EDFER2r}) becomes
\be
\ba{lll}
-\star\rd\star\!\left(e^{-2\phi}H_{\scriptscriptstyle{(3)}}\right)\!\!&\!\!=\!\!&\!-\star\,\rd \!\left(e^{-4\phi}A^{2}BC^{-1}\rd t+e^{-4\phi}hC^{-1}\rd r\right)\!= A^{-1}C\sin\vartheta\frac{\rd~}{\rd r}\!\big(e^{-4\phi}A^{2}BC^{-1}\big)
\rd \vartheta\wedge\rd \varphi\\
{}&\!\!=\!\!&16\pi Ge^{-2\phi}\EK_{\scriptscriptstyle{(2)}}=16\pi Ge^{-2\phi}\big(
E(r)\rd t\wedge\rd r+F(r)\sin\vartheta\rd\vartheta\wedge\rd\varphi\big)\,,
\ea
\label{EOMB3}
\ee
which gives
\be
\ba{ll}
\EK_{[tr]}(r)=E(r)=0\,,\quad&\quad
16\pi Ge^{-2\phi}AC^{-1}F(r)=\frac{\rd~}{\rd r}\big(e^{-4\phi}A^{2}BC^{-1}\big)\,.
\ea
\label{Ezero}
\ee
The former result of (\ref{Ezero}) satisfies   the conservation relation~(\ref{conKT21}) trivially,  while the latter combined with (\ref{eorm}) implies  the conservation relation~(\ref{conKT11}).  Integrating the latter, we get
\be
e^{-4\phi}A^{2}BC^{-1}=q+\cV(r)\,,
\label{ABC}
\ee
where  we set
\be
\cV(r):=-16\pi G{\dis{\int^{\infty}_{r}}\rd r^{\prime}~e^{-2\phi(r^{\prime})}}A(r^{\prime})F(r^{\prime})/C(r^{\prime})=-16\pi G\dis{\int_{r}^{\infty}}\rd r~e^{-2d}\EK^{\vartheta\varphi}\,,
\label{cVdef}
\ee
and $q$ is a constant of  integration.  From our assumption~(\ref{BCT}), when $r\geq\rc$,   both $F(r)$ and $\cV(r)$ vanish,   and consequently    
$e^{-4\phi}A^{2}BC^{-1}$ assumes the  constant value, $q$.    Now, substituting (\ref{ABC}) into the second formula in (\ref{dgn}), we get
\be
\ba{lll}
8\pi G\EK_{rr}&=&\half {A^{\prime}}A^{-1}\frac{\rd~}{\rd r}\ln({A^{\prime}}A^{-1}C)-\half{A^{\prime}}^{2}A^{-2}
-C^{-1/2}\frac{\rd~}{\rd r}\left({C^{\prime}}C^{-1/2}\right)-2{\phi^{\prime}}^{2}-{\phi^{\prime}}\frac{\rd~}{\rd r}\ln({\phi^{\prime}}C)\\
{}&{}&-\half  e^{4\phi}A^{-2}\left(q+\cV\right)^{2}\,.
\ea
\ee
The infinite radius  limit of this expression   implies, with the conditions of  (\ref{BCT}) and (\ref{BCinfty}),  that  actually $q$  must be    trivial: $q=0$. Therefore, from (\ref{ABC}),  we are able to fix $B(r)$ and hence $H_{r\vartheta\varphi}$, 
\be
\ba{ll}
B=e^{4\phi}A^{-2}C\cV\,,\qquad&\qquad
H_{r\vartheta\varphi}=e^{4\phi}A^{-2}C\cV\sin\vartheta\,,
\ea
\label{fixB}
\ee
which vanish when  $r\geq\rc$,  in agreement with   the  known vacuum solution~\cite{Ko:2016dxa}. \\

\noindent The remaining Einstein Double Field Equations~(\ref{scalarphipp}), (\ref{dgn}), (\ref{eorm}) reduce, with $\intdn\equiv
4\pi e^{2\phi}A^{-1}C$ from (\ref{emd}),  to
\begin{eqnarray}
h\cV&=&-16\pi G\EK_{(tr)}C\,,\label{hV}\\
C^{\prime\prime}&=&2+4G(\EK_{r}{}^{r}+\EK_{\vartheta}{}^{\vartheta}-\To){\intd}+e^{4\phi}A^{-2}C\cV^{2}\,,\label{Cpp}\\
\frac{\rd~}{\rd r}({A^{\prime}}A^{-1}C)&=&
2G(\EK_{\mu}{}^{\mu}-2\EK_{t}{}^{t}-\To)\intd+e^{4\phi}A^{-2}C\cV^{2}\,,\label{AAC}\\
\frac{\rd~}{\rd r}({\phi^{\prime}}C)&=&\half h^{2}e^{-4\phi}C^{-1}- G(\EK_{\mu}{}^{\mu}-\To)\intd-\half e^{4\phi}A^{-2}C\cV^{2}\,,\label{phiC}
\end{eqnarray}
\be
4(\phi^{\prime}C)^{2}+(A^{\prime}A^{-1}C)^{2}+4C-{C^{\prime}}^{2}
+h^{2}e^{-4\phi}+e^{4\phi}A^{-2}C^{2}\cV^{2}
+4CG(2\EK_{r}{}^{r}-\To){\intd}=0\,.\label{phip2}
\ee
The first equation~(\ref{hV}) indicates that $h$ is a proportionality constant relating    $\cV=e^{-4\phi}A^{2}BC^{-1}$ to $\EK_{(tr)}C$. On the other hand,  the radial derivative of the entire expression in the last formula~(\ref{phip2}),  after substitution of   (\ref{fixB}), (\ref{Cpp}), (\ref{AAC}), and (\ref{phiC}),  implies the remaining   conservation relation~(\ref{conKT12}):
\be\dis{
\ba{l}
\frac{\rd~}{\rd r}\left[4(\phi^{\prime}C)^{2}+(A^{\prime}A^{-1}C)^{2}+4C-{C^{\prime}}^{2}
+h^{2}e^{-4\phi}+e^{4\phi}A^{-2}C^{2}\cV^{2}
+4CG(2\EK_{r}{}^{r}-\To){\intd}\right]\\
=\Big[\,
\frac{\rd~}{\rd r}(\EK_{r}{}^{r}-\half\To)-\half A^{\prime}A^{-1}(\EK_{t}{}^{t}+\EK_{r}{}^{r}-2\EK_{\vartheta}{}^{\vartheta})-\phi^{\prime}(\EK_{t}{}^{t}-\EK_{r}{}^{r}+2\EK_{\vartheta}{}^{\vartheta})+
C^{\prime}C^{-1}(\EK_{r}{}^{r}-\EK_{\vartheta}{}^{\vartheta})\\
\qquad\,+e^{-4\phi}A^{2}BC^{-2}F\,\Big]\times 8CG\intd\,.
\ea}
\ee
This provides  a consistency check for the  equations~(\ref{Cpp})--(\ref{phip2}), and completes our  concrete verification  that  all the  Energy-Momentum conservation laws  indeed follow   from the Einstein Double Field Equations.\\

\noindent In order to solve or integrate the second and the third  equations,   (\ref{Cpp}), (\ref{AAC}), we prepare the following definitions,
\be
\ba{lc}
\cW(r):=\dis{\int_{0}^{r}\!\rd r}~e^{4\phi}A^{-2}C\,\cV^{2}=\textstyle{\frac{1}{4\pi}}
\dis{\int_{0}^{r}}\!\rd r\intdn H_{r\vartheta\varphi}H^{r\vartheta\varphi}\,,\quad&\quad
\hcW(r):=\dis{\int_{0}^{r}\!\rd r}~\cW\,,\\
\cX(r):=G\dis{\int_{0}^{r}\!\rd r}\intdn\,(\EK_{r}{}^{r}+\EK_{\vartheta}{}^{\vartheta}-\To)\,,\quad&\quad
\hcX(r):=\dis{\int_{0}^{r}\!\rd r}~\cX\,,\\
\cY(r):=G\dis{\int_{0}^{r}\!\rd r}\intdn\,(\EK_{r}{}^{r}+\EK_{\vartheta}{}^{\vartheta}+\EK_{\varphi}{}^{\varphi}-\EK_{t}{}^{t}-\To)\,,\quad&
\ea
\label{WXY}
\ee
which all vanish at the origin, $r=0$. Further, when  $r\geq\rc$, the un-hatted functions, $\cW(r)$, $\cX(r)$, $\cY(r)$, become constant---for example, $\cX(r)=\cX(\rc)\equiv\cX_{\rmc}$ for $r\geq\rc$\,.
 Consequently,  the hatted functions become linear in the outside region, 
\be
\ba{llll}
\hcW(r)=\hcW_{\rmc}+(r-\rc)\cW_{\rmc}\,,\quad&\quad\hcX(r)=\hcX_{\rmc}+(r-\rc)\cX_{\rmc}&\quad\mbox{\bf{for}}\quad& r\geq\rc\,.
\ea
\ee

\noindent  Integrating ({\ref{Cpp}) twice, we can solve for    $C(r)$. There are  two  constants of integration which  we  fix  by imposing  the boundary conditions at the origin:  firstly we set ${C(0)=0}$ directly from (\ref{BCo})   and secondly, with  $\phio\equiv\phi(0)$,  we fix $C^{\prime}(0)=\pm \hephi$   from  the consideration of the small $r$ limit  of (\ref{phip2}).   We get
\be
C(r)=r^{2}\pm \hephi r+4\hcX(r)+\hcW(r)\,.
\label{Cr1}
\ee
Outside the matter this reduces to a quadratic equation,
\be
\ba{lll}
C(r)=(r-\alpha)(r+\beta)=r^{2}+(4\cX_{\rmc}+\cW_{\rmc}\pm \hephi) r +4\hcX_{\rmc}-4\rc\cX_{\rmc}+\hcW_{\rmc}-\rc\cW_{\rmc}
&\quad\!
\mbox{\bf{for}}~~~~&
r\geq\rc\,,
\ea
\label{Cr2}
\ee
where we set two constants, 
\be
\ba{l}
\alpha:=\half\left[{-(4\cX_{\rmc}+\cW_{\rmc}\pm \hephi)+
\sqrt{(4\cX_{\rmc}+\cW_{\rmc}\pm \hephi)^{2}+16\rc\cX_{\rmc}-16\hcX_{\rmc}
+4\rc\cW_{\rmc}-4\hcW_{\rmc}}\,}\right]\,,\\
\beta:=\half\left[4\cX_{\rmc}+\cW_{\rmc}\pm \hephi+\sqrt{(4\cX_{\rmc}+\cW_{\rmc}\pm\hephi)^{2}
+16\rc\cX_{\rmc}-16\hcX_{\rmc}+4\rc\cW_{\rmc}-4\hcW_{\rmc}}\,\right]\,.
\ea
\label{alphabeta}
\ee
\noindent Similarly, (\ref{AAC}) gives
\be
\ba{ll}
A^{\prime}A^{-1}C=2\cY+\cW\,,\qquad&\quad
\dis{\lim_{r\rightarrow 0}A^{\prime}A^{-1}C=0\,,}
\ea
\label{AAC2}
\ee
for which the trivial constant of integration (zero) has been chosen  to meet the boundary condition  at the origin~(\ref{BCo}).   Eq.(\ref{AAC2}) can be further  integrated to determine   $A(r)$  with    the boundary  condition, this time at infinity~(\ref{BCinfty}),
\be
\ba{ll}
A(r)=\exp\left[-\dis{\int_{r}^{\infty}\rd r~(2\cY+\cW)C^{-1}}\right]\,,\qquad&\quad
\dis{\lim_{r\rightarrow\infty}A(r)=1\,.}
\ea
\label{Ar1}
\ee
Away from the matter,  with (\ref{Cr2}),  this reduces to a closed  form,  
\be
\ba{lll}
A(r)=\left(\frac{r-\alpha}{r+\beta}\right)^{\frac{a}{\alpha+\beta}}&\qquad
\mbox{\bf{for}}\qquad&
r\geq\rc\,,
\ea
\label{Ar2}
\ee
where we have introduced another constant,
\be
\dis{a:=\lim_{r\rightarrow\infty}A^{\prime}A^{-1}C
=2\cY(\rc)+\cW(\rc)=2\cY_{\rmc}+\cW_{\rmc}\,,}
\label{fixa}
\ee
such that outside the matter,
\be
\ba{lll}
A^{\prime}A^{-1}C=a\qquad&\mbox{{\bf{for}}}&\qquad r\geq\rc\,.
\ea
\label{AACa}
\ee
For $A(r)$ to be real and positive,  it is necessary to restrict the range of $r$. Since $\alpha+\beta$ is positive semi-definite  from (\ref{alphabeta}), we are lead to require the cutoff radius to be greater than $\alpha$, such that
\be
\ba{lll}
r\geq\rc>\alpha\geq-\beta\,,\qquad&\qquad
\dis{\frac{r-\alpha}{r+\beta}~>~0\,.}
\ea
\label{restr1}
\ee
However, note that the signs of $\alpha$ and $\beta$ are not yet fixed: in the next subsection~\ref{SECEC} we shall  assume  some energy conditions which will ensure both  $\alpha$ and $\beta$ are positive.\\

\noindent We now turn to the last differential equation~(\ref{phip2}).   Upon substitution of   (\ref{Cr2}) and  (\ref{AAC2}),    it takes the form
\be
\ba{lll}
4(\phi^{\prime}C)^{2}
+h^{2}e^{-4\phi}&=&\h^{2}e^{-4\phio}+8\cX\cW-4(\cY+\cW)\cY
+16(\cX^{2}-\hcX)-4\hcW
-e^{4\phi}A^{-2}C^{2}\cV^{2}\\
{}&{}&+4\left(r\pm\half\hephi\right)(4\cX+\cW)
-4CG(2\EK_{r}{}^{r}-\To){\intd}\,,
\ea
\ee
which, from  (\ref{Ar2}), (\ref{fixa}),  reduces  outside the matter   to
\be
\ba{lll}
\,~4(\phi^{\prime}C)^{2}+h^{2}e^{-4\phi}\equiv b^{2}&\qquad
\mbox{\bf{for}}\quad&
r\geq\rc\,.
\ea
\label{ppC2}
\ee
This new constant, $b$, meets
\be
b^{2}:=(\alpha+\beta)^{2}-a^{2}=16\rc\cX_{\rmc}-16\hcX_{\rmc}
+4\rc\cW_{\rmc}-4\hcW_{\rmc}+
\big(4\cX_{\rmc}+\cW_{\rmc}\pm \hephi\big)^{2}-\big(2\cY_{\rmc}+\cW_{\rmc}\big)^{2}\,.
\label{fixb}
\ee
Since the left-hand side of (\ref{ppC2}) is positive,  $b$ should be real, and    from (\ref{alphabeta}),  $\alpha+\beta$ is   positive since
\be
\alpha+\beta=\sqrt{a^{2}+b^{2}}=\sqrt{(4\cX_{\rmc}+\cW_{\rmc}\pm \hephi)^{2}+16\rc\cX_{\rmc}-16\hcX_{\rmc}
+4\rc\cW_{\rmc}-4\hcW_{\rmc}}\,.
\ee
Therefore,  outside the matter we have
\be
\ba{lll}
\,\qquad\qquad\qquad2\phi^{\prime}C=\pm\sqrt{b^{2}-h^{2}e^{-4\phi}\,}~&~\qquad
\mbox{\bf{for}}\qquad&
r\geq\rc\,,
\ea
\label{2phipC}
\ee
such that
\be
\ba{lll}
\dis{
\pm\int\frac{2\rd\phi}{\sqrt{b^{2}-h^{2}e^{-4\phi}}\,}=\int\frac{\rd r}{(r-\alpha)(r+\beta)}}
&\qquad
\mbox{\bf{for}}\qquad&
r\geq\rc\,,
\ea
\label{phirINT}
\ee
of which both sides can be integrated to give
\be
\dis{\pm\frac{1}{\left|b\right|}\ln\left(e^{2\phi}+\sqrt{e^{4\phi}-h^{2}/b^{2}}\right)=\frac{1}{\alpha+\beta}\ln\left(\frac{r-\alpha}{r+\beta}\right)\,+\,\mbox{constant}\,.}
\label{1b}
\ee
We can determine the constant of integration in (\ref{1b}) from  the boundary condition at infinity~(\ref{BCinfty}), 
\be
\dis{\lim_{r\rightarrow\infty}e^{2\phi}=1\,,}
\ee
to  obtain the profile of the string dilaton outside the matter,
\be
\ba{lll}
e^{2\phi}=\gamma_{+}\left(\frac{r-\alpha}{r+\beta}\right)^{\frac{b}{\sqrt{a^{2}+b^{2}}}}+\gamma_{-}\left(\frac{r+\beta}{r-\alpha}\right)^{\frac{b}{\sqrt{a^{2}+b^{2}}}}&\qquad
\mbox{\bf{for}}\qquad&
r\geq\rc\,,
\ea
\label{finalphi}
\ee
where $b$ can be either positive or negative, and   $\gamma_{+}$, $\gamma_{-}$ denote two positive semi-definite   constants, 
\be
\gamma_{\pm}:=\half(1\pm\sqrt{1-{h^{2}/b^{2}}})\,.
\label{constantg}
\ee
For the sake of reality, we require\footnote{In fact, when ${b=0}$,  from (\ref{ppC2}), we get ${h=0}$ and  ${\phi^{\prime}=0}$.  Although  (\ref{phirINT}) and  (\ref{1b}) would be problematic if $b=0$, the final result~(\ref{finalphi}) is still valid, as $e^{2\phi}=1$.}
\be
b^{2}\geq h^{2}\,.
\ee
This completes our derivation of the spherically symmetric, static, regular solution  to ${D=4}$ Stringy Gravity with a localized stringy matter distribution.\\

\noindent We conclude this subsection by summarizing and analyzing our results.
\begin{itemize}
\item Outside the cutoff radius,  $r\geq\rc$, we recover the spherically symmetric static   vacuum solution~\cite{Ko:2016dxa}:
\be
{\ba{cc}
e^{2\phi}=\gamma_{+}\left(\frac{r-\alpha}{r+\beta}\right)^{\frac{b}{\sqrt{a^{2}+b^{2}}}}+\gamma_{-}\left(\frac{r+\beta}{r-\alpha}\right)^{\frac{b}{\sqrt{a^{2}+b^{2}}}}\,,\quad&\quad
B_{\scriptscriptstyle{(2)}}=h\cos\vartheta\,\rd t\wedge\rd\varphi\,,\\
\multicolumn{2}{c}{
\rd s^{2}=e^{2\phi}\left[-\left(\frac{r-\alpha}{r+\beta}\right)^{\frac{a}{\sqrt{a^{2}+b^{2}}}}\rd t^{2}
+\left(\frac{r+\beta}{r-\alpha}\right)^{\frac{a}{\sqrt{a^{2}+b^{2}}}}\big\{\rd r^{2}+(r-\alpha)(r+\beta)
\rd\Omega^{2}\big\}\right]\,.}
\ea}
\label{SOL2}
\ee
\item  Moreover, the  constants, $\alpha,\beta,a,b,h$, are now all determined by the stringy Energy-Momentum tensor of the matter localized inside the cutoff radius: from   (\ref{alphabeta}), (\ref{fixa}), (\ref{fixb}),
\be
\ba{c}
\alpha=\half\left[
\sqrt{\left(\cZ_{\rmc}\pm \hephi\right)^{2}+4\widetilde{\cZ}_{\rmc}\,}\,-\left(\cZ_{\rmc}\pm \hephi\right)\right]\,,\\
\beta=\half\left[
\sqrt{\left(\cZ_{\rmc}\pm \hephi\right)^{2}+4\widetilde{\cZ}_{\rmc}\,}\,+\left(\cZ_{\rmc}\pm \hephi\right)\right]\,,\\
a=\dis{\int_{0}^{\infty}\!\rd r\int_{0}^{\pi}\!\rd\vartheta\int_{0}^{2\pi}\!\rd\varphi}~e^{-2d}\left[{\frac{1}{4\pi}}H_{r\vartheta\varphi}H^{r\vartheta\varphi}+2 G\left(\EK_{r}{}^{r}+\EK_{\vartheta}{}^{\vartheta}+\EK_{\varphi}{}^{\varphi}-\EK_{t}{}^{t}-\To\right)\right]\,,\\
b^{2}=\left(\cZ_{\rmc}\pm \hephi\right)^{2}+4\widetilde{\cZ}_{\rmc}\,-a^{2}\,,\qquad\qquad\qquad
\dis{h=\frac{\EK_{tr}C}{\dis{\int_{r}^{\infty}\rd r~e^{-2d}\EK^{\vartheta\varphi}}}\,,}
\ea
\label{ababh}
\ee
where,  with   (\ref{WXY}),
\be
\ba{l}
\cZ(r):=\dis{\int_{0}^{r}\!\rd r\int_{0}^{\pi}\!\rd\vartheta\int_{0}^{2\pi}\!\rd\varphi}~e^{-2d}\left[{\frac{1}{4\pi}}H_{r\vartheta\varphi}H^{r\vartheta\varphi}+4 G\left(\EK_{r}{}^{r}+\EK_{\vartheta}{}^{\vartheta}-\To\right)\right]=\cW(r)+4\cX(r)\,,\\
\widetilde{\cZ}_{\rmc}:=\dis{\int_{0}^{\rc}\rd r}~\big[\cZ_{\rmc}-\cZ(r)\big]\,.
\ea
\label{Zdef}
\ee
As before, the subscript index, $\rmc$, denotes the position at $r=\rc$, such that $\cZ_{\rmc}=\cZ(\rc)$\,.

\item Some further comments are in order.
\end{itemize}
\begin{itemize}

\item[--] There are two classes of solutions: $b=\sqrt{(\alpha+\beta)^{2}-a^{2}}\geq 0~$ or $~b=-\sqrt{(\alpha+\beta)^{2}-a^{2}}<0$.

\item[--] Direct computation from (\ref{finalphi}) shows
\be
2\phi^{\prime}C e^{2\phi}
=b\left[\gamma_{+}\left(\textstyle{\frac{r-\alpha}{r+\beta}}\right)^{\frac{b}{\sqrt{a^{2}+b^{2}}}}-\gamma_{-}\left(\textstyle{\frac{r+\beta}{r-\alpha}}\right)^{\frac{b}{\sqrt{a^{2}+b^{2}}}}\right]
=\sgn b\sqrt{e^{4\phi}-h^{2}/b^{2}}\,,
\label{2phipCe}
\ee
where we define a sign factor, 
\be
\sgn:=\left\{\ba{lllll}
+1\quad&\mbox{{\bf{if}}}\quad& b>0& \mbox{{\bf{and}}}  &r\geq \rphi\\
-1\quad&\mbox{{\bf{if}}}\quad& b>0&\mbox{{\bf{and}}}  & \rphi>r\geq \alpha\\
+1\quad&\mbox{{\bf{if}}}\quad&b<0& \mbox{{\bf{and}}} & r\geq\alpha\,,
\ea\right.
\label{sgn}
\ee
with  the zero of $\phi^{\prime}$ given by
\be
\ba{lll}
\dis{\rphi:=\frac{\,\alpha\,+\,\beta\left(\frac{\gamma_{-}}{\gamma_{+}}\right)^{\frac{\sqrt{a^{2}+b^{2}}}{2b}}}{1-\left(\frac{\gamma_{-}}{\gamma_{+}}\right)^{\frac{\sqrt{a^{2}+b^{2}}}{2b}}}\,,}\quad&\quad
\phi^{\prime}(\rphi)=0\,,\quad&\quad
\phi(\rphi)=\half\ln\left|h/b\right|<0\,.
\ea
\label{rplus}
\ee
When $h$ is trivial, we have  $\gamma_{-}=0$, $\rphi=\alpha$ and hence $\sgn$ is fixed to be $+1$. 
If $h\neq 0$ and $b>0$, then $\rphi>\alpha$. Otherwise ($h\neq0$ and $b<0$)  we have the opposite,  $\rphi<\alpha$. In fact, when  $b$ is negative,  $\rphi$ also becomes negative and thus  unphysical.  That is to say,  for large enough $r$,   \textit{i.e.}  either $r\geq\alpha$ with  negative ${b}\,$  or $\,{r\geq \rphi}$ with positive $b$,  the sign of $\phi^{\prime}$ coincides with that of $b$, but when $b$ is positive, $\phi^{\prime}$ becomes  negative in the finite interval $\,\alpha\leq r<\rphi$.

\item[--] Since $b$ is real,   the following inequality must be met:
\be
\left(\cZ_{\rmc}\pm \hephi\right)^{2}+4\widetilde{\cZ}_{\rmc}\,\geq\, a^{2}\,,
\label{realb}
\ee
which imposes a constraint on the stringy Energy-Momentum tensor through (\ref{ababh}).

\item[--] From (\ref{emd}), outside the matter the integral measure in Stringy Gravity  reads
\be
\ba{lll}
e^{-2d}=R^{2}\sin\vartheta
=\left[
\gamma_{+}\left(\frac{r+\beta}{r-\alpha}\right)^{\frac{a-b}{\sqrt{a^{2}+b^{2}}}}+\gamma_{-}\left(\frac{r+\beta}{r-\alpha}\right)^{\frac{a+b}{\sqrt{a^{2}+b^{2}}}}\right](r-\alpha)(r+\beta)\sin\vartheta
\quad&\!\!\mbox{\bf{for}}\!\!&\quad r\geq\rc\,.
\ea
\ee
\item[--] With the boundary condition at the origin~(\ref{BCo}),  the Einstein Double Field Equations~(\ref{AAC}), (\ref{phiC}) enable us to  evaluate,  for arbitrary $r\geq 0$,
\be
\ba{lll}
2\phi^{\prime}C+A^{\prime}A^{-1}C
&=&\dis{\int_{0}^{r}\rd r~{\frac{\rd~}{\rd r}}\left(2\phi^{\prime}C+A^{\prime}A^{-1}C\right)}\\
{}&=&\dis{\int_{0}^{r}\rd r~\left(
 h^{2}e^{-4\phi}C^{-1}- 16\pi G\EK_{t}{}^{t}e^{2\phi}A^{-1}C \right)}\\
{}&=&\dis{\int_{0}^{r}\rd r\int_{0}^{\pi}\rd\vartheta\int_{0}^{2\pi}\rd\varphi~e^{-2d}\left(\frac{1}{4\pi}\left|H_{t\vartheta\varphi}H^{t\vartheta\varphi}\right|
- 4 G\EK_{t}{}^{t} \right)}\,,
\ea
\label{ab}
\ee
where $H_{t\vartheta\varphi}H^{t\vartheta\varphi}=-h^{2}e^{-6\phi}AC^{-2}$. 

\item[--] Combining (\ref{2phipCe}) and (\ref{ab})  with the boundary condition at infinity~(\ref{BCinfty}), (\ref{AACa}), we acquire
\be
\dis{a+b\sqrt{1-h^{2}/b^{2}}=\int_{0}^{\infty}\rd r\int_{0}^{\pi}\rd\vartheta\int_{0}^{2\pi}\rd\varphi~e^{-2d}\left(\frac{1}{4\pi}\left|H_{t\vartheta\varphi}H^{t\vartheta\varphi}\right|
- 4 G\EK_{t}{}^{t} \right)\,.}
\label{abINT}
\ee
We stress that this result  is valid irrespective of the sign of   $b$.

\item[--] The constant parameter, $h$, corresponding to the electric $H$-flux, is given by the formula~(\ref{ababh})
\be
h=\EK_{tr}(r)C(r)\left/\left[\dis{\int_{r}^{\infty}\rd r~e^{-2d}\EK^{\vartheta\varphi}}\right]\right.\,.
\label{htr}
\ee
While this is a nontrivial relation, as the right-hand side of the equality must be constant independent of $r$, it is less informative compared to the integral expressions of $a$ in (\ref{ababh})  or $a+b\sqrt{1-h^{2}/b^{2}}$ in (\ref{abINT}). We expect a fuller  understanding  of the  $h$ parameter will arise if we solve for  the   time-dependent dynamical Einstein Double Field Equations, allowing $h$ to be time-dependent, $h\rightarrow h(t)$.  In any case, (\ref{htr}) implies that if $\EK_{tr}$ is nontrivial somewhere in the interior, there must be electric $H$-flux everywhere, including outside the matter.

\item[--] The small-$r$ radial derivative of the areal radius $R$~(\ref{arealR}) outside the matter reads,  with the sign factor $\sgn$~(\ref{sgn}),
\be
\ba{lll}
\frac{\rd R}{\rd r}=
e^{2\phi}A^{-1}R^{-1}\left[r-\alpha+\half\sqrt{a^{2}+b^{2}}-\half   a
+\half\sgn
b\sqrt{1-(h^{2}/b^{2})e^{-4\phi}}\,\right]\quad&\mbox{{\bf{for}}}&\quad r\geq\rc\,.
\ea
\label{Rr}
\ee

\item[--] From (\ref{emd}), (\ref{EMxi}), (\ref{Ezero}),  (\ref{hV}),  the Noether charge~(\ref{cQxi})  for a generic Killing vector reads  
\be
\cQ[\xi]=\dis{\int_{\Sigma}}~e^{-2d}T^{t}{}_{A}\xi^{A}=\dis{\int_{0}^{\infty}\!\rd r\int_{0}^{\pi}\!\rd\vartheta\int_{0}^{2\pi}\!\rd\varphi}~\left[e^{-2d}(\EK_{t}{}^{t}-\half\To)\xi^{t}+\frac{1}{16\pi G}h\cV A^{-2}\xi^{r}\sin\vartheta\right]\,.
\label{cQxi2}
\ee
\end{itemize}


\subsection{Energy conditions\label{SECEC}}
In this subsection we   assume  that the stringy Energy-Momentum tensor and the stringy graviton  fields satisfy  the following three conditions: 
\begin{itemize}
\item[{{\textit{i)}}}] the {\textit{strong energy condition},} with magnetic $H$-flux,
\be
\dis{\int_{0}^{\infty}\rd r\int_{0}^{\pi}\rd\vartheta\int_{0}^{2\pi}\rd\varphi\,\,\,e^{-2d}\left(
-\EK_{t}{}^{t}+\EK_{r}{}^{r}+\EK_{\vartheta}{}^{\vartheta}+\EK_{\varphi}{}^{\varphi}-\To+\frac{1}{8\pi G}\left|H_{r\vartheta\varphi}H^{r\vartheta\varphi}\right|\right)~\geq~0\,;}
\label{strong}
\ee
\item[{{\textit{ii)}}}] the {\textit{weak energy condition},} with electric $H$-flux,
\be
\dis{\int_{0}^{\infty}\rd r\int_{0}^{\pi}\rd\vartheta\int_{0}^{2\pi}\rd\varphi\,\,\,
e^{-2d}\left(
-\EK_{t}{}^{t}+\frac{1}{16\pi G}\left|H_{t\vartheta\varphi}H^{t\vartheta\varphi}\right|\right)~\geq~0\,;}
\label{weak}
\ee
\item[{{\textit{iii)}}}] the {\textit{pressure condition},} with magnetic $H$-flux and  without integration,  
\be
\EK_{r}{}^{r}+\EK_{\vartheta}{}^{\vartheta}-\To+\frac{1}{16\pi G}\left|H_{r\vartheta\varphi}H^{r\vartheta\varphi}\right|
~\geq~0\,.
\label{pressure}
\ee
\end{itemize}
While the nomenclatures   are  in analogy with those  in General Relativity, the precise  expression  in each inequality,  including  the  $H$-flux,  is   what we shall need    in our  discussion.  Since the magnetic $H$-flux vanishes outside the matter along with the stringy Energy-Momentum tensor,  (\ref{cVdef}), (\ref{fixB}), the radial integration  in the strong energy condition (\ref{strong}) is  taken  effectively   from zero to the  cutoff radius, \textit{i.e.}~$\int_{0}^{\rc}$. In contrast, the electric $H$-flux has a long tail and the weak energy condition genuinely   concerns   the infinite volume  integral.

 If the matter comprises  point particles only as in (\ref{particleL}), the above conditions are all clearly met, since $\EK_{r}{}^{r}$, $\EK_{\vartheta}{}^{\vartheta}$, $\EK_{\varphi}{}^{\varphi}$ and  $-\EK_{t}{}^{t}$ are individually  positive semi-definite, while $\To$ is trivial.  In such cases, not only the integrals but also the integrands themselves are positive semi-definite, which would imply
\be
\ba{rcl}
\dis{-\EK_{t}{}^{t}+\frac{1}{16\pi G}\left|H_{t\vartheta\varphi}H^{t\vartheta\varphi}\right|~\geq~0\,}&\quad:\quad&\textit{weak~energy~density~condition}?
\ea
\label{weakdensity}
\ee
and similarly for the strong energy density condition.  However, for stringy matter, such as fermions~(\ref{fermionL}) or fundamental strings~(\ref{stringL}),  the diagonal components, $\EK_{\mu\mu}$, may not be positive definite, and so the above inequalities appear not to be guaranteed (hence the  question mark in (\ref{weakdensity})).  If the diagonal components are negative,   the   positively  squared   $H$-fluxes need to compete with them.  In fact, while  we take the  energy and the  pressure  conditions~(\ref{strong}), (\ref{weak}), (\ref{pressure}) for granted,\footnote{Strictly speaking, we might relax the pressure condition~(\ref{pressure}) and require $\widetilde{\cZ}_{\rmc}\geq 0$ only. However  we do not pursue this here. }  we shall distinguish the energy condition from the energy density condition, and in particular investigate  the implications of the   relaxation of   the latter~(\ref{weakdensity}). \\

\noindent Obviously,  from (\ref{ababh}),  the strong energy condition sets the constant, $a$,  to be positive semi-definite.  Similarly, the pressure condition ensures $\cZ(r)$  is a non-decreasing   positive  function,  reaching  its maximum value  at the cutoff radius, such that   $0\leq\cZ(r)\leq\cZ_{\rmc}$ and $\cZ^{\prime}\geq 0$.  Consequently, $\widetilde{\cZ}_{\rmc}$ is positive semi-definite and thus $\alpha$ and $\beta$ are both safely  real and positive. In this way,  the strong energy and the pressure   conditions ensure
\be
\ba{lll}
\alpha=\left|\alpha\right|\geq 0\,,\quad&\qquad
\beta=\left|\beta\right|\geq 0\,,\quad&\qquad
a=\left|a\right|\geq 0\,.
\ea
\label{abapos}
\ee
Given this positiveness, the string dilaton, $\phi$,  outside the matter~(\ref{finalphi}), 
\be
\ba{lll}
\phi=\frac{1}{2}\ln\left[\gamma_{+}\left(\frac{r-\alpha}{r+\beta}\right)^{\frac{b}{\sqrt{a^{2}+b^{2}}}}+\gamma_{-}\left(\frac{r+\beta}{r-\alpha}\right)^{\frac{b}{\sqrt{a^{2}+b^{2}}}}\right]\qquad&\mbox{{\bf{for}}}&\qquad r>\rc\,,
\ea
\label{philn}
\ee
diverges  to plus infinity  when  $r\rightarrow\alpha^{+}$ and $h\neq0$, irrespective of the sign of $b$.   Specifically, if $b>0$ and $h\neq0$,  $\phi$ decreases from $\infty$ over the finite  interval of $\alpha<r<\rphi$, crossing the horizontal axis of ${\phi=0}$  at $r=\frac{\alpha+\beta (\gamma_{-}/\gamma_{+})^{\sqrt{1+a^{2}/b^{2}}}}{1-(\gamma_{-}/\gamma_{+})^{\sqrt{1+a^{2}/b^{2}}}}$,   reaches its minimum, $\phi_{\scriptscriptstyle{\mathrm{min}}}=\half\ln\left|h/b\right|<0$ at $r=\rphi$ (\ref{rplus}), and then increases to  converge   to   zero over the semi-infinite range, $\,\rphi<r\leq\infty$.  On the other hand, when  $b>0$ and $h=0$,  the dilaton $\phi$ increases  monotonically  from $-\infty$ to zero   over the whole range, $\alpha<r\leq\infty$, and  if $b<0$, for all $h$,  it is the opposite: $\phi$ decreases monotonically  from $\infty$ to zero   over  $\alpha<r\leq\infty$.

Now we turn to the weak energy condition.  To see its implications, we   consider  the  circular  geodesic motion of a point particle~(\ref{particleL}),  which  orbits around the central matter with fixed $r$ larger than $\rc$, 
\be
\ba{lll}
\dis{\frac{\rd^{2}x^{\lambda}}{\rd\tau^{2}}+\gamma^{\lambda}_{\mu\nu}\frac{\rd x^{\mu}}{\rd\tau}\frac{\rd x^{\nu}}{\rd\tau}=0}\,,\qquad&\quad\dis{\frac{\rd r}{\rd\tau}=0}\,,\qquad&\quad
\dis{\vartheta=\frac{\pi}{2}\,.}
\ea
\ee
With the nontrivial Christoffel symbols~(\ref{RCS}),  the radial `$\lambda=r$' component of the geodesic equation,  $\frac{\rd^{2}r}{\rd\tau^{2}}+\gamma^{r}_{tt}\big(\frac{\rd t}{\rd\tau}\big)^{2}+\gamma^{r}_{\varphi\varphi}\big(\frac{\rd \varphi}{\rd\tau}\big)^{2}=0$, determines the angular velocity~\cite{Ko:2016dxa},
\be
\dis{\left(\frac{\rd\varphi}{\rd t}\right)^{2}=-\left(\frac{\rd g_{tt}}{\rd r}\right)\left(\frac{\rd g_{\varphi\varphi}}{\rd r}\right)^{-1}=-\frac{g_{tt}^{\prime}}{2RR^{\prime}}
=\frac{2\phi^{\prime}A^{2}+A^{\prime}A}{2\phi^{\prime}C-A^{\prime}A^{-1}C+C^{\prime}}}\,.
\ee
Associating this with the centripetal acceleration measured by  the areal radius   through  Newtonian gravity,  we define and analyze an effective mass, $M(r)$, as a function of the radius, 
\be
\dis{\frac{GM(r)}{R^{2}}\equiv R\left(\frac{\rd\varphi}{\rd t}\right)^{2}\,.}
\label{Newton}
\ee
From  (\ref{2phipCe}) and (\ref{ab}), we have explicitly, with the sign factor, $\sgn$~(\ref{sgn}),
\be
\ba{lll}
M(r)&=&\cI(r)\times\frac{1}{2G} \left[(a+b)\gamma_{+}\left(\frac{r-\alpha}{r+\beta}\right)^{\frac{2b}{\sqrt{a^{2}+b^{2}}}}+(a-b)\gamma_{-}\right]\left[\gamma_{+}\left(\frac{r-\alpha}{r+\beta}\right)^{\frac{2b}{\sqrt{a^{2}+b^{2}}}}+\gamma_{-}\right]^{-1}\\
{}&=&\cI(r)\times \frac{1}{2G}\dis{\left(a+\sgn b\sqrt{1-e^{-4\phi\,}h^{2}/b^{2}}\,\right)}\\
{}&=&\cI(r)\times
\dis{\int_{0}^{r}\rd r\int_{0}^{\pi}\rd\vartheta\int_{0}^{2\pi}\rd\varphi~e^{-2d}\left(\frac{1}{8\pi G}\left|H_{t\vartheta\varphi}H^{t\vartheta\varphi}\right|-2 \EK_{t}{}^{t} \right)}\,,
\ea
\label{MIint}
\ee
where we have set
\be
\cI(r):=\dis{\frac{e^{3\phi}\sqrt{AC}}{\phi^{\prime}C-\half A^{\prime}A^{-1}C+\half C^{\prime}}=
\frac{\textstyle{\sqrt{(r-\alpha)(r+\beta)}\left[\gamma_{+}\left(\frac{r-\alpha}{r+\beta}\right)^{\frac{b+a/3}{\sqrt{a^{2}+b^{2}}\,}}+\gamma_{-}\left(\frac{r+\beta}{r-\alpha}\right)^{\frac{b-a/3}{\sqrt{a^{2}+b^{2}}\,}}\right]^{\frac{3}{2}}\,}}{r-\alpha+\half\sqrt{a^{2}+b^{2}}-\half   a+\half\sgn 
b\sqrt{1-e^{-4\phi\,}h^{2}/b^{2}\,}}}\,,
\label{cI}
\ee
which is positive  definite for sufficiently large $r$,  having  the limit,
\be
\dis{\lim_{r\rightarrow\infty}}\,\cI(r)=1\,.
\ee
In fact, for $r>\alpha$, $h\neq 0$,\footnote{For the case of $h=0$, see (\ref{Mhzero}).} the numerator in (\ref{cI}) is positive-definite,  while the denominator,
\be
\Omega(r):=r-\alpha+\half\sqrt{a^{2}+b^{2}}-\half   a+\half\sgn 
b\sqrt{1-e^{-4\phi\,}h^{2}/b^{2}}\,,
\label{Omega}
\ee
is a monotonically increasing function over $\alpha<r\leq\infty$,   taking values from a non-positive  number    to plus   infinity: with (\ref{2phipCe}),
\be
\ba{lll}
\dis{\Omega^{\prime}=1+\frac{h^{2}e^{-4\phi}}{2(r-\alpha)(r+\beta)}~>~0\,,}\quad&
\dis{\lim_{r\rightarrow\alpha}\Omega(r)=\half\left(
\sqrt{a^{2}+b^{2}}-\left|a\right|-\left|b\right|\right)~\leq~0\,,}
\quad&\dis{\lim_{r\rightarrow\infty}\Omega(r)=\infty\,.}
\ea
\label{Omegap}
\ee
If and only if $ab=0$,  we have strictly $\dis{\lim_{r\rightarrow\alpha}\Omega(r)=0}$. Otherwise,     $\cI(r)$   diverges generically  for some finite $r=\rO>\alpha$, as $\Omega(\rO)=0$.\\

We consider taking  the  large $r$ limit of  (\ref{MIint}), with the boundary condition~(\ref{BCinfty}),  to obtain
\be
M_{\infty}\equiv\lim_{r\rightarrow\infty}M(r)=\frac{\,a+ b\sqrt{1-h^{2}/b^{2}}\,}{2G}=
\dis{\int_{0}^{\infty}\!\rd r\int_{0}^{\pi}\!\rd\vartheta\int_{0}^{2\pi}\!\rd\varphi~e^{-2d}\left(\frac{1}{8\pi G}\left|H_{t\vartheta\varphi}H^{t\vartheta\varphi}\right|
- 2 \EK_{t}{}^{t} \right)}\,.
\label{Minfty}
\ee
While the first equality, or $GM_{\infty}=\half(a+b\sqrt{1-h^{2}/b^{2}})$, is the confirmation of the known result in \cite{Ko:2016dxa}, the second equality    reveals   the relationship of  the mass, $M_{\infty}$, to   the infinite-volume integral of the stringy Energy-Momentum tensor and the electric $H$-flux.  The weak energy condition~(\ref{weak})  then  precisely corresponds to the  sufficient and necessary condition for  the mass  to be positive semi-definite, $M_{\infty}\geq0$.

In fact, the mass, $M_{\infty}$,  appears in the  expansion of the metric, specifically the temporal component, $g_{tt}$,  in  the inverse of the  areal radius,\footnote{However, the mass, $M_{\infty}$, does not coincide with  the time-translational Noether  charge~(\ref{cQxi2}) for the Killing vector $\xi^{\mu}\partial_{\mu}=\partial_{t}$, nor the ADM mass \textit{a la} Wald,  $GM_{{\rm{ADM}}}=\quarter\!\left[\textstyle{a+\left(\frac{a-b}{a+b}\right)\sqrt{a^{2}+b^{2}}}\right]$~\cite{Ko:2016dxa} ( \textit{c.f.~}\cite{Park:2015bza,Blair:2015eba}).}
\be
\ba{rll}
\rd s^{2}&=&g_{tt}\rd t^{2}
+g_{RR}\rd R^{2}+R^{2}\rd\Omega^{2}\,,\\
g_{tt}(R)&=& -\left(1-\frac{2GM_{\infty}}{R}+\frac{\,b^{2}+ab\sqrt{1-h^{2}/b^{2}}-\frac{1}{2}h^{2}\,}{R^{2}}\,+\,\cdots~ \right),\\
g_{RR}(R)&=&1+\frac{\,a-b\sqrt{1-h^{2}/b^{2}}\,}{R} +\frac{a^{2}+b^{2}-\frac{5}{2}ab\sqrt{1-h^{2}/b^{2}}-\frac{1}{4} h^{2}\,}{R^{2}}\,+\,\cdots~\,,
\ea
\label{largeRmetric}
\ee 
and similarly,  
\be
2GM(R)=2GM_{\infty}+\frac{\,h^{2}-2b^{2}-2ab\sqrt{1-h^{2}/b^{2}}\,}{R}\,+\,\cdots~\,.
\ee 
Thus, effectively,   the   circular geodesic  becomes   Keplerian for large  enough  $R$: that is to say,  Stringy Gravity tends to agree with GR at long distances (${R>>GM_{\infty}}$). However, if $M(r)$ is ever  negative for some finite $r$, it means that   the gravitational ``force"    \textit{a la} (\ref{Newton})  is repulsive!   With (\ref{Omegap}),  if $M(\rO)$ diverges to plus infinity, the gravitational force is attractive and singular  at ${r=\rO}$. On the other hand, if $M(\rO)=-\infty$,  there appears  an infinite `wall' of repulsive force at the surface of ${r=\rO}$. \\

\noindent We proceed to investigate if $M(r)$ can be negative. For this, we look for the zero of $M(r)$,  denoted by $\rM$, which from (\ref{restr1}) should be greater than $\alpha$,
\be
\ba{ll}
M(\rM)=0\,,\qquad&\qquad \rM\geq\rc>\alpha\,.
\ea
\label{zMrM}
\ee
Firstly, for the special case of $h=0$ (and thus ${\gamma_{+}=1}$, ${\gamma_{-}=0}$), we have outside the matter,
\be
2GM(r)=\dis{\frac{(a+b)\sqrt{(r-\alpha)(r+\beta)}}{r-\alpha+\frac{1}{2}(\sqrt{a^{2}+b^{2}}+b-a)}\left(\frac{r-\alpha}{r+\beta}\right)^{\frac{a+3b}{2\sqrt{a^{2}+b^{2}}}}\,.}
\label{Mhzero}
\ee
This does not admit any zero which is greater than $\rc$. In particular, if $b=0$, we get $2GM=a$ constant. Henceforth we focus on nontrivial $h\neq0$. In this case, if and only if $\,\left|b\right|\geq\left|a\right|$,  $\,M(r)$ admits a  zero, $\rM$,  which is   uniquely fixed \textit{a priori}  from the first equality   of (\ref{MIint}):   with  ${a=\left|a\right|}$ (\ref{abapos}),  
\be
\rM=\dis{\frac{\,\alpha+\beta\left[\frac{\gamma_{-}(b-a)}{\gamma_{+}(a+b)}\right]^{\frac{\sqrt{a^{2}+b^{2}}}{2b}}}{1-\left[\frac{\gamma_{-}(b-a)}{\gamma_{+}(a+b)}\right]^{\frac{\sqrt{a^{2}+b^{2}}}{2b}}}}
=\left\{\!\!\ba{r}
\dis{\frac{\,\alpha+\beta\left[\frac{\gamma_{-}(\left|b\right|-\left|a\right|)}{\gamma_{+}(\left|b\right|+\left|a\right|)}\right]^{\frac{1}{2}\sqrt{1+a^{2}/b^{2}}}}{1-\left[\frac{\gamma_{-}(\left|b\right|-\left|a\right|)}{\gamma_{+}(\left|b\right|+\left|a\right|)}\right]^{\frac{1}{2}\sqrt{1+a^{2}/b^{2}}}}}=
\dis{\frac{\,\alpha+\beta\left[\frac{\gamma_{-}(\left|b\right|-\left|a\right|)}{\gamma_{-}(\left|b\right|-\left|a\right|)+2GM_{\infty}}\right]^{\frac{1}{2}\sqrt{1+a^{2}/b^{2}}}}{1-\left[\frac{\gamma_{-}(\left|b\right|-\left|a\right|)}{\gamma_{-}(\left|b\right|-\left|a\right|)+2GM_{\infty}}\right]^{\frac{1}{2}\sqrt{1+a^{2}/b^{2}}}}}
\\
\mbox{{\bf{for}}}\quad b> 0\,\quad\\
\dis{\frac{\,\alpha+\beta\left[\frac{\gamma_{+}(\left|b\right|-\left|a\right|)}{\gamma_{-}(\left|b\right|+\left|a\right|)}\right]^{\frac{1}{2}\sqrt{1+a^{2}/b^{2}}}}{1-\left[\frac{\gamma_{+}(\left|b\right|-\left|a\right|)}{\gamma_{-}(\left|b\right|+\left|a\right|)}\right]^{\frac{1}{2}\sqrt{1+a^{2}/b^{2}}}}}
=\dis{\frac{\,\alpha+\beta\left[\frac{\gamma_{+}(\left|b\right|-\left|a\right|)}{\gamma_{+}(\left|b\right|-\left|a\right|)+2GM_{\infty}}\right]^{\frac{1}{2}\sqrt{1+a^{2}/b^{2}}}}{1-\left[\frac{\gamma_{+}(\left|b\right|-\left|a\right|)}{\gamma_{+}(\left|b\right|-\left|a\right|)+2GM_{\infty}}\right]^{\frac{1}{2}\sqrt{1+a^{2}/b^{2}}}}}\\
\mbox{{\bf{for}}}\quad b<0\,.\quad
\ea
\right.
\label{rM}
\ee
Clearly  in the case of  $\left|b\right|\geq\left|a\right|$ (and $\h\neq0$),  $\rM$ is real valued  and thus can be   physical. The second equality of (\ref{MIint}) implies that  the zero of  $M(r)$  necessarily meets 
\be
\left|a\right|-\sqrt{b^{2}-h^{2}e^{-4\phi(\rM)}}=0\,,
\label{lastzero}
\ee
such that, from (\ref{sgn}),  when $b>0$, the zero must be between $\alpha$ and $\rphi$,
\be
\ba{lll}
\alpha<\rM<\rphi\qquad&\mbox{{\bf{for}}}&\qquad b>0\,.
\ea
\label{arr}
\ee
Indeed, this  can be verified directly as follows.  Provided $\left|b\right|>\left|a\right|$, the root is greater than  $\alpha$, since
\be
\rM-\alpha=\left\{
\ba{lll}
\dis{\frac{\sqrt{a^{2}+b^{2}}}{\left[\frac{\gamma_{-}(\left|b\right|-\left|a\right|)+2GM_{\infty}}{\gamma_{-}(\left|b\right|-\left|a\right|)}\right]^{\frac{1}{2}\sqrt{1+a^{2}/b^{2}}}-1}}\quad&\mbox{{\bf{for}}}&\quad b> 0\\
\dis{\frac{\sqrt{a^{2}+b^{2}}}{\left[\frac{\gamma_{+}(\left|b\right|-\left|a\right|)+2GM_{\infty}}{\gamma_{+}(\left|b\right|-\left|a\right|)}\right]^{\frac{1}{2}\sqrt{1+a^{2}/b^{2}}}-1}}\quad&\mbox{{\bf{for}}}&\quad b< 0\,,\\
\ea
\right.
\ee
and $2GM_{\infty}$ is  positive definite,  irrespective of the sign of $b$, owing to the weak energy condition~(\ref{weak}).
Further, with $b>0$, we have  from (\ref{rplus}),
\be
{\rphi-\rM=\frac{\sqrt{a^{2}+b^{2}}\left[
\left(\frac{\gamma_{-}}{\gamma_{+}}\right)^{\frac{1}{2}\sqrt{1+a^{2}/b^{2}}}
-\left(\frac{\gamma_{-}(\left|b\right|-\left|a\right|)}{\gamma_{+}(\left|b\right|+\left|a\right|)}\right)^{\frac{1}{2}\sqrt{1+a^{2}/b^{2}}}\right]
}{\left[1-\left(\frac{\gamma_{-}}{\gamma_{+}}\right)^{\frac{1}{2}\sqrt{1+a^{2}/b^{2}}}\right]\left[1-\left(\frac{\gamma_{-}(\left|b\right|-\left|a\right|)}{\gamma_{+}(\left|b\right|+\left|a\right|)}\right)^{\frac{1}{2}\sqrt{1+a^{2}/b^{2}}}\right]}}\,,
\ee
which is positive  since    $
0<\frac{\gamma_{-}(\left|b\right|-\left|a\right|)}{\gamma_{+}(\left|b\right|+\left|a\right|)}<\frac{\gamma_{-}}{\gamma_{+}}<1$.  This completes our direct  verification of (\ref{arr}).  \\

\noindent In order to see the behaviour of $M(r)$, or its sign change around the zero, ${r=\rM}$,  we need to analyze $\cI(r)$. As mentioned earlier, the numerator is (harmlessly) positive. Therefore, we focus on  the denominator, $\Omega(r)$~(\ref{Omega}). At $r=\rM$, with $\left|b\right|\geq\left|a\right|$ it reads 
\be
\ba{lll}
\Omega(\rM)&=&\rM-\alpha+\half\sqrt{a^{2}+b^{2}}-\left|a\right|\\
{}&=&\left\{\ba{lll}
{\frac{\sqrt{a^{2}+b^{2}}}{2}\left[\frac{1+\left(\frac{\gamma_{-}(\left|b\right|-\left|a\right|)}{\gamma_{-}(\left|b\right|-\left|a\right|)+2GM_{\infty}}\right)^{\frac{1}{2}\sqrt{1+a^{2}/b^{2}}}}{
\,1-\left[\frac{\gamma_{-}(\left|b\right|-\left|a\right|)}{\gamma_{-}(\left|b\right|-\left|a\right|)+2GM_{\infty}}\right]^{\frac{1}{2}\sqrt{1+a^{2}/b^{2}}}}\right]-\left|a\right|}\quad&\mbox{{\bf{for}}}&\quad b>\left|a\right|\\
{\frac{\sqrt{a^{2}+b^{2}}}{2}\left[\frac{1+\left[\frac{\gamma_{+}(\left|b\right|-\left|a\right|)}{\gamma_{+}(\left|b\right|-\left|a\right|)+2GM_{\infty}}\right]^{\frac{1}{2}\sqrt{1+a^{2}/b^{2}}}}{
\,1-\left(\frac{\gamma_{+}(\left|b\right|-\left|a\right|)}{\gamma_{+}(\left|b\right|-\left|a\right|)+2GM_{\infty}}\right)^{\frac{1}{2}\sqrt{1+a^{2}/b^{2}}}}\right]-\left|a\right|}\quad&\mbox{{\bf{for}}}&\quad b<-\left|a\right|\,.
\ea\right.
\ea
\label{OrM}
\ee
Clearly,  $\Omega(\rM)$ is negative when $\left|b\right|=\left|a\right|$,  as it reads  $\Omega(\rM)=(\frac{1}{\sqrt{2}}-1)\left| a\right|$. On the other hand,   when $\left|b\right|$ is sufficiently large it becomes positive, for example,  $\left|b\right|\geq\sqrt{3}\left|a\right|$.  In order to locate   the exact critical value of $\left|b\right|$, we let $\left|b\right|=\left|a\right|\sinh\upsilon$. For  $\left|b\right|\geq\left|a\right|$,  we restrict the range of the parameter $\upsilon\geq\ln(1+\sqrt{2})$.  Now, the precise critical value of $\upsilon$ for which  $\Omega(\rM)$~(\ref{OrM}) is trivial is determined by the following relation:
\be
\dis{
(\sinh\upsilon -1)\left[\left(\frac{2+\cosh\upsilon}{2-\cosh\upsilon}\right)^{2\tanh\upsilon}-1\right]}=\left\{\ba{lll}
\dis{\frac{2GM_{\infty}}{\gamma_{-}\left|a\right|}=\frac{1+(\gamma_{+}-\gamma_{-})\sinh\upsilon}{\gamma_{-}}}\quad
&\mbox{{\bf{for}}}&\quad b>\left|a\right|\\
\dis{\frac{2GM_{\infty}}{\gamma_{+}\left|a\right|}=\frac{1-(\gamma_{+}-\gamma_{-})\sinh\upsilon}{\gamma_{+}}}\quad
&\mbox{{\bf{for}}}&\quad b<-\left|a\right|\,.
\ea
\right.
\label{upsilon}
\ee
The  left-hand side of the first   equality, viewed  as a function of $\upsilon$, increases monotonically from zero to infinity  over the range  $\ln(1+\sqrt{2})\leq\upsilon\leq\ln(2+\sqrt{3})$, corresponding  to $1\leq\sinh\upsilon\leq\sqrt{3}$.  For the  right-hand side, we treat the two cases,  $b>\left|a\right|$ and $b<-\left|a\right|$, separately.  When $b>\left|a\right|$,  the far right-hand side increases from zero to  a finite positive value, $1/\gamma_{-}+\sqrt{3}(\gamma_{+}/\gamma_{-}-1)>0$.  Hence there must be one critical value of $\upsilon$, say $\upsilon_{\rmc}$, that meets the above equality. For the opposite case of $b<-\left|a\right|$, since the right-hand side or $M_{\infty}$ should be positive, the range of $\sinh\upsilon$ is  further  restricted to $1\leq\sinh\upsilon\leq\min\big[\sqrt{3},(\gamma_{+}-\gamma_{-})^{-1}\big]$. In either case,  $\sqrt{3}\geq\frac{1}{\gamma_{+}-\gamma_{-}}$ or  $\sqrt{3}<\frac{1}{\gamma_{+}-\gamma_{-}}$,   it is easy to see,  from the boundary values, that there must also exist  $\upsilon_{\rmc}$ satisfying  the above equality for $b<-\left|a\right|$. 

Lastly,   when $\left|b\right|=\left|a\right|\sinh\upsilon_{\rmc}$, we have $\rM=\rO$ and  $GM(\rM)$ turns out be positive-finite:
\be
\ba{ll}
\dis{\frac{2a-\sqrt{a^{2}+b^{2}}}{2a+\sqrt{a^{2}+b^{2}}}=\left[\frac{\gamma_{-}(b-a)}{\gamma_{+}(b+a)}\right]^{\frac{\sqrt{a^{2}+b^{2}}}{2b}}\,,}&\quad
GM(\rM)=\left|h\right|^{\frac{3}{2}}\frac{\left(2a+\sqrt{a^{2}+b^{2}}\right)\left(b^{2}-a^{2}\right)^{\frac{1}{4}}}{a^{2}+b^{2}}
\left[\frac{\gamma_{-}(b-a)}{\gamma_{+}(b+a)}\right]^{\frac{2a+\sqrt{a^{2}+b^{2}}}{4b}}\,.
\ea
\ee
In this case, it diverges at  ${r=\alpha\,}$ as  $\dis{\lim_{\,~r\rightarrow\alpha^{+}}GM(r)=\infty}$.

\begin{center}
\begin{figure}[t]
\centering
\includegraphics[width=75mm]{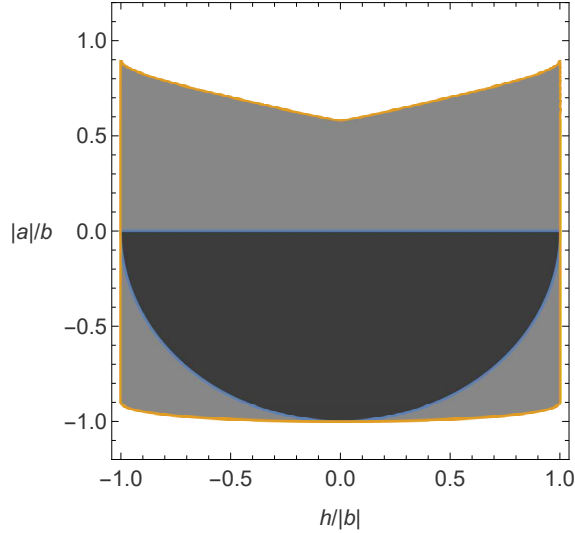}
\caption{\small{The parameter space for $(h/\left|b\right|,\left|a\right|/b)$.  In the gray region we have $\rM>\rO\,$, whereas this is not satisfied in the outer white region.  The black region corresponds to violation of the weak energy condition and is thus excluded.}}
\label{param}
\end{figure}
\end{center}

\noindent To summarize, there exists a critical value of $\left|b\right|$ given by $\left|a\right|\sinh\upsilon_{\rmc}$ which is located between $\left|a\right|$ and $\sqrt{3}\left|a\right|$, such that if $\left|b\right|>\left|a\right|\sinh\upsilon_{\rmc}$,   we have $\rO<\rM$ and hence, 
crossing  ${r=\rM}$ from  outside (${r>\rM}$)  to inside (${r<\rM}$), the gravity changes from  being   attractive, $M(r)>0$,  to   repulsive,   $M(r)<0$.  Further deep inside at ${r=\rO}$, there is an infinite repulsive wall.    On the other hand, if $\left|b\right|<\left|a\right|\sinh\upsilon_{\rmc}$, we have $\rM<\rO$, such that upon approaching from infinity towards  the center, the  effective mass, $M(r)$, or the attractive gravitational force,  increases and  eventually   diverges   at  $r=\rO$.   In the special case of $\left|b\right|=\left|a\right|\sinh\upsilon_{\rmc}$,  $M(r)$ is positive-definite for $r\geq\rc>\alpha$.

In either the case of $\left|b\right|>\left|a\right|\sinh\upsilon_{\rmc}$ or $\left|b\right|<\left|a\right|\sinh\upsilon_{\rmc}$, in order to avoid  (crossing) the  singularity at $\rO$,  it seems  physically reasonable to  have
\be
\rc\geq\rO\,.
\label{rcrO} 
\ee
Specifically when  $\left|b\right|$ is large enough, or $\left|b\right|>\left|a\right|\sinh\upsilon_{\rmc}$,  and thus there is an infinite repulsive wall at ${r=\rO}$, the above assumption~(\ref{rcrO}) appears even more  physically natural, as no falling body can penetrate the wall.  In contrast, if $b$ is small such that $\left|b\right|<\left|a\right|\sinh\upsilon_{\rmc}$,  it may be hard to maintain the above condition since  the gravitational attraction may become too strong.  
For this reason,  it appears necessary to postulate the large $\left|b\right|$   condition, \textit{i.e.}
\be
\left|b\right|>\left|a\right|\sinh\upsilon_{\rmc}\,,
\ee  
along with   the strong energy~(\ref{strong}), weak energy~(\ref{weak}), and pressure~(\ref{pressure}) conditions.  The allowed parameter region is shown in Figure~\ref{param}.\\



\section{Conclusion\label{SECFINAL}}
In this work we have proposed that Double Field Theory is Stringy Gravity, \textit{i.e.}~the upgrade   of General Relativity which is in accordance with the symmetries of string theory.  To this end  we developed a definition for the stringy Energy-Momentum tensor and presented the Einstein Double Field Equations.  As an example, we focused on    ${D=4}$ regular solutions.   Our main results  are summarized below with comments.

\begin{itemize}
\item[$\ast$] For a generic  action of Stringy Gravity~(\ref{SGaction}),
\be
\dis{\int_{\Sigma}e^{-2d}\Big[\,\textstyle{\frac{1}{16\pi G}}\So+L_{\rm{matter}}\,\Big]\,,}
\label{Caction}
\ee
the  stringy Energy-Momentum tensor is defined by~(\ref{EMSG}), 
\be
\EM_{AB}:=4V_{[A}{}^{p}\brV_{B]}{}^{\brq}\EK_{p\brq}-\half\cJ_{AB}\To\,,
\ee
which contains   ${D^{2}+1}$ components  (\ref{cK}),
\be
\ba{ll}
\EK_{p\brq}:=\dis{\frac{1}{2} \left(V_{Ap}\frac{\delta L_{\rm{matter}}}{\delta \brV_{A}{}^{\brq}}-\brV_{A\brq}\frac{\delta L_{\rm{matter}}}{\delta V_{A}{}^{p}}\right)\,,}\qquad&\qquad
\To:= e^{2d}\times\dis{\frac{\delta\left(e^{-2d} L_{\rm{matter}}\right)}{\delta d}\,.}
\ea
\ee
The general covariance of the action~(\ref{Caction}) under  doubled-yet-gauged diffeomorphisms   implies  the on-shell conservation law,  
\be
\cD_{A}\EM^{AB}\equiv 0\,,
\ee
which holds up to the equations of motion of the matter fields and  consists of  ${D+D}$ components.

\item[$\ast$] The Einstein Double Field Equations~(\ref{EDFE}) equate the stringy Einstein tensor and  the stringy   Energy-Momentum tensor as
\be
G_{AB}=8\pi G\EM_{AB}\,.
\ee
They comprise  the full  set of equations of motion for the closed string massless sector, \textit{i.e.~}the stringy graviton fields, $\{V_{Ap},\brV_{B\brq},d\}$,  which may reduce to the conventional fields, $\{g_{\mu\nu},B_{\mu\nu},\phi\}$,   upon reduction to Riemannian backgrounds.  They can also be applied readily to non-Riemannian spacetimes~(\ref{cHFINAL}), \cite{Morand:2017fnv}.

\item[$\ast$]  The further-generalized Lie derivative is defined as (\ref{fgL})
\be
\ba{lll}
\fcL_{\xi}T_{Mp\brp}{}^{\alpha}{}_{\beta}{}^{\bralpha}{}_{\brbeta}\!\!\!&:=\!\!&\xi^{N}\cD_{N}T_{Mp\brp}{}^{\alpha}{}_{\beta}{}^{\bralpha}{}_{\brbeta}+\omega\cD_{N}\xi^{N}T_{Mp\brp}{}^{\alpha}{}_{\beta}{}^{\bralpha}{}_{\brbeta}+2\cD_{[M}\xi_{N]}T^{N}{}_{p\brp}{}^{\alpha}{}_{\beta}{}^{\bralpha}{}_{\brbeta}\\
{}&{}&+2\cD_{[p}\xi_{q]}T_{M}{}^{q}{}_{\brp}{}^{\alpha}{}_{\beta}{}^{\bralpha}{}_{\brbeta}+
\half\cD_{[r}\xi_{s]}(\gamma^{rs})^{\alpha}{}_{\delta}T_{Mp\brp}{}^{\delta}{}_{\beta}{}^{\bralpha}{}_{\brbeta}- \half\cD_{[r}\xi_{s]}(\gamma^{rs})^{\delta}{}_{\beta}T_{Mp\brp}{}^{\alpha}{}_{\delta}{}^{\bralpha}{}_{\brbeta}\\
{}&{}&+2\cD_{[\brp}\xi_{\brq]}T_{Mp}{}^{\brq\alpha}{}_{\beta}{}^{\bralpha}{}_{\brbeta}+
\half\cD_{[\brr}\xi_{\brs]}(\brgamma^{\brr\brs})^{\bralpha}{}_{\brdelta}T_{Mp\brp}{}^{\alpha}{}_{\beta}{}^{\brdelta}{}_{\brbeta}-
\half\cD_{[\brr}\xi_{\brs]}(\brgamma^{\brr\brs})^{\brdelta}{}_{\brbeta}T_{Mp\brp}{}^{\alpha}{}_{\beta}{}^{\bralpha}{}_{\brdelta}\,,
\ea
\label{CfgL}
\ee
which is completely covariant for doubled-yet-gauged diffeomorphisms, twofold local Lorentz symmetries, and $\ODD$ rotations. It is closed by the C-bracket plus twofold local Lorentz transformations, (\ref{fclosed}).  Specifically,  acting on DFT vielbeins, they read (\ref{fcLVV})
\be
\ba{ll}
\fcL_{\xi}V_{Ap}=\brP_{A}{}^{B}\left(\hcL_{\xi}P_{BC}\right)V^{C}{}_{p}
\,,\qquad&\qquad
\fcL_{\xi}\brV_{A\brp}=P_{A}{}^{B}\left(\hcL_{\xi}\brP_{BC}\right)\brV^{C}{}_{\brp}\,.
\ea
\ee
Thus, by setting these  further-generalized Lie derivatives to vanish,  it becomes  possible to  characterize the isometry of the doubled-yet-gauged spacetime within the DFT-vielbein formalism.

\item[$\ast$] The most general ${D=4}$   spherically symmetric ansatzes  for both  the (Riemannian) stringy graviton fields  and  the  stringy Energy-Momentum tensor are identified, (\ref{ansatzgB}),  (\ref{ansatzcK}),  which  enjoy $\so(3)$ isometries~(\ref{so3Hd}), (\ref{Killing}), (\ref{so3VV}).  In particular, with  $\EK_{\mu\nu}\equiv2 e_{\mu}{}^{p}\bre_{\nu}{}^{\brq}\EK_{p\brq}$, the spherically symmetric Energy-Momentum tensor may possess  not only  diagonal but also    off-diagonal components, such as generic $\EK_{tr}=\EK_{(tr)}+\EK_{[tr]}$ and skew-symmetric $\EK_{[\vartheta\varphi]}$ components. The skew-symmetry  is a genuine feature  of    fermionic or stringy matter, (\ref{spinorEK}), (\ref{stringEK}), which is induced by their coupling to the $B$-field. 

\item[$\ast$] The ${D=4}$ Einstein Double Field equations were  solved for the most general spherically symmetric, asymptotically flat, static `regular' configurations in which the stringy Energy-Momentum tensor of the `matter' vanishes   completely  outside a  cutoff radius $\rc$: $T_{AB}=0$ for $r\geq\rc$ (\ref{BCT}).

Outside the matter,  we recover the known vacuum geometry~\cite{Ko:2016dxa} with  four constant parameters including  electric $H$-flux, $\{\alpha,\beta,a,h\}$~(\ref{SOL2}),
\be
\ba{ccc}
{\ba{lc}
e^{2\phi}=\gamma_{+}\left(\frac{r-\alpha}{r+\beta}\right)^{\frac{b}{\sqrt{a^{2}+b^{2}}}}+\gamma_{-}\left(\frac{r+\beta}{r-\alpha}\right)^{\frac{b}{\sqrt{a^{2}+b^{2}}}}\,,\quad&
H_{\scriptscriptstyle{(3)}}=h\sin\vartheta\,\rd t\wedge\rd\vartheta\wedge\rd\varphi\,,\\
\multicolumn{2}{c}{
\rd s^{2}=e^{2\phi}\left[-\left(\frac{r-\alpha}{r+\beta}\right)^{\frac{a}{\sqrt{a^{2}+b^{2}}}}\rd t^{2}
+\left(\frac{r+\beta}{r-\alpha}\right)^{\frac{a}{\sqrt{a^{2}+b^{2}}}}\big\{\rd r^{2}+(r-\alpha)(r+\beta)
\rd\Omega^{2}\big\}\right]}
\ea}&\mbox{\bf{for}}&r\geq\rc\,,
\ea
\ee
where $b=\sqrt{(\alpha+\beta)^{2}-a^{2}}\geq 0\,$ or $\,b=-\sqrt{(\alpha+\beta)^{2}-a^{2}}<0$, and $\gamma_{\pm}=\half(1\pm\sqrt{1-h^{2}/b^{2}})$.  Inside the matter, \textit{i.e.~}$r<\rc$, while the electric $H$-flux assumes  the same form as  outside, magnetic $H$-flux, \textit{i.e.~}$H_{r\vartheta\varphi}$, may be present,  being  sourced by the  skew-symmetric angular  component, $\EK_{[\vartheta\varphi]}$,  \textit{c.f.~}(\ref{cVdef}), (\ref{fixB}).  Crucially, we derive integral expressions for  all  the constants,  $\{\alpha,\beta,a,h\}$, with the integrands given by the stringy Energy-Momentum tensor~(\ref{ababh}), which thus  reveal the physical meanings of the ``free" constant parameters in the vacuum geometry.  In particular, while in General Relativity a diagonal metric implies a diagonal Energy-Momentum tensor~(\ref{TcompGR}) through the Einstein Field Equations, in Stringy Gravity we can have a non-zero off-diagonal component $\EK_{(tr)}$ which sources nontrivial electric $H$-flux and thus modifies the Schwarzschild geometry.  This reflects the general nature of Stringy Gravity: the richer stringy Energy-Momentum tensor---$(D^{2}+1)$ degrees of freedom \textit{vs.} $D(D+1)/2$---enhances the geometry beyond General Relativity.

\item[$\ast$] We spell out the strong energy condition~(\ref{strong}) and the pressure condition~(\ref{pressure}), which  make $\alpha$, $\beta$ and $a$ positive semi-definite~(\ref{abapos}). For the solution to be real and non-degenerate,   we postulate that    $\rc$ should exceed $\alpha$ (\ref{restr1}).   If $b>0$ and $h
\neq 0$, $\phi$ decreases from $\infty$ to a single negative minimum value  and then starts to increase monotonically converging to zero as $r\rightarrow\infty$ (\ref{philn}). On the other hand, if $b<0$, $\phi$ decreases monotonically from $\infty$ to  zero over the whole range of $\alpha\leq r\leq\infty$.

\item[$\ast$] We consider  an effective mass, $M(r)$~(\ref{Newton}), which is responsible for the  centripetal acceleration  of  circular geodesics, 
\be
\dis{\frac{GM(r)}{R^{2}}=R\left(\frac{\rd\varphi}{\rd t}\right)^{2}\,.}
\ee
Taking the large-$r$ limit, we derive an integral expression for  the asymptotic mass~(\ref{Minfty}),
\be
M_{\infty}=\lim_{r\rightarrow\infty}M(r)=\frac{a+ b\sqrt{1-h^{2}/b^{2}}}{2G}=
\dis{\int_{0}^{\infty}\!\rd r\!\int_{0}^{\pi}\!\rd\vartheta\!\int_{0}^{2\pi}\!\rd\varphi~e^{-2d}\left(\frac{1}{8\pi G}\left|H_{t\vartheta\varphi}H^{t\vartheta\varphi}\right|
- 2 \EK_{t}{}^{t} \right)}\,,
\label{MinftyCON}
\ee
where the integrand comprises the electric $H$-flux squared and the stringy Energy-Momentum component, $\EK_{t}{}^{t}$.  Requiring  $M_{\infty}$ to be positive-definite amounts  to the weak energy condition~(\ref{weak}). 

Compared to the  mass formula in GR~(\ref{MassGR2}),  (\ref{sqrtgneq}),  the above result~(\ref{MinftyCON}) is in a sense more satisfactory, as the integration is equipped  with the proper   integral measure in Stringy Gravity, $e^{-2d}$, while from (\ref{EKEKEK}),  $\EK_{t}{}^{t}=-2\EK_{0}{}^{\bar{0}}$ which is a diffeomorphism scalar.

Rather than requiring the integrand of (\ref{MinftyCON}) to be strictly  positive-definite, namely, the weak energy density condition~(\ref{weakdensity}),  by demanding this only of the whole integral, \textit{i.e.~}the weak energy condition,  we may  allow  $M(r)$ to  possibly  assume  negative values,  and hence for gravity to become repulsive  for some range of the radius.

While we view Stringy Gravity as a unique string-theory-based alternative to General Relativity,  from (\ref{MinftyCON}) one may regard the $H$-flux, especially the electric component, as dark matter  
(\textit{c.f.~}\cite{Cheung:2008fz}) and $-2\EK_{t}{}^{t}$ as the baryonic energy (mass) density.

\item[$\ast$] As long as $ab\neq0$ and $h\neq 0$, $M(r)$ becomes singular at one finite radius greater than $\alpha$, ${r=\rO}>\alpha$. If $M(\rO)=+\infty$,  the gravitational ``force" remains attractive and diverges.   If  the opposite is true,  ${M(\rO)=-\infty}$, then  there exists  an infinite repulsive `wall' located at $r=\rO$.  Furthermore, when $\left|b\right|>\left|a\right|$, $M(r)$ admits a single real root which is greater than $\alpha$: $M(\rM)=0$, $\rM>\alpha$. Thus, with $M_{\infty}>0$, if $\rM$ is either  complex (non-physical) or less than $\rO$,   the gravity is attractive in the entire region from $r=\infty$ to the singular surface of    $r=\rO$. On the other hand, if $\rM>\rO$,  the gravity is attractive for $\rM<r\leq\infty$, vanishes at $r=\rM$, and becomes repulsive for  $\rO\leq r<\rM$, encountering the infinite wall at $r=\rO$.

\item[$\ast$] Indeed, when $\left|b\right|$ is sufficiently large, such as $\left|b\right|\gtrsim\sqrt{3}\left|a\right|$, the zero of $M(r)$ is guaranteed to exceed the singular radius, $\rM>\rO$. In such cases, gravity  \textit{i)~}remains  attractive over the semi-infinite range, $\rM<r\leq\infty$,  featuring   Keplerian limiting behavior towards infinity,  \textit{ii)~}vanishes at $r=\rM$, and  \textit{iii)~}becomes  repulsive in the finite interval $\rM<r\leq\rO$,  with the infinite wall  at $r=\rO$.   It is then physically reasonable to assume that the cutoff radius should be larger than or equal to  the singular radius  of the  infinite wall,  $\rc\geq\rO$. Specifically,  for non-interacting dust, $r=\rc$ may well coincide with the surface of zero gravity: $\rc=\rM$. This seems to provide a dynamical mechanism to avoid the singularity. 
\end{itemize}

\noindent Possible future work includes exhaustive exploration of the parameter space,   exploring the causal structure,  direct verification of the energy and pressure  conditions including the large-$\left|b\right|$ condition,  \textit{c.f.}~(\ref{realb}),      extensions  to non-Riemannian spacetimes, stringy thermodynamics,  applications to cosmology,  and last but not least,  tests against observations.


\section*{Acknowledgements}
We wish to thank    Chris Hull, Hyun Gyu Kang,   Kevin Morand for helpful discussions, and Chaiho Rim for generous support.  This work  was  supported by  the National Research Foundation of Korea   through  the Grant 2016R1D1A1B01015196.\\
\newpage
\appendix

\section{C20 GR prior to C21 DFT\label{SECAPPENDIX}}
In this Appendix,  after reviewing  the Energy-Momentum tensor and  Einstein Field Equations in GR, we discuss the most general  static spherically symmetric regular solution in four-dimensional spacetime.

\subsection{Energy-Momentum tensor and Einstein Field Equations in GR}
Let us consider the  Einstein--Hilbert action coupled to generic matter,
\be
\int\rmd^{D\!}x~\sqrt{-g}\left(\textstyle{\frac{1}{16\pi G}}R+L_{\rm{matter}}\right)\,,
\label{GRaction}
\ee
where $L_{\rm{matter}}$  is the Lagrangian for  various  matter fields, which we denote by  $\Upsilon_{a}$. The Energy-Momentum tensor  is  conventionally defined  by
\be
T_{\mu\nu}:=\frac{-2}{\sqrt{-g}}\frac{\,\delta(\sqrt{-g}L_{\rm{matter}})\,}{\delta g^{\mu\nu}}=-2\frac{\delta L_{\rm{matter}}}{\delta g^{\mu\nu}}+g_{\mu\nu} L_{\rm{matter}}\,,
\label{TGRdef}
\ee
where $\frac{\delta\quad}{\delta g^{\mu\nu}}$ is the standard functional derivative which is best  computed from the infinitesimal variation of the Lagrangian density induced by $\delta g^{\mu\nu}$,  as
\be
\delta\!\left(\sqrt{-g}L_{\rm{matter}}\right)+\half\sqrt{-g}\,\delta g^{\mu\nu}g_{\mu\nu}L_{\rm{matter}}=\sqrt{-g}\,\delta g^{\mu\nu}\frac{\,\delta L_{\rm{matter}}\,}{\delta g^{\mu\nu}}\,+\,\mbox{total~derivatives}\,,
\label{deltaLd}
\ee
where the  disregarded total derivatives appear generically,  since    $L_{\rm{matter}}$     includes   covariant derivatives of the form  $\trd_{\mu}\!\Upsilon_{a}$,   and the connections  contain    the derivatives of the metric.  The right-hand side of (\ref{deltaLd}) then gives, or operationally  defines, the functional derivative, $\frac{\,\delta L_{\rm{matter}}\,}{\delta g^{\mu\nu}}$, as the  functional `coefficient' of $\sqrt{-g}\,\delta g^{\mu\nu}$. It is also useful to note that if we vary not the Lagrangian density but the  weightless scalar, $L_{\rm{matter}}$, covariant total derivatives of  the form `$\trd_{\mu}J^{\mu}$' will  appear and should be neglected,
\be
\delta L_{\rm{matter}}=\delta g^{\mu\nu}\frac{\,\delta L_{\rm{matter}}\,}{\delta g^{\mu\nu}}\,+\,\mbox{covariant~total~derivatives}\,.
\label{deltaLd2}
\ee

 Of course, when $ L_{\rm{matter}}$ involves  fermions,
\be
\sqrt{-g}L_{\rm{fermion}}=|e|\,\bar{\rho}(\gamma^{\mu}\trd_{\mu}\rho+m\rho)\,,
\label{Lfermion}
\ee
one should consider \textit{a priori}   the variation of    the  vielbein  rather than the metric, 
\be
\ba{ll}
\delta\gamma^{\rho}_{\mu\nu}=\half g^{\rho\sigma}\left(\trd_{\mu}\delta g_{\sigma\nu}+\trd_{\nu}\delta g_{\mu\sigma}-\trd_{\sigma}\delta g_{\mu\nu}\right)\,,\quad&\quad \delta g_{\mu\nu}=2e_{(\mu}{}^{a}\delta e_{\nu) a}\,,\\
\multicolumn{2}{c}{
\delta\omega_{\mu ab}=e_{\mu}{}^{c}\left(e^{-1}_{[a}\trd_{c]}\delta e_{\nu b}-e^{-1}_{[b}\trd_{c]}\delta e_{\nu a}+e^{-1}_{[a}\trd_{b]}\delta e_{\nu c}\right)\,.}
\ea
\ee
Up  to  the fermionic equations of motion,  
\be
\gamma^{\mu}\trd_{\mu}\rho+m\rho\equiv 0\,,
\ee
we note that
\be
\trd_{\lambda}\!\left(\bar{\rho}\gamma^{\lambda\mu\nu}\rho\right)
\equiv -2\left(\bar{\rho}\gamma^{[\mu}\trd^{\nu]}\rho
+\trd^{[\mu}\bar{\rho}\gamma^{\nu]}\rho\right)\,.
\ee
Here the equivalence symbol, `$\equiv$', denotes  on-shell equality which holds up to the  equations of motion of the matter fields.   
Using this relation and  disregarding any  total derivatives ($\simeq$),   we   derive    the variation of  the fermionic Lagrangian, $\sqrt{-g}L_{\rm{fermion}}$~(\ref{Lfermion}), given by
\be
\delta\Big[|e|\,\bar{\rho}(\gamma^{\mu}\trd_{\mu}\rho+m\rho)\Big]\simeq\half |e|\,\delta e_{\mu p}\left(\trd^{(\mu}\bar{\rho}\gamma^{p)}\rho
-\bar{\rho}\gamma^{(\mu}\trd^{p)}\rho\right)=-\quarter 
|e|\,\delta g^{\mu\nu}\left(\trd_{(\mu}\bar{\rho}\gamma_{\nu)}\rho-\bar{\rho}\gamma_{(\mu}\trd_{\nu)}\rho\right)\,.
\label{GRspinor}
\ee
This result shows that even for fermions,  the definition of the  Energy-Momentum tensor through the `metric' variation~(\ref{TGRdef}) is in a sense still valid, leading  to the following `symmetric'  contribution, 
\be
T^{\rm{fermion}}_{\mu\nu}=\half\!\left(\trd_{(\mu}\bar{\rho}\gamma_{\nu)}\rho-\bar{\rho}\gamma_{(\mu}\trd_{\nu)}\rho\right)\,.
\ee
That is to say,  the Energy-Momentum tensor  is always symmetric in 
GR.\footnote{\textit{c.f.~}(\ref{EMSG}), (\ref{TH}) for Stringy Gravity, \textit{i.e.}~DFT.}  In fact, this can be shown in a  more general setup. The arbitrary variation of the vielbein  decomposes  into two parts,
\be
\delta e_{\mu}{}^{a}=e_{\mu b}\delta e_{\nu}{}^{(a}e^{b)\nu}+
\delta e_{\nu}{}^{[a}e^{b]\nu}e_{\mu b}=\half e^{a\nu}\delta g_{\mu\nu}+\delta e_{\nu}{}^{[a}e^{b]\nu}e_{\mu b}\,,
\label{deltae}
\ee
of which the last term can be viewed  as the infinitesimal local Lorentz transformation of the vielbein. Since $L_{\rm{matter}}$ is supposed to be a singlet of  local Lorentz symmetry,  using (\ref{deltae})  the generic variation of $L_{\rm{matter}}$ can be written  as
\be
\delta L_{\rm{matter}}=\delta e_{\mu}{}^{a}\frac{\delta L_{\rm{matter}}}{\delta e_{\mu}{}^{a}}+\delta\Upsilon_{a}\frac{\delta L_{\rm{matter}}}{\delta\Upsilon_{a}}+\cdots=
\half \delta g_{\mu\nu}\left(e^{a\nu}\frac{\delta L_{\rm{matter}}}{\delta e_{\mu}{}^{a}}\right)+\delta^{\prime}\Upsilon_{a}\frac{\delta L_{\rm{matter}}}{\delta\Upsilon_{a}}+\cdots\,.
\label{deltamatter}
\ee
Here the dots, `$\cdots$',  denote  any disregarded  covariant total derivatives;  $\delta^{\prime}\Upsilon_{a}$ is the  arbitrary variation  of the matter field supplemented by  the  infinitesimal local Lorentz rotation, $\delta e_{\nu}{}^{[a}e^{b]\nu}$. Further,  $\frac{\delta L_{\rm{matter}}}{\delta \Upsilon_{a}}$ corresponds to the Euler--Lagrange equation  for each  matter field, $\Upsilon_{a}$.  
Thus, the Energy-Momentum tensor~(\ref{TGRdef}) is `on-shell' equivalent to 
\be
T_{\mu\nu}=-\half e^{-1}\left[e_{\mu a}\frac{\delta(e L_{\rm{matter}})}{\delta e_{a}{}^{\nu}}+e_{\nu a}\frac{\delta (eL_{\rm{matter}})}{\delta e_{a}{}^{\mu}}\right]=-\half\left(e_{\mu a}\frac{\delta L_{\rm{matter}}}{\delta e_{a}{}^{\nu}}+e_{\nu a}\frac{\delta L_{\rm{matter}}}{\delta e_{a}{}^{\mu}}\right)+g_{\mu\nu} L_{\rm{matter}}\,.
\label{eTGRdef}
\ee

\noindent Now for  the Einstein--Hilbert term,   it is useful to remember  that the variation of the Riemann curvature is given by covariant total derivatives,
\be
\delta R^{\rho}{}_{\sigma\mu\nu}=\trd_{\mu}(\delta\gamma^{\rho}_{\nu\sigma})-
\trd_{\nu}(\delta\gamma^{\rho}_{\mu\sigma})\,,
\label{deltaRiemann}
\ee
such that
\be
\delta R=\delta(g^{\mu\nu}R_{\mu\nu})=\delta g^{\mu\nu}R_{\mu\nu}+
\trd_{\rho}(g^{\mu\nu}\delta\gamma^{\rho}_{\nu\mu}-
g^{\rho\mu}\delta\gamma^{\nu}_{\nu\mu})\,,
\label{deltaR}
\ee
and
\be
\delta(\sqrt{-g} R)=\delta g^{\mu\nu}\sqrt{-g}(R_{\mu\nu}-\half g_{\mu\nu}R)+
\partial_{\rho}(\sqrt{-g}g^{\mu\nu}\delta\gamma^{\rho}_{\nu\mu}-\sqrt{-g}
g^{\rho\mu}\delta\gamma^{\nu}_{\nu\mu})\,.
\ee
Bringing this all together, while disregarding any  surface integral,   the arbitrary variation of the action  reads
\be
\ba{l}
\delta\dis{\int\rmd^{D\!}x~\sqrt{-g}\left(\textstyle{\frac{1}{16\pi G}}R+L_{\rm{matter}}\right)}\\
\equiv
\dis{\int\rmd^{D\!}x~\sqrt{-g}\left[
{\textstyle{\frac{1}{16\pi G}}}\delta g^{\mu\nu}\!\left(R_{\mu\nu}-\half g_{\mu\nu}R-8\pi G\, T_{\mu\nu}\right)+\delta^{\prime}\Upsilon_{a}\frac{\delta L_{\rm{matter}}}{\delta \Upsilon_{a}}\right]\,.}
\ea
\label{deltaaction}
\ee
Clearly the equation of motion of the metric leads to the (undoubled) Einstein Field Equations,
\be
R_{\mu\nu}-\half g_{\mu\nu}R=8\pi G T_{\mu\nu}\,,
\label{EFE}
\ee
which are equivalent to
\be
R_{\mu\nu}=8\pi G\left(T_{\mu\nu}-\textstyle{\frac{1}{\,D-2\,}}g_{\mu\nu}T^{\lambda}{}_{\lambda}\right)\,.
\label{EFE2}
\ee
In particular, when the variation is caused by diffeomorphisms,  \textit{i.e.~}$\delta g^{\mu\nu}=\cL_{\xi}g^{\mu\nu}=-2\trd^{(\mu}\xi^{\nu)}$,  the action should be invariant.  From (\ref{deltaaction}),     we get,   up to the equations of motion,   $\frac{\delta L_{\rm{matter}}}{\delta \Upsilon_{a}}\equiv0$, and up to surface integrals, 
\be
0\equiv
\dis{\int\rmd^{D\!}x~\partial_{\mu}\left[\xi^{\mu}\sqrt{-g}\left(\textstyle{\frac{1}{16\pi G}}R+L_{\rm{matter}}\right)\right]\equiv
{\textstyle{\frac{1}{8\pi G}}}\dis{\int\rmd^{D\!}x~\sqrt{-g}
\xi_{\nu}\trd_{\mu}\left(R^{\mu\nu}-\half g^{\mu\nu}R-8\pi G\, T^{\mu\nu}\right)\,.}}
\ee
This should hold for arbitrary $\xi^{\mu}$, \textit{e.g.~}highly localized delta-function-type vector fields.  Therefore  the Einstein curvature  and   the Energy-Momentum tensor should be conserved  off-shell and on-shell, respectively,
\be
\ba{cc}
\trd_{\mu}(R^{\mu\nu}-\half g^{\mu\nu}R)=0\,,\qquad&\qquad
\trd_{\mu}T^{\mu\nu}\equiv0\,.
\ea
\label{consET}
\ee
As is well known, the former can also be obtained directly  from the Bianchi identity, $\trd_{[\lambda}R_{\mu\nu]\rho\sigma}=0$.\\

\noindent For any  Killing vector satisfying the isometric  condition, 
\be
\cL_{\xi}g_{\mu\nu}=\trd_{\mu}\xi_{\nu}+\trd_{\nu}\xi_{\mu}=0\,,
\ee
the conservation of the Energy-Momentum tensor~(\ref{consET}) implies the existence of  a  conserved Noether current,
\be
\partial_{\mu}(\sqrt{-g}T^{\mu}{}_{\nu}\xi^{\nu})=
\sqrt{-g}\trd_{\mu}(T^{\mu\nu}\xi_{\nu})\equiv\sqrt{-g}T^{\mu\nu}\trd_{(\mu}\xi_{\nu)}\equiv0\,,
\ee
which in turn defines a Noether charge,
\be
Q[{\xi}]:=\int\rd x^{D-1}~\sqrt{-g}T^{t}{}_{\mu}\xi^{\mu}\,.
\label{NoetherchargeGR}
\ee


\subsection{Regular spherically symmetric  solution in ${D=4}$ GR}
Here we derive the most general,   asymptotically flat,  spherically symmetric, static, regular solution to the  $D=4$ Einstein Field Equations.  We require   the   metric and Energy-Momentum tensor to be spherically symmetric, 
\be
\ba{ll}
\cL_{\xi_{a}}g_{\mu\nu}=0\,,\qquad&\qquad
\cL_{\xi_{a}}T_{\mu\nu}=0\,,
\ea
\label{Killingeq}
\ee
with three Killing vectors,  $\xi^{\mu}_{a}$, $a=1,2,3$,  corresponding to  the usual  angular momentum differential operators, 
\be
\ba{lll}
\xi_{1}=\sin\varphi\partial_{\vartheta}+\cot\vartheta\cos\varphi\partial_{\varphi}\,,\quad&\quad
\xi_{2}=-\cos\varphi\partial_{\vartheta}+
\cot\vartheta\sin\varphi\partial_{\varphi}\,,\quad&\quad
\xi_{3}=-\partial_{\varphi}\,.
\ea
\label{KillingA}
\ee
They  satisfy the $\so(3)$ commutation relation,
\be
\big[\xi_{a},\xi_{b}\big]=\sum_{c}\epsilon_{abc\,}\xi_{c}\,.
\ee
It follows from (\ref{Killingeq}) that   
\be
\ba{llll}
g_{\vartheta\varphi}=g_{\varphi\vartheta}=0\,,\quad&\quad 
g_{\varphi\varphi}=\sin^{2}\vartheta g_{\vartheta\vartheta}\,,\quad&\quad
T_{\vartheta\varphi}=T_{\varphi\vartheta}=0\,,\quad&\quad 
T_{\varphi\varphi}=\sin^{2}\vartheta T_{\vartheta\vartheta}\,.
\ea
\ee
Furthermore, without loss of generality,  we can put ${g_{tr}=0}$, $\,{g_{\vartheta\vartheta}=r^{2}}$, and set the metric to be diagonal, utilizing diffeomorphisms (see \textit{e.g.~}\cite{Weinberg:1972kfs}),
\be
\rd s^{2}=-B(r)\rd t^{2} +A(r)\rd r^{2}+r^2\rd\Omega^{2}\,,
\label{metric}
\ee
where we put as  shorthand  notation, 
\be
\rd\Omega^{2}:=\rd\vartheta^{2}+\sin^{2}\vartheta\,\rd\varphi^{2}\,.
\ee
The nonvanishing  Christoffel symbols  are then exhaustively, 
\be
\ba{lll}
\gamma^{t}_{tr}=\gamma^{t}_{rt}
=\frac{~B^{\prime}}{2B}\,,\quad&\quad
\gamma^{r}_{tt}=\frac{~B^{\prime}}{2A}\,,\quad&\quad
\gamma^{r}_{rr}=\frac{~A^{\prime}}{2A}\,,\\
\gamma^{r}_{\vartheta\vartheta}=-\frac{r}{A}\,,\quad&\quad
\gamma^{r}_{\varphi\varphi}
=-\sin^{2}\vartheta\, \frac{r}{A}\,,\quad&\quad
\gamma^{\vartheta}_{r\vartheta}=\gamma^{\vartheta}_{\vartheta r}=\frac{1}{r}\,,\\
\gamma^{\vartheta}_{\varphi\varphi}=-\sin\vartheta\cos\vartheta\,,\qquad&\quad
\gamma^{\varphi}_{r\varphi}=\gamma^{\varphi}_{\varphi r}=
\frac{1}{r}\,,\quad&\quad
\gamma^{\varphi}_{\vartheta\varphi}=\gamma^{\varphi}_{\varphi\vartheta}
=\cot\vartheta\,,
\ea
\ee
and subsequently the Ricci curvature, $R_{\mu\nu}$, becomes  diagonal, with components
\be
\ba{ll}
R_{tt}=\frac{~B^{\prime\prime}}{2A}-\frac{~B^{\prime}}{4A}\left(\frac{~A^{\prime}}{A}+\frac{~B^{\prime}}{B}\right)+\frac{1}{r}\frac{~B^{\prime}}{A}\,,\quad&\qquad
R_{rr}=-\frac{~B^{\prime\prime}}{2B}+\frac{~B^{\prime}}{4B}\left(\frac{~A^{\prime}}{A}+\frac{~B^{\prime}}{B}\right)+\frac{1}{r}\frac{~A^{\prime}}{A}\,,\\
R_{\vartheta\vartheta}=1+\frac{~r}{2A}\left(\frac{~A^{\prime}}{A}-\frac{~B^{\prime}}{B}\right)-\frac{1}{A}\,,\quad&\qquad
R_{\varphi\varphi}=\sin^{2\!}\vartheta\,R_{\vartheta\vartheta}\,.
\ea
\ee
Since both the Ricci curvature and the metric are diagonal, the Einstein Field Equations imply that the Energy-Momentum tensor must also be diagonal, thus fixing {$T_{tr}=0$},
\be
T_{\mu\nu}=\mbox{diag}\big(\,T_{tt}\,,~~~\, T_{rr}\,,~~~\,
T_{\vartheta\vartheta}\,,~~~\,
T_{\varphi\varphi}=\sin^{2\!}\vartheta\, T_{\vartheta\vartheta}\,\big)\,.
\label{TcompGR}
\ee
Now the conservation of the Energy-Momentum tensor, $\trd_{\mu}T^{\mu}{}_{\nu}$,  boils down to   a single equation:
\be
\frac{\rd~}{\rd r}\left(T^{r}{}_{r}\right)+\frac{2}{r}\left(T^{r}{}_{r}-T^{\vartheta}{}_{\vartheta}\right)+\frac{~B^{\prime}}{2B}\left(T^{r}{}_{r}-T^{t}{}_{t}\right)=0\,.
\label{conEM}
\ee
The Einstein Field Equations, or (\ref{EFE2}),  reduce to
\be
\ba{llcll}
R_{tt}&=&\frac{~B^{\prime\prime}}{2A}-\frac{~B^{\prime}}{4A}\left(\frac{~A^{\prime}}{A}+\frac{~B^{\prime}}{B}\right)+\frac{1}{r}\frac{~B^{\prime}}{A}
&=&-4\pi G B\left(T^{t}{}_{t}-T^{r}{}_{r}-2T^{\vartheta}{}_{\vartheta}\right)\,,\\
R_{rr}&=&-\frac{~B^{\prime\prime}}{2B}+\frac{~B^{\prime}}{4B}\left(\frac{~A^{\prime}}{A}+\frac{~B^{\prime}}{B}\right)+\frac{1}{r}\frac{~A^{\prime}}{A}
&=&-4\pi G A\left(T^{t}{}_{t}-T^{r}{}_{r}+2T^{\vartheta}{}_{\vartheta}\right)\,,\\
R_{\vartheta\vartheta}&=&1+\frac{~r}{2A}\left(\frac{~A^{\prime}}{A}-\frac{~B^{\prime}}{B}\right)-\frac{1}{A}&=&-4\pi G r^{2}\left(T^{t}{}_{t}+T^{r}{}_{r}\right)\,,
\ea
\ee
which are linearly  equivalent to
\be
\ba{l}
\frac{\rd~}{\rd r}\left[r\left(1-\frac{1}{A}\right)\right]=\frac{rA^{\prime}}{A^{2}}+1-\frac{1}{A}=-8\pi G  r^{2}T^{t}{}_{t}\,,\\
\frac{\rd~}{\rd r}\ln(AB)=\frac{A^{\prime}}{A}+\frac{B^{\prime}}{B}=-8\pi G A r(T^{t}{}_{t}-T^{r}{}_{r})\,,\\
\frac{~B^{\prime\prime}}{B}-\frac{~B^{\prime}}{2B}\left(\frac{A^{\prime}}{A}+\frac{B^{\prime}}{B}\right)-\frac{1}{r}\left(\frac{A^{\prime}}{A}-\frac{B^{\prime}}{B}\right)=16\pi G A T^{\vartheta}{}_{\vartheta}\,.
\ea
\label{three}
\ee
The first equation can be integrated to give
\be
\half r\left(1-\frac{1}{A}\right)=-G\int_{0}^{r}\rd r^{\prime} 4\pi r^{\prime 2} T^{t}{}_{t}(r^{\prime})\,,
\ee
where we have assumed `regularity' at the origin,
\be
\lim_{r\rightarrow 0}  r\left(1-\frac{1}{A}\right)=0\,.
\label{regularGR}
\ee
This fixes the function $A(r)$,
\be
A(r)=\frac{1}{1-\frac{\,2GM(r)}{r}}\,,
\label{solA}
\ee
for which we have defined
\be
M(r):=-\int_{0}^{r}\rd r^{\prime} 4\pi r^{\prime}{}^{2\,} T^{t}{}_{t}(r^{\prime})\,.
\label{MassGR}
\ee
The regularity condition~(\ref{regularGR}) is then equivalent to
\be
\lim_{r\rightarrow0}M(r)=0\,.
\label{regularGRM}
\ee 
Furthermore,  the positive energy (density) condition implies $T^{tt}\geq 0$, such that,  owing to the convention of the mostly plus signature of the metric~(\ref{metric}),  $M(r)$ is generically positive. 
 Similarly, assuming the `flat' boundary condition at infinity,
\be
\lim_{r\rightarrow \infty}A(r)B(r)=1\,,
\label{AB1}
\ee
the second equation in (\ref{three}) can be integrated to fix $B(r)$,
\be
B(r)=\left[1-\frac{\,2GM(r)}{r}\right]\exp\left[8\pi G\int_{r}^{\infty}\rd r^{\prime} r^{\prime}A(r^{\prime})\left\{T^{t}{}_{t}(r^{\prime})-T^{r}{}_{r}(r^{\prime})\right\}\right]\,,
\label{solB}
\ee
such that the metric takes the final form:
\be
\rd s^{2}=-e^{-2\Delta(r)}\normalsize{\left(1-\frac{\,2GM(r)}{r}\right)\rd{t}^{2}}+\frac{{\rd r}^{2}}{\,1-\frac{\,2GM(r)}{r}\,}+r^{2}\rd\Omega^{2}\,.
\ee
Here $M(r)$ is given by (\ref{MassGR}) and $\Delta(r)$ is defined by 
\be
\Delta(r):=4\pi G\int_{r}^{\infty}\rd r^{\prime}\, 
\frac{\big\{T^{r}{}_{r}(r^{\prime})-T^{t}{}_{t}(r^{\prime})\big\}r^{\prime}}{1-\frac{\,2GM(r^{\prime})}{r^{\prime}}}\,.
\ee
Finally, we show that, up to the first and the second relations in (\ref{three}) and their solutions, (\ref{solA}), (\ref{solB}),  the third  relation is equivalent to  the conservation of the Energy-Momentum tensor~(\ref{conEM}). For this, we solve for $T^{r}{}_{r}$ and $T^{r}{}_{r}-T^{t}{}_{t}$ from the first and the second relations in (\ref{three}),
\be
\ba{ll}
T^{r}{}_{r}=\frac{1}{8\pi G}\left(\frac{B^{\prime}}{ABr}+\frac{1}{Ar^{2}}-\frac{1}{r^{2}}\right)\,,\quad&\qquad
T^{r}{}_{r}-T^{t}{}_{t}=\frac{1}{8\pi GAr}\left(\frac{A^{\prime}}{A}+\frac{B^{\prime}}{B}\right)\,,
\ea
\ee
and substitute these two expressions into the left-hand side of the conservation relation~(\ref{conEM}), to obtain
\be
\ba{l}
\frac{\rd~}{\rd r}\left(T^{r}{}_{r}\right)+\frac{2}{r}\left(T^{r}{}_{r}-T^{\vartheta}{}_{\vartheta}\right)+\frac{~B^{\prime}}{2B}\left(T^{r}{}_{r}-T^{t}{}_{t}\right)\\=
\frac{1}{8\pi GAr}\left[\frac{~B^{\prime\prime}}{B}-\frac{~B^{\prime}}{2B}\left(\frac{A^{\prime}}{A}+\frac{B^{\prime}}{B}\right)-\frac{1}{r}\left(\frac{A^{\prime}}{A}-\frac{B^{\prime}}{B}\right)-16\pi G A T^{\vartheta}{}_{\vartheta}\right]\,.
\ea
\ee
This  result  clearly establishes the equivalence between  the Energy-Momentum conservation equation~(\ref{conEM}) and the third relation in (\ref{three}).\\

\newpage
\noindent Some comments are in order.  
\begin{itemize}
\item[--]
When the matter is localized up to a finite radius $r_{c}$, such that outside this radius, ${r>r_{c}}$,    we have    $T_{\mu\nu}(r)=0$ and $\Delta(r)=0$, we   recover the Schwarzschild solution, in which the mass 
  agrees with the ADM mass~\cite{Arnowitt:1962hi} and     from (\ref{MassGR})  is further   given by the volume integral,
\be
M=-\int_{0}^{r_{c}}\rd r\, 4\pi r^{2\,} T^{t}{}_{t}(r)=-\int_{0}^{\infty}\rd r\, 4\pi r^{2\,} T^{t}{}_{t}(r)\,.
\label{MassGR2}
\ee
However, this differs from the Noether charge~(\ref{NoetherchargeGR}) of the time translational Killing vector,
\be
M\neq -Q[\partial_{t}]=-\int_{0}^{\infty}\rd r\, 4\pi r^{2\,}e^{-\Delta(r)\,} T^{t}{}_{t}(r)\,,
\ee
since  from (\ref{regularGR}) the integral measure is nontrivial,
\be
\sqrt{-g}=e^{-\Delta(r)\,}r^{2}\sin\vartheta\neq r^{2}\sin\vartheta\,.
\label{sqrtgneq}
\ee
This discrepancy and its  remedy by an extra surface integral    are rather well known, see~\cite{Wald:1993nt,Iyer:1994ys,Iyer:1995kg,Kim:2013zha,Hyun:2014kfa}.

\item[--]If there is a spherical void in which $T_{\mu\nu}=0$  for $r_{1}<r<r_{2}$,  both $M(r)$ and $\Delta(r)$ become   constant inside the void as $M(r)=M(r_{1})$ and $\Delta(r)=\Delta(r_{2})$.  After a constant rescaling of the time, $t_{\rm{new}}=e^{-\Delta(r_{2})}t$, the local geometry inside the void coincides  precisely with the Schwarzschild solution.  We note that  the mass is determined through  the integral over $0<r<r_{1}$ only and is independent of the matter  distribution outside, $r>r_{2}$. While this is certainly  true in Newtonian gravity (namely the iron sphere theorem),  if we solved the vacuum Einstein Field Equations with vanishing Energy-Momentum tensor inside the void, we would merely recover  the Schwarzschild geometry in accordance with Birkhoff's theorem. Nevertheless it would be hard to conclude that  the constant mass parameter  is unaffected by the outer region.

\item[--]The radial derivative of $B(r)$  amounts to   the gravitational acceleration for a circular geodesic~\cite{Ko:2016dxa},
\be
r\left(\frac{\rd\vartheta}{\rd t}\right)^{\!2}=-\frac{1}{2}\frac{\rd g_{tt}(r)}{\rd r}=\half B^{\prime}=\left[\frac{GM(r)}{r^{2}}+4\pi Gr T^{r}{}_{r}(r)\right]e^{-2\Delta(r)}\,.
\label{acc}
\ee
Again, inside a void or the outer vacuum region,   we may absorb the constant factor of $e^{-2\Delta(r)}$ into the rescaled time,  and thus recover   the Keplerian acceleration, 
\be
r\left(\frac{\rd\vartheta}{\rd t}\right)^{2}=\frac{GM}{r^{2}}\,.
\ee

\end{itemize}




{{

}}

\end{document}